\documentclass[12pt]{article}
\pdfoutput=1

\usepackage{amsmath,amssymb,amsfonts,graphicx,amsfonts}
\usepackage{textcomp,mathcomp}
\usepackage{xcolor}
\usepackage[hidelinks]{hyperref}
\usepackage{url}
\usepackage{ascmac}
\usepackage{physics}
\usepackage{comment}

\allowdisplaybreaks[4]

\date{}
\begin{document}
\begin{flushright}
YITP-23-125
%\\
%\today\\
\end{flushright}

\vspace{0.1cm}

\begin{center}

{\Large On thermal transition in QCD}

\end{center}
\vspace{0.1cm}
\vspace{0.1cm}
\begin{center}

Masanori Hanada$^a$ and Hiromasa Watanabe$^b$

\end{center}
\vspace{0.3cm}

\begin{center}

{\small

$^a$School of Mathematical Sciences, Queen Mary University of London\\
Mile End Road, London, E1 4NS, United Kingdom

$^b$Yukawa Institute for Theoretical Physics, Kyoto University\\
Kitashirakawa Oiwakecho, Sakyo-ku, Kyoto 606-8502, Japan

\vspace{5pt}
Email: 
\url{m.hanada@qmul.ac.uk},\;
\url{hiromasa.watanabe@yukawa.kyoto-u.ac.jp}
}
\end{center}

\vspace{0.5cm}

\begin{center}
  {\bf Abstract}
\end{center}

We describe how the general mechanism of partial deconfinement applies to large-$N$ QCD and the partially-deconfined phase inevitably appears between completely-confined and completely-deconfined phases. Furthermore, we propose how the partial deconfinement can be observed in the real-world QCD with the SU(3) gauge group. For this purpose, we employ lattice configurations obtained by the WHOT-QCD collaboration and examine our proposal numerically. In the discussion, the Polyakov loop plays a crucial role in characterizing the phases, without relying on center symmetry, and hence, we clarify the meaning of the Polyakov loop in QCD at large $N$ and finite $N$.

Both at large $N$ and finite $N$, the complete confinement is characterized by the Haar-random distribution of the Polyakov line phases. Haar-randomness, which is stronger than unbroken center symmetry, indicates that Polyakov loops in any nontrivial representations have vanishing expectation values, and deviation from the Haar-random distribution at higher temperatures is quantified with the loops. We discuss that the transitions separating the partially-deconfined phase are characterized by the behaviors of Polyakov loops in various representations. The lattice QCD data provide us with the signals exhibiting two different characteristic temperatures: deconfinement of the fundamental representation and deconfinement of higher representations. As a nontrivial test for our proposal, we also investigate the relation between partial deconfinement and instanton condensation and confirm the consistency with the lattice data.

To make the presentation more easily accessible, we provide a detailed review of the previously known aspects of partial deconfinement.

\newpage
\tableofcontents

%%%%%%%%%%%%
%%%%%%%%%%%%
\section{Introduction}\label{sec:introduction}
\hspace{0.51cm}
%%%%%%%%%%%%
%%%%%%%%%%%%
Confinement/deconfinement transition in gauge theory~\cite{Polyakov:1978vu,Susskind:1979up} has been investigated for a long time for the understanding of the quark-gluon plasma in QCD under extreme conditions and the description of black holes via gauge/gravity duality. 
Focusing on real-world QCD with three colors, it is widely believed that there is no phase transition at zero chemical potential; rather, confined and deconfined phases are believed to be connected by a rapid crossover. This belief is based on lattice QCD simulations, e.g., Ref.~\cite{Aoki:2006we}, and the apparent lack of symmetry that could characterize a possible phase transition. In this paper, we argue that this belief may be wrong and there might be a phase transition of third or higher order. 

\begin{figure}[htbp]
\begin{center}
\scalebox{0.4}{
\includegraphics{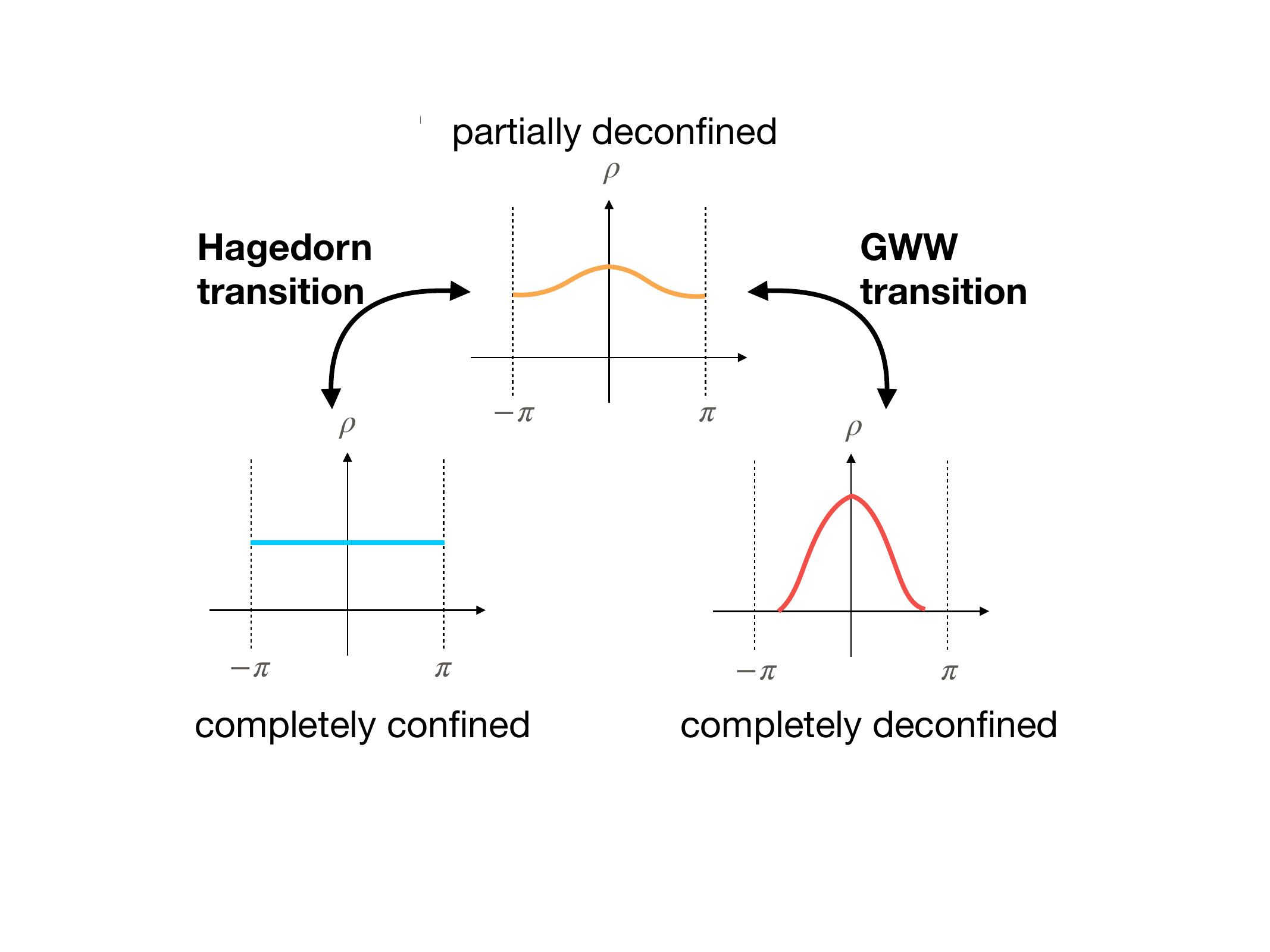}}
\end{center}
\caption{Completely-confined, partially-deconfined, and completely-deconfined phases correspond to uniform, non-uniform but jointed, and disjointed distributions of Polyakov line phases, respectively. 
This figure is a slight modification of the one from Ref.~\cite{Hanada:2019rzv}.
}\label{fig:Polyakov_distribution}
\end{figure}

\begin{figure}[htbp]
\begin{center}
\scalebox{0.4}{
\includegraphics{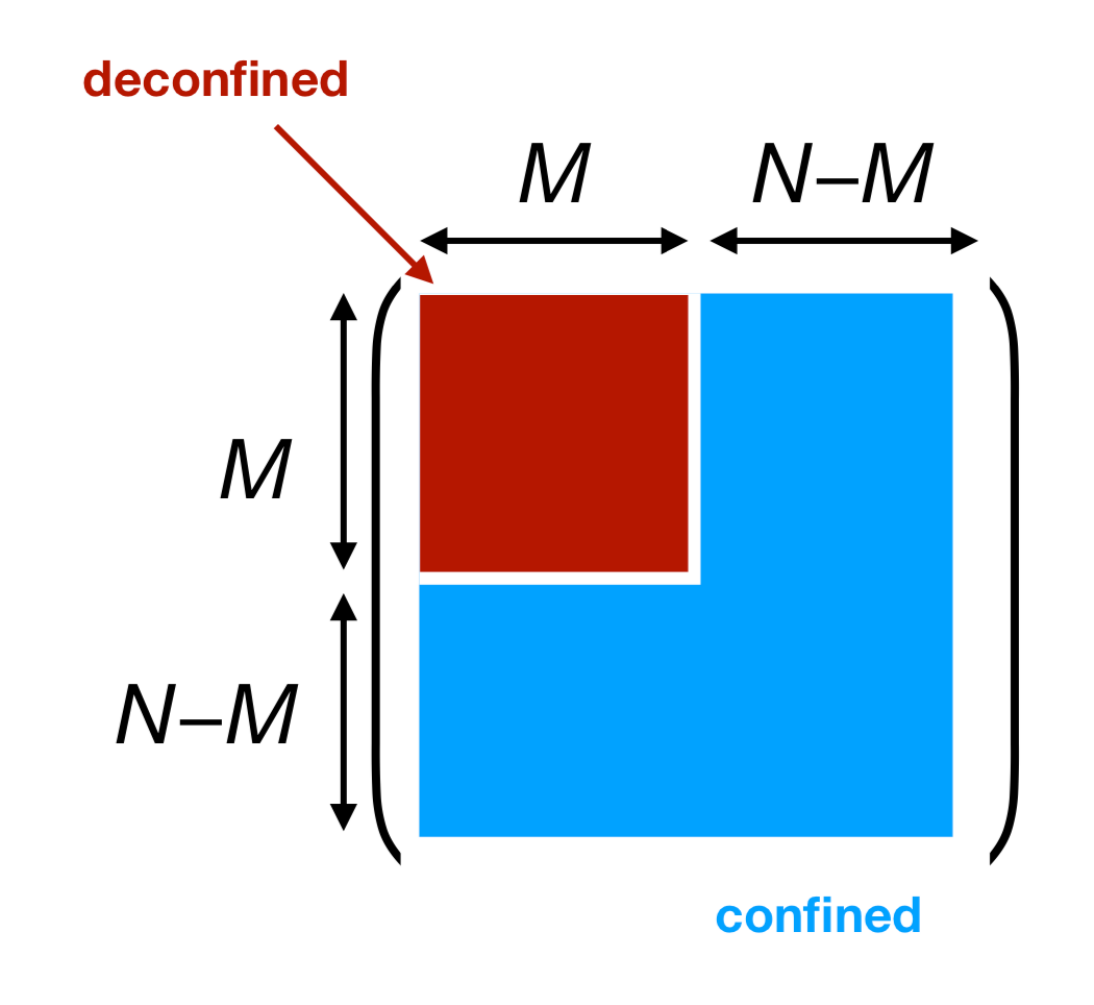}}
\end{center}
\caption{Partial deconfinement in the gauge sector and adjoint matters. 
Only the $M\times M$-block shown in red is deconfined. 
The consistency of this cartoon picture with the gauge singlet constraint is discussed in the main text. 
This figure is taken from Ref.~\cite{Hanada:2019czd}. 
}\label{fig:matrix}
\end{figure}

\begin{figure}[htbp]
\begin{center}
\scalebox{0.4}{
\includegraphics{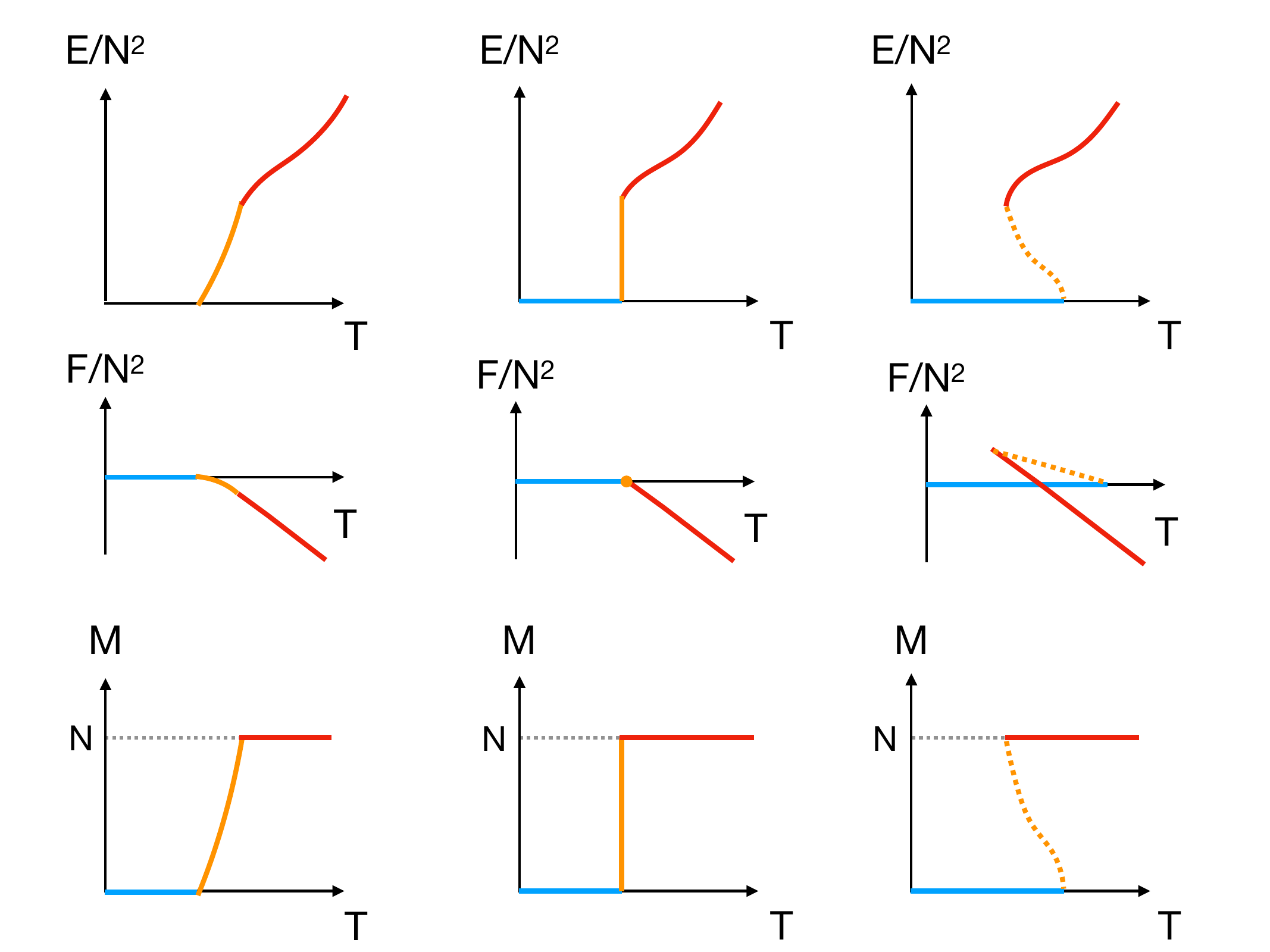}}
\end{center}
\caption{Basic three patterns of confinement/deconfinement transition. 
Blue, red, and orange lines correspond to completely-confined, completely-deconfined, and partially-deconfined phases, respectively. Solid and dotted lines are the minimum and maximum of free energy at each temperature.
}\label{fig:three=patterns}
\end{figure}

To motivate this seemingly bizarre proposal, let us consider the 't Hooft large-$N$ limit of SU($N$) gauge theory.
Many large-$N$ gauge theories exhibit confinement/deconfinement transition. In theories such as pure Yang-Mills or 4d $\mathcal{N}=4$ super Yang-Mills theory, the Polyakov loop is commonly used as an order parameter associated with the $\mathbb{Z}_N$ center symmetry. The $N$ phases of Polyakov loop $\theta_1,\cdots,\theta_N$ form a continuous distribution $\rho(\theta)$ between $-\pi$ and $+\pi$ in the large-$N$ limit. The $\mathbb{Z}_N$ center acts on $\theta$ as a constant shift, and hence the confined phase is characterized by the uniform distribution $\rho(\theta)=\frac{1}{2\pi}$, while the deconfinement phase is characterized by non-uniform distribution. An interesting observation~\cite{Aharony:2003sx,Sundborg:1999ue} is that there are two types of non-uniform distributions --- jointed ($\rho(\theta)>0$ everywhere) and disjointed ($\rho(\theta)=0$ in a finite interval) --- separated by the `Gross-Witten-Wadia (GWW) transition.'~\footnote{The original setup of GWW is the two-dimensional lattice gauge theory at large $N$, which has a phase transition associated with the formation of a gap in the distribution of phases of the plaquette. The GWW model can describe confinement/deconfinement transition in a class of weakly-coupled systems including pure Yang-Mills on S$^3$, modulo the difference of the interpretation of the unitary variable (plaquette or Polyakov loop). Even when the confinement/deconfinement transition is not precisely described in the GWW model, we use the word `GWW transition' for the formation of a gap, although it may be a little bit of abuse of terminology.} 
Because the jointed and disjointed phases cannot be distinguished in terms of the center symmetry, it is reasonable to expect that the Polyakov loop has some meaning that is independent of the center symmetry. 
This expectation is strengthened by noticing that the Polyakov loop can characterize the phase transition in QCD despite the explicit breaking of center symmetry due to quarks in the fundamental representation~\cite{Schnitzer:2004qt}. As we will explain in detail, the answer is given by \textit{partial deconfinement}; see Fig.~\ref{fig:Polyakov_distribution}, Fig.~\ref{fig:matrix}, and Fig.~\ref{fig:three=patterns}. 

Another convenient way to characterize confined and deconfined phases is to use the $N$-dependence of energy and entropy. In the confined phase, because individual color degrees of freedom are not visible (they are `confined'), the energy and entropy typically scale as $N^0$ up to zero-point contributions. On the other hand, in the deconfined phase, the scaling is typically of order $N^2$ because color degrees of freedom are `deconfined.' 
What if the energy is set to $\epsilon N^2$, where $\epsilon$ is small but of order $N^0$? The energy is too high to be in the confined phase, but not high enough to deconfine all color degrees of freedom. In many cases including large-$N$ QCD, \textit{partial deconfinement} takes place, i.e., an SU($M$) subgroup of SU($N$) deconfines, where, roughly speaking, $M\sim\sqrt{\epsilon}N$ (see Fig.~\ref{fig:matrix}). Originally, partial deconfinement was introduced for 4d maximal super Yang-Mills theory compactified on S$^3$, in order to explain the QFT-dual of the small black hole in AdS$_5\times$S$^5$~\cite{Hanada:2016pwv}. Later, it was suggested that partial deconfinement is a rather generic phenomenon in many large-$N$ gauge theories~\cite{Berenstein:2018lrm,Hanada:2018zxn}. Subsequent investigations provided analytic~\cite{Hanada:2019czd,Hanada:2019kue,Hanada:2021ksu} and numerical~\cite{Bergner:2019rca,Watanabe:2020ufk,Hanada:2021ksu,Gautam:2022exf} evidence, and clarified the underlying mechanism~\cite{Hanada:2020uvt,Hanada:2021ipb,Hanada:2021swb}.  

\begin{figure}[htbp]
\begin{center}
\scalebox{0.3}{
\includegraphics{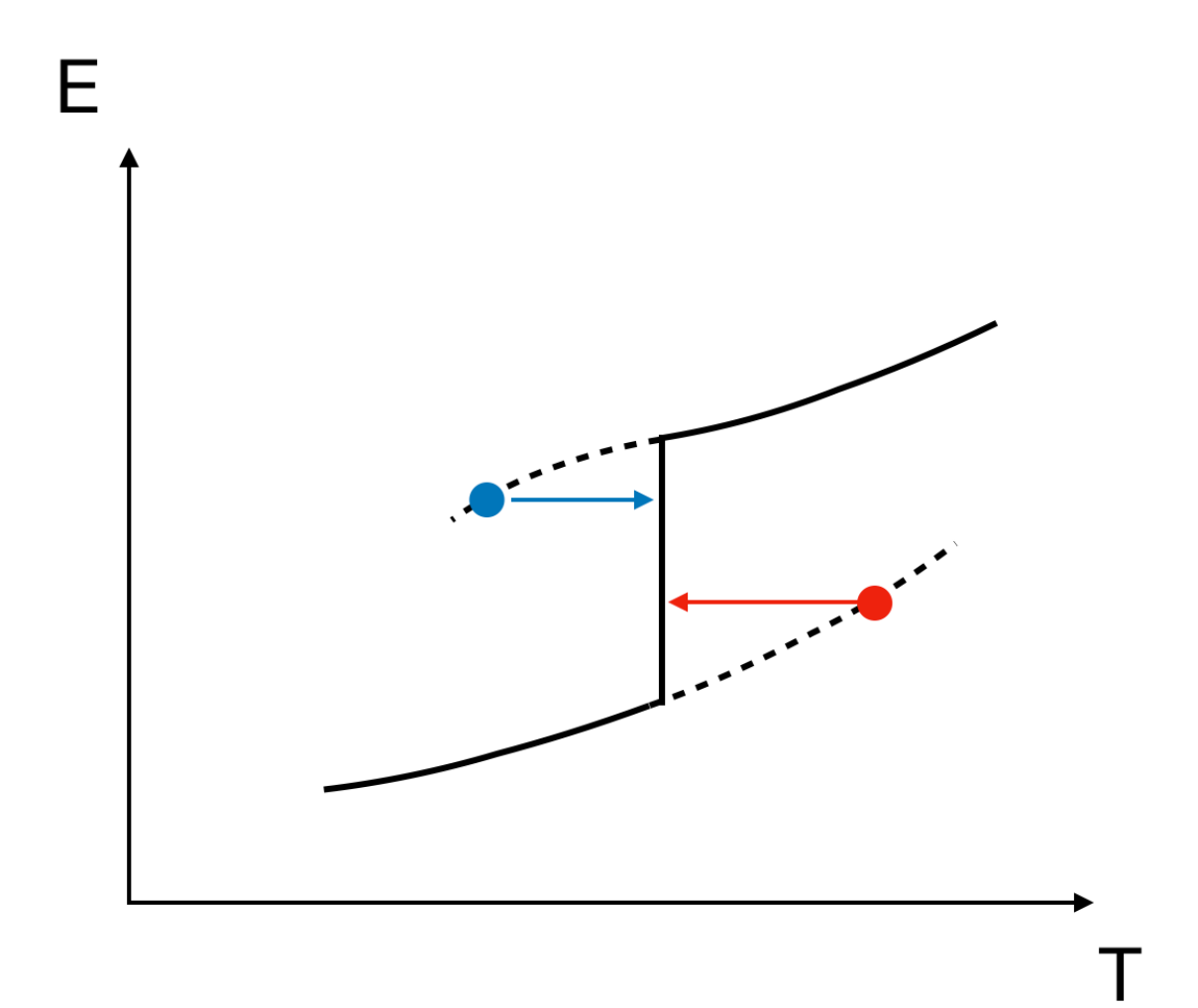}}
\end{center}
\caption{First-order transition between water (liquid water) and ice (solid water) at 1 atm. At 0\textcelsius, water and ice coexist. 
Supercooled water below 0\textcelsius\  and superheated ice above 0\textcelsius\ are not stable, and they turn to a mixture of water and ice just by a small perturbation as shown by arrows in Fig.~\ref{fig:water}. (Energy is fixed because the change takes place quickly.)
This figure is taken from Ref.~\cite{Hanada:2018zxn}.
}\label{fig:water}
\end{figure}

\begin{figure}[htbp]
\begin{center}
\scalebox{0.3}{
\includegraphics{matrix.pdf}}
\scalebox{0.35}{
\includegraphics{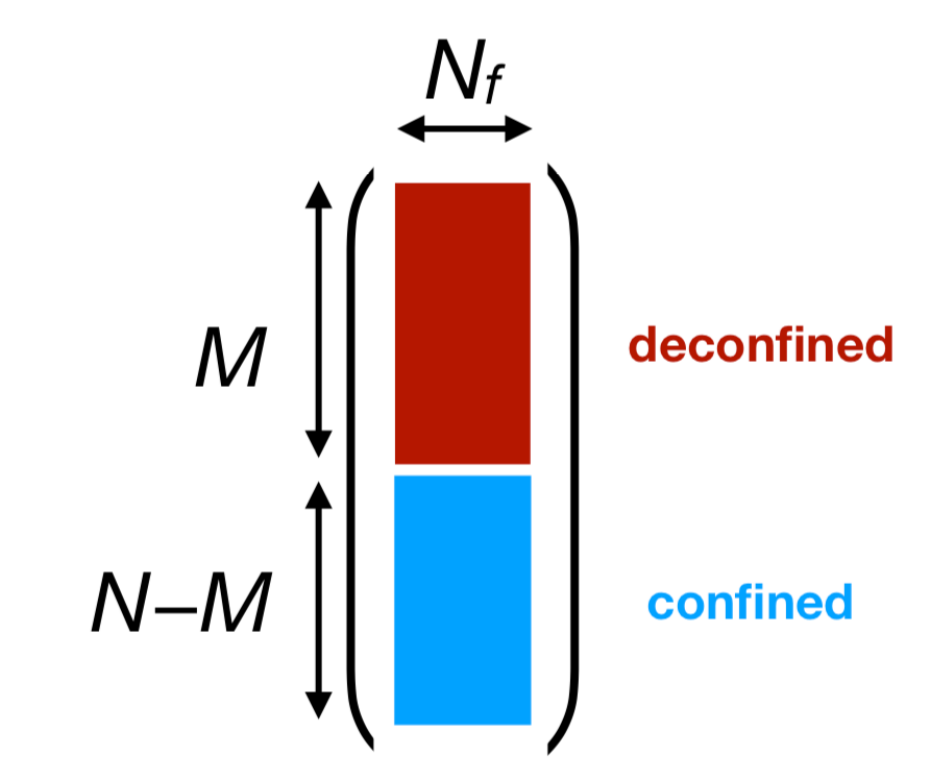}}
\end{center}
\caption{Partial deconfinement in weakly-coupled QCD on S$^3$~\cite{Hanada:2019czd}. 
[Left] the gauge sector and matters in the adjoint representation, [Right] matters in the fundamental representation. $M$ colors shown in red are deconfined. 
The consistency of this cartoon picture with the gauge singlet constraint is discussed in the main text. 
This figure is taken from Ref.~\cite{Hanada:2019czd}
}\label{fig:QCD-deconf-pattern}
\end{figure}

Intuitively, partial deconfinement can be phrased as the coexistence of two phases (confined phase and deconfined phase) in the internal space (specifically, color space), rather than the coordinate space. A common example of the coexistence of two phases in coordinate space is observed in the first-order transition between water (liquid water) and ice (solid water) at 1 atm (Fig.~\ref{fig:water}).
Supercooled water below 0\textcelsius\  and superheated ice above 0\textcelsius\ are not stable, and they turn to a mixture of water and ice just by a small perturbation as shown by arrows in Fig.~\ref{fig:water}. The temperature of the mixture is 0\textcelsius\ due to the local nature of the interaction.\footnote{Because of the locality, each small region is described by the canonical thermodynamics. Free energy is defined locally. They are minimized such that water or ice is favored. The interaction at the interface of two phases can be ignored if the interaction is local and the spatial volume is large.}
However, we can recognize that the coexistence occurs in color space as well and the above picture does not work there anymore.
The interactions in the color space are all-to-all, and this nonlocal feature makes the system robust against the small perturbation.
Furthermore, the way of splitting into two sectors depends on the representations, as depicted in Fig.~\ref{fig:QCD-deconf-pattern} for the case of adjoint and fundamental representations in large-$N$ QCD. 
Taking these into account, the temperature is not necessarily fixed during the process of transition as depicted in Fig.~\ref{fig:three=patterns}, and a variety of equations of states is allowed in the intermediate region. 
In retrospect, such two-phase-coexistence in gauge theory has been known for a long time: a system of $N$ indistinguishable bosons is described by gauge theory with S$_N$-gauge symmetry that permutes bosons, and two phases --- superfluid (Bose-Einstein condensate) and normal fluid --- coexist in the internal space. In fact, these two kinds of coexistence take place due to the same mechanism~\cite{Hanada:2020uvt}.\footnote{
In this sense, Einstein was the first person who considered non-Abelian gauge theory, took the large-$N$ limit, and took the weak-coupling limit. We can also see that Feynman introduced an equivalent of the Polyakov loop and characterized the Bose-Einstein condensation by the uniform distribution of the phases. See Sec.~\ref{sec:underlying_mechanism} for details. 
}   

Due to partial deconfinement, there are three phases: completely-confined phase ($M=0$), partially-deconfined phase ($0<M<N$), and completely-deconfined phase ($M=N$). Even if the theory possesses the center symmetry or not, three phases can be distinguished by the distribution of the phases of Polyakov line~\cite{Hanada:2020uvt,Hanada:2018zxn}: completely-confined, partially-deconfined, and completely-deconfined phases correspond to uniform, non-uniform but jointed, and disjointed distributions of Polyakov line phases, respectively, as depicted in Fig.~\ref{fig:Polyakov_distribution}. When the system possesses center symmetry, the transition from the completely-confined phase to the partially-deconfined phase is associated with the breaking of center symmetry, which is a traditional definition of deconfinement for theories in this class. Since, at the critical point, the formation of long strings is involved, we call it the Hagedorn transition. The transition from the partially-deconfined phase to the completely-deconfined phase is associated with the formation of a gap in the distribution of Polyakov line phases. We call it the Gross-Witten-Wadia (GWW) transition~\cite{Gross:1980he,Wadia:2012fr}.

In this paper, we explain how partial deconfinement can take place in large-$N$ QCD. Furthermore, we suggest that partial deconfinement takes place in SU(3) QCD that describes our universe. To make the presentation as self-contained as possible, we provide a detailed review of the previously known aspects of partial deconfinement.

The organization of this paper is as follows. In Sec.~\ref{sec:Hermitian-matrix-model}, we use gauged Hermitian matrix models as the simplest examples to explain the essence of partial deconfinement. Then, we apply the same idea to QFT in Sec.~\ref{sec:QFT}. 
Potential generalization to finite $N$ is discussed in Sec.~\ref{sec:Finite-N}. 
The majority of these three sections review the results provided in several references in the past. Because crucial results appeared in different papers and there was no systematic review, we hope that the review parts make the topic more accessible. 
Sec.~\ref{sec:YM-continuum-limit} and Sec.~\ref{sec:Finite-N} contain new materials.
In Sec.~\ref{sec:SU(3)-QCD}, we analyze the configurations from lattice gauge theory simulations and discuss the consistency with our conjecture. 

We have included several appendices in this paper to supplement the main text.
Appendix~\ref{sec:Haar-random-distribution} and Appendix~\ref{sec:character} have large overlap with the letter version~\cite{Hanada:2023krw}. They are included here to make this paper self-contained. 

An important part of this paper is based on collaboration with Hiroki Ohata and Hidehiko Shimada. They contributed significantly to the analysis of lattice QCD data, including a key idea of the use of the character. 
%%%%%%%%%%%%
%%%%%%%%%%%%
\section{Partial deconfinement in large-$N$ Hermitian matrix model}\label{sec:Hermitian-matrix-model}
\hspace{0.51cm}
%%%%%%%%%%%%
%%%%%%%%%%%%
In this section, we review partial deconfinement by taking a large-$N$ Hermitian matrix model as an example. 
This example contains all the essences that can also be applied to other theories. 
As a concrete example, we start with a bosonic SU($N$) matrix model with the following Hamiltonian:
\begin{align}
\hat{H}
=
{\rm Tr}\left(
\frac{1}{2}\hat{P}_I^2
+
\frac{1}{2}\hat{X}_I^2
-
\frac{g_{\rm YM}^2}{4}[\hat{X}_I,\hat{X}_J]^2
\right).  
\label{Bosonic-MM-Hamiltonian}
\end{align}
Here $I,J=1,2,\cdots,D$, where $D\ge 2$. 
Each $\hat{X}_I$ has $N^2$ components $\hat{X}_{I,ij}$, where $i,j=1,2,\cdots,N$, that satisfy the Hermiticity condition $(\hat{X}_{I,ij})^\dagger=\hat{X}_{I,ji}$. 
We do not impose the traceless condition.
The operator $\hat{P}_{I,ij}$ is the conjugate momentum of $\hat{X}_{I,ji}$. 

We introduce the adjoint index running from 1 to $N^2$ by using the generators $\tau_\alpha$ of U($N$) normalized as ${\rm Tr}(\tau_\alpha\tau_\beta)=\delta_{\alpha\beta}$, $\sum_\alpha\tau_{\alpha,ij}\tau_{\alpha,kl}=\delta_{il}\delta_{jk}$. Then, the operators $\hat{P}_I$ and $\hat{X}_I$ are expressed as
\begin{align}
\hat{P}_{I,ij}
=
\sum_{\alpha=1}^{N^2}
\hat{P}_I^\alpha\tau_{\alpha,ij}, 
\qquad
\hat{X}_{I,ij}
=
\sum_{\alpha=1}^{N^2}
\hat{X}_I^\alpha\tau_{\alpha,ij}.  
\end{align}
These $\hat{X}_I^\alpha$ and $\hat{P}_I^\alpha$ are real, i.e., $(\hat{X}_I^\alpha)^\dagger = \hat{X}_I^\alpha$ and $(\hat{P}_I^\alpha)^\dagger = \hat{P}_I^\alpha$. 
The canonical commutation relation is\footnote{
Equivalently, 
\begin{align}
[\hat{X}_{I,ij},\hat{P}_{J,kl}]
=
i\delta_{IJ}\delta_{il}\delta_{jk}\, . 
\nonumber
\end{align}
}
\begin{align}
[\hat{X}^\alpha_I,\hat{P}^\beta_J]
=
i\delta_{IJ}\delta^{\alpha\beta}\, . 
\end{align}

The Hamiltonian \eqref{Bosonic-MM-Hamiltonian} is invariant under the adjoint action of SU($N$) defined by
\begin{align}
\hat{X}_{I,ij}
\to
(U\hat{X}_{I}U^{-1})_{ij}
=
\sum_{k,l}U_{ij}\hat{X}_{I,kl}U^{-1}_{lj}, 
\qquad
\hat{P}_{I,ij}
\to
(U\hat{P}_{I}U^{-1})_{ij}
=
\sum_{k,l}U_{ik}\hat{P}_{I,kl}U^{-1}_{lj}. 
\end{align}
We consider \textit{gauged} matrix model, i.e., this SU($N$) symmetry is gauged.
One way of gauging SU($N$) is to restrict the Hilbert space to SU($N$)-singlet states. 
By using the singlet Hilbert space (gauge-invariant Hilbert space) ${\cal H}_{\rm inv}$, the canonical partition function at temperature $T$ can be written as 
\begin{align}
Z(T)
&=
{\rm Tr}_{\mathcal{H}_{\rm inv}}\left(
e^{-\hat{H}/T}
\right). 
\label{eq:Z-H-inv-MM}
\end{align}
Equivalently, we can select the \textit{extended} Hilbert space with non-singlet states, which we denote by ${\cal H}_{\rm ext}$. 
The extended space is spanned by, for example, the coordinate eigenstates $\ket{X}$ that satisfy $\hat{X}_{I,ij}\ket{X}=X_{I,ij}\ket{X}$:
\begin{align}
\mathcal{H}_{\rm ext}
=
\textrm{Span}\{
\ket{X}|X\in\mathbb{R}^{DN^2}
\}. 
\end{align}
Note that $X$ consists of $DN^2$ real numbers $X_{I=1,\cdots,D}^{\alpha=1,\cdots,N^2}$. 
We can also use the momentum eigenstates $\ket{P}$ that satisfy $\hat{P}_{I,ij}\ket{P}=P_{I,ij}\ket{P}$:
\begin{align}
\mathcal{H}_{\rm ext}
=
\textrm{Span}\{
\ket{P}|P\in\mathbb{R}^{DN^2}
\}. 
\end{align}
The coordinate basis and momentum basis are related to each other by Fourier transformation. 
Gauge transformation acts on $\mathcal{H}_{\rm ext}$ as
\begin{align}
\ket{X}
\to
\ket{U^{-1} XU}\, , 
\qquad
\ket{P}
\to
\ket{U^{-1} PU}\, . 
\end{align}
The extended Hilbert space is convenient because the operators $\hat{X}$ and $\hat{P}$ are defined on this space. 
On the other hand, these operators are not defined on the singlet Hilbert space, because they map a singlet to a non-singlet.
In the extended Hilbert space, the gauged SU($N$) enforces that the states related by an SU($N$) transformation should be identified.
This is the same as in the usual treatment in the classical theory, and also in path-integral formalism: Field configurations that are related by gauge transformation should be identified.
The canonical partition function can be written as
\begin{align}
Z(T)
&=
\frac{1}{{\rm vol}G}\int_Gdg
{\rm Tr}_{{\cal H}_{\rm ext}}\left(
\hat{g}
e^{-\hat{H}/T}
\right). 
\label{eq:Z-H-ext-MM}
\end{align}
Here $G=\textrm{SU}(N)$ is the gauge group, $g$ is a group element, and $\hat{g}$ is the representation of $g$ acting on the extended Hilbert space ${\cal H}_{\rm ext}$ (in the case under consideration, the adjoint representation). The integral is taken with respect to the Haar measure.\footnote{
Some readers may be familiar with a similar expression for the case of the discrete gauge group. The insertion of $\hat{g}$ is regarded as a twisted boundary condition, and the discrete group is gauged by taking a sum over all twisted boundary conditions.  
}   
As shown in Appendix~\ref{sec:Hamiltonian-to-Lagrangian}, this expression is equivalent to the other expression \eqref{eq:Z-H-inv-MM} and this is directly related to the path integral with temporal gauge field $A_t$. 
The group element $g$ can be translated into the Polyakov line. 
It is easy to see that
\begin{align}
\hat{\pi}
\equiv
\frac{1}{\textrm{vol}G}\int_G dg
\hat{g}
\label{eq:projector}
\end{align}
is a projection operator from ${\cal H}_{\rm ext}$ to ${\cal H}_{\rm inv}$. 
By using this, \eqref{eq:Z-H-ext-MM} can also be written as 
\begin{align}
Z(T)
&=
{\rm Tr}_{{\cal H}_{\rm ext}}\left(
\hat{{\pi}}
e^{-\hat{H}/T}
\right). 
\end{align}
From a state $|\phi\rangle\in\mathcal{H}_{\rm ext}$, which can be non-singlet, we can obtain a singlet $\hat{\pi}|\phi\rangle\in\mathcal{H}_{\rm inv}$. 
By using this correspondence, we can describe the same physics by using $\mathcal{H}_{\rm ext}$ or $\mathcal{H}_{\rm inv}$. 
As we will see, the use of $\mathcal{H}_{\rm ext}$ can make physics much more transparent. 
%%%%%%%%%%%%
%%%%%%%%%%%%
\subsection{Partial deconfinement at weak coupling}\label{sec:Gaussian-Matrix-Model}
\hspace{0.51cm}
%%%%%%%%%%%%
%%%%%%%%%%%%
We start with the gauged Gaussian matrix model, i.e., the free limit ($g_{\rm YM}^2=0$) of the theory with the Hamiltonian \eqref{Bosonic-MM-Hamiltonian}, following Ref.~\cite{Hanada:2019czd}. 
Despite the lack of interactions, confinement/deconfinement transition can take place in the large-$N$ limit due to the SU($N$)-singlet constraint, and some qualitative features of strong coupling region can be captured~\cite{Sundborg:1999ue,Aharony:2003sx}.\footnote{
The same applies to Bose-Einstein condensation~\cite{Hanada:2020uvt}: the transition can take place in the large-$N$ limit due to the S$_N$-singlet constraint as pointed out by Einstein, and some qualitative features of strong coupling region can be captured as pointed out by Feynman. 
}  
The Hamiltonian is 
\begin{align}
\hat{H}_{\rm free}
=
{\rm Tr}
\sum_{I=1}^D
\left(
\frac{1}{2}\hat{P}_I^2
+
\frac{1}{2}\hat{X}_I^2
\right)
=
 {\rm Tr}
 \sum_{I=1}^D
 \left(
\hat{A}_I^\dagger\hat{A}_I
\right)
+
\frac{DN^2}{2}\, , 
\end{align}
where $\hat{A}_I^\dagger=\frac{\hat{X}_I-i\hat{P}_I}{\sqrt{2}}$ and $\hat{A}_I=\frac{\hat{X}_I+i\hat{P}_I}{\sqrt{2}}$ are creation and annihilation operators.
Equivalently, 
\begin{align}
\hat{H}_{\rm free}
=
{\rm Tr}
\sum_{I=1}^D
\left(
\frac{1}{2}(\hat{P}_I^\alpha)^2
+
\frac{1}{2}(\hat{X}_I^\alpha)^2
\right)
=
\sum_{I=1}^D\sum_{\alpha=1}^{N^2}
\hat{A}_I^{\alpha\dagger}\hat{A}_I^\alpha
+
\frac{DN^2}{2}. 
\end{align}
There are $DN^2$ harmonic oscillators that are subject to the gauge-singlet constraint or, equivalently, the identification of non-singlets that transform to each other via SU($N$) transformation. 
The ground state is the Fock vacuum $\ket{0}$ that is annihilated by any annihilation operator:
\begin{align}
\hat{A}_I^\alpha\ket{0}=0. 
\end{align} 

Generic states in the extended Hilbert space $\mathcal{H}_{\rm ext}$ can be obtained by acting creation operators. 
States in the singlet Hilbert space $\mathcal{H}_{\rm inv}$ can be obtained by acting the projection operator $\hat{\pi}$, or equivalently, by multiplying the traces of products of creation operators to the ground state. 

Due to the above construction in $\mathcal{H}_{\rm ext}$, SU($M$)-deconfined states can be obtained by acting $\hat{A}_{I,ij}^\dagger$ with $1\le i,j\le M$ to the ground state. Such states are invariant under $\textrm{SU}(N-M)\subset\textrm{SU}(N)$ that does not act nontrivially on the SU($M$)-sector. This unbroken symmetry in the extended Hilbert space plays a crucial role in partial deconfinement, as we will see in Sec.~\ref{sec:underlying_mechanism}. By definition, SU($M$)-deconfined states explain at least a part of the entropy of the system. The nontrivial point is that these states can explain the entire entropy, and hence, thermodynamically dominant. We can show this as follows:
\begin{enumerate}
\item
We calculate the entire entropy at energy $E$ without assuming partial deconfinement.  

\item
Besides, we calculate the entropy of the SU($M$)-deconfined states at energy $E$. 

\item
We confirm that the entire entropy comes from the SU($M$)-deconfined states.  

\end{enumerate}

\begin{figure}[htbp]
\begin{center}
\scalebox{0.12}{
\includegraphics{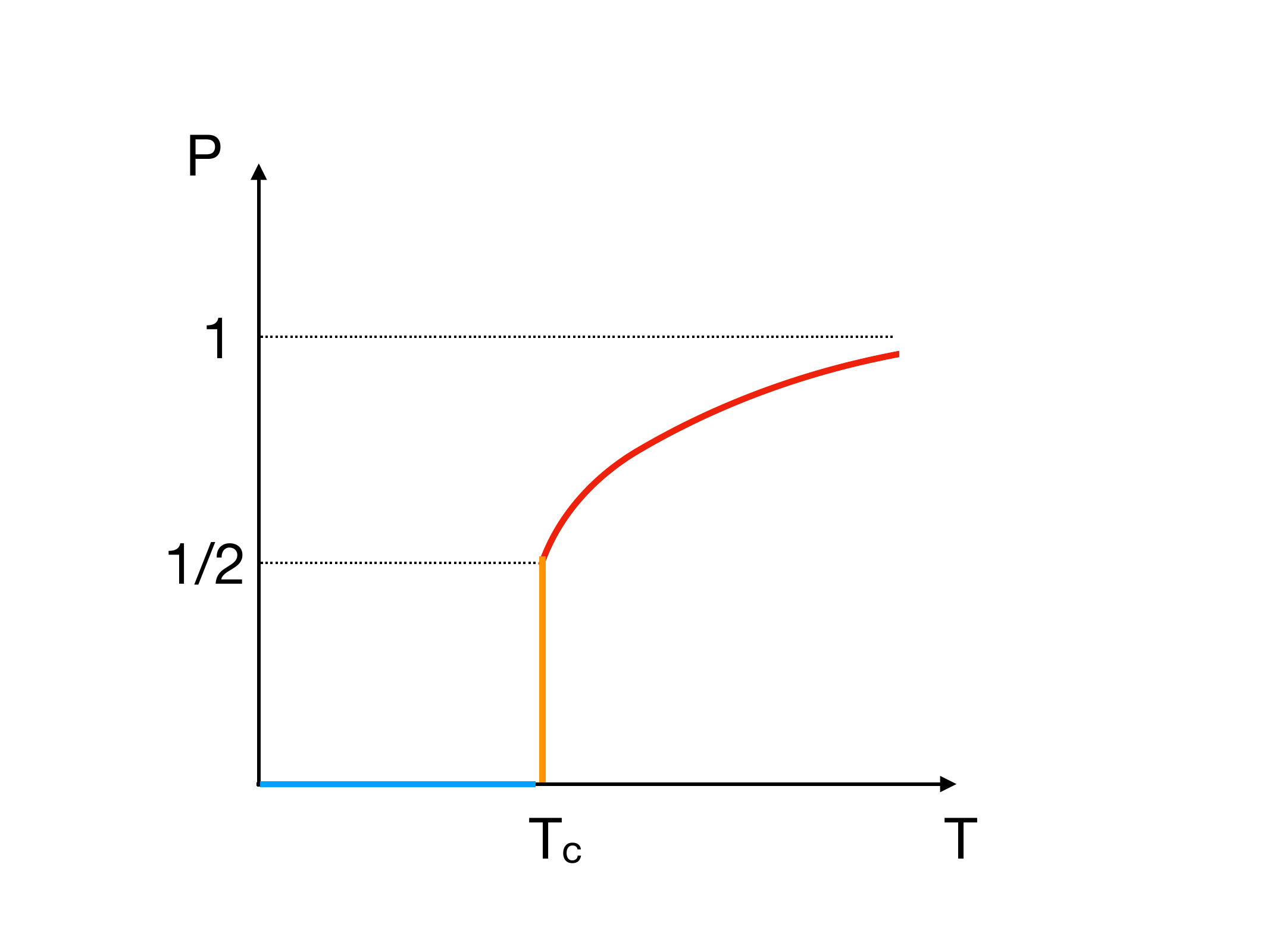}}
\scalebox{0.12}{
\includegraphics{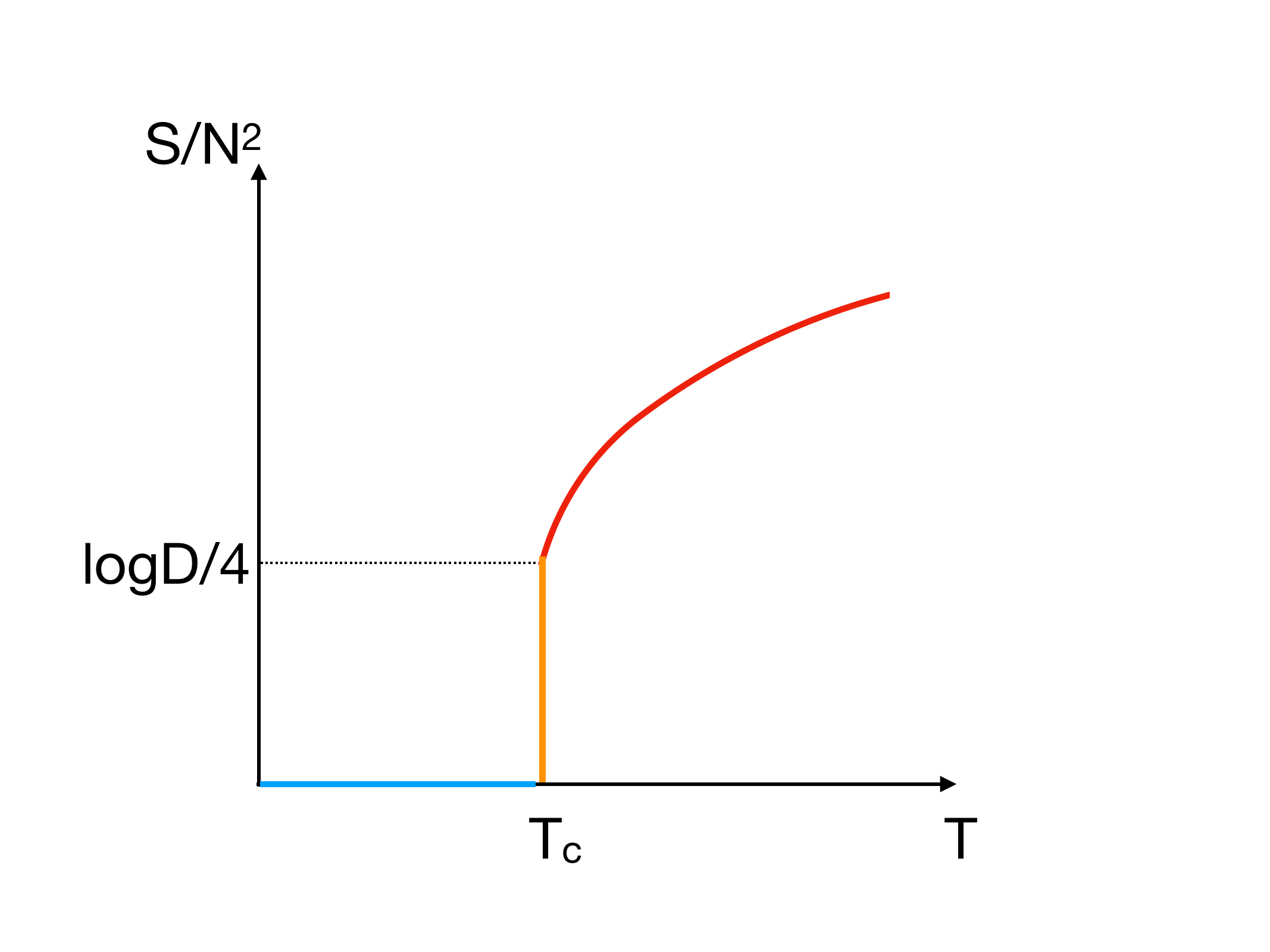}}
\scalebox{0.12}{
\includegraphics{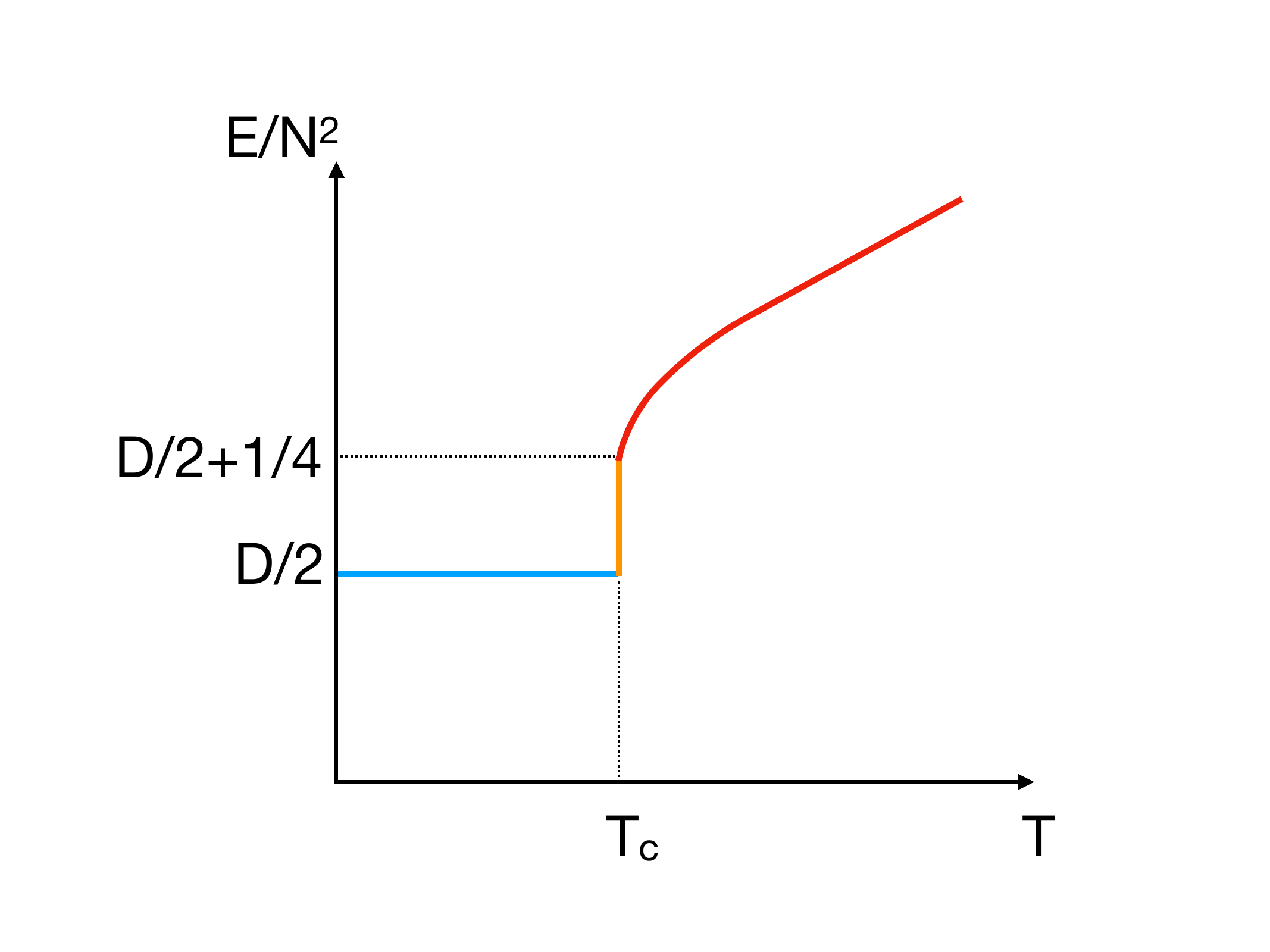}}
\end{center}
\caption{
Cartoon pictures of the temperature dependence of the Polyakov loop $P$, entropy $S$, and energy $E$ in the gauged Gaussian matrix model. 
Blue, orange, and red lines are identified with the completely-confined, partially-deconfined, and completely-deconfined phases, respectively. 
These pictures appeared originally in Ref.~\cite{Hanada:2019czd}.
}\label{fig:transition_GMM}
\end{figure}

Let us start with the step 1. The free energy can be calculated by using the path-integral formulation
and repeating similar computations in Refs.~\cite{Sundborg:1999ue,Aharony:2003sx}. The free energy is expressed  as~\cite{Hanada:2019czd}
\begin{eqnarray}
\beta F 
=
\frac{DN^2\beta}{2}
+
N^2\sum_{n=1}^\infty
\frac{1-Dx^n}{n}|u_n|^2,
\end{eqnarray}
in terms of the Polyakov loops $u_n=\frac{1}{N}{\rm Tr}{\cal P}^n$, where ${\cal P}$ is the Polyakov line, and $x=e^{-\beta}$. 
According to the saddle point condition, the stable configuration of $u_n$'s of the model should minimize the free energy. 
At $T<T_{\rm c}=\frac{1}{\log D}$, $|u_n|=0$ ($n\ge 1$) is favored. 
At $T=T_{\rm c}$, $|u_1|$ can take any value from 0 to $\frac{1}{2}$, 
while $u_2,u_3,\cdots$ remain zero, without changing the free energy.  
This is a first-order deconfinement transition.
In Fig.~\ref{fig:transition_GMM}, we have shown how the Polyakov loop $P$, entropy $S$ and energy $E$ depend on temperature. Notice that the energy and entropy are determined by the values of $u_n$'s, just as the free energy is.
The point $P=0$ is the onset of deconfinement in the sense that center symmetry breaks and entropy and energy (excluding the zero-point energy) become of order $N^2$. The point $P=\frac{1}{2}$ is called the Gross-Witten-Wadia (GWW) transition point.  

Let us focus on the deconfinement transition at $T=T_{\rm c}$. 
\begin{itemize}
\item
The Polyakov loop $P=|\frac{1}{N}{\rm Tr}{\cal P}|=|u_1|$ can take any value between 0 and $\frac{1}{2}$.\footnote{
Strictly speaking, there is an ambiguity associated with the center symmetry, 
namely the constant shift of all the phases. We fixed the center symmetry such that the Polyakov loop becomes 
real and non-negative, i.e. $P=|P|$. 
} 
Each value of $P$ corresponds to the different saddle point in the path integral.
From the microcanonical viewpoint, $P=0$ is the transition point from confinement to deconfinement, 
and $P=\frac{1}{2}$ is the GWW transition point. We will interpret them as the transitions from confinement to 
partial deconfinement, and from partial deconfinement to complete deconfinement.

\item
The energy can be written as 
\begin{eqnarray}
E(T=T_{\rm c}, P,N)
=
\frac{DN^2}{2}
+
N^2P^2. 
\label{eq:energy_MM}
\end{eqnarray}
The first term is the zero-point energy. 

\item
The entropy $S=\beta(E-F)$ is 
\begin{eqnarray}
S(T=T_{\rm c}, P,N)
=
N^2P^2\log D. 
\label{eq:entropy_MM}
\end{eqnarray}
\item
The distribution of the Polyakov line phases at the critical temperature is
\begin{eqnarray}
\rho(\theta)=
\frac{1}{2\pi}\left(
1+2P\cos\theta\right)
=
\left(
1-2P
\right) 
\cdot
\frac{1}{2\pi}
+
2P\cdot
\frac{1}{2\pi}\left(
1+\cos\theta
\right),
\label{eq:Gaussian_Polyakov_loop}
\end{eqnarray}
up to $1/N$ corrections. In particular, the distribution at the GWW transition is 
\begin{eqnarray}
\rho_{\rm GWW}(\theta; N)
=
\frac{1}{2\pi}
\left(
1+\cos\theta
\right). 
\end{eqnarray}

\end{itemize}

\begin{figure}[htbp]
\begin{center}
\scalebox{0.3}{
\includegraphics{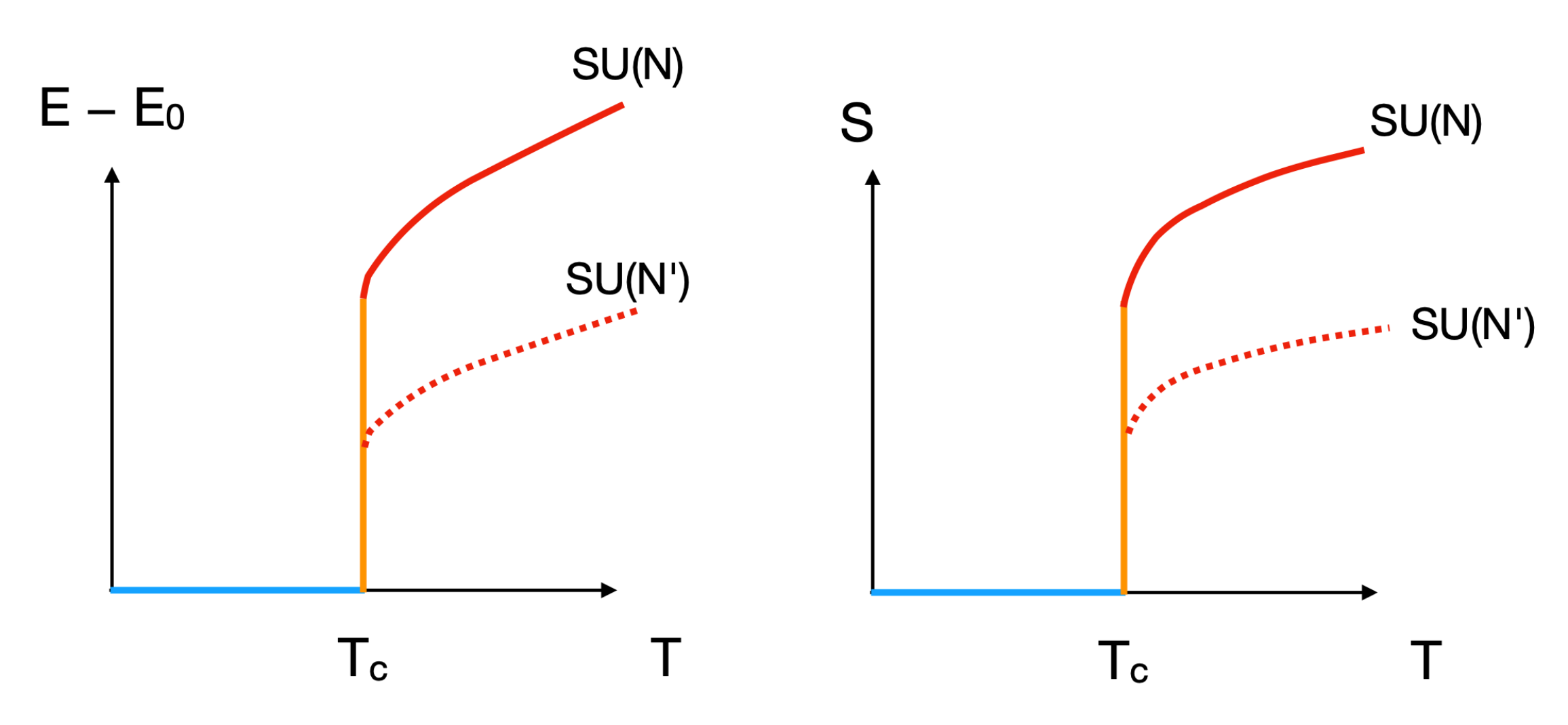}}
\end{center}
\caption{
With the absence of the interaction, SU($N$)-theory and SU($N'$)-theory behave in the same way until the GWW point of the SU($N'$), up to zero-point energy $E_0$.
%$E-E_0=\frac{N^{\prime 2}}{4}$ (here, $E_0$ is the zero-point energy) and $S=\frac{\log D}{4}N^{\prime 2}$. 
From this, it follows that the SU($M$)-deconfined sector in the SU($N$)-theory corresponds to the GWW point of the SU($M$)-theory.
}\label{fig:why_GWW_free_1}
\end{figure}

Let us move on to step 2. 
Let us consider SU($N$) theory and SU($N'$) theory, with $N'<N$. 
To obtain the energy and entropy in the SU($N'$)-theory, we only have to replace $N$ with $N'$ in \eqref{eq:energy_MM} and \eqref{eq:entropy_MM}. 
The plots of the energy and entropy are shown in Fig.~\ref{fig:why_GWW_free_1}. A seemingly trivial but crucial observation is that the SU($N$)-theory and SU($N'$)-theory behave in the same way below the GWW point of the SU($N'$)-theory, $E-E_0=\frac{N^{\prime 2}}{4}$ (here, $E_0$ is the zero-point energy) and $S=\frac{\log D}{4}N^{\prime 2}$. Indeed, 
\begin{align}
S
=
\log D\cdot (E-E_0)
\end{align}
holds for both SU($N$)- and SU($N'$)-theories. What does it mean?

Suppose, for a given state in the SU($N$)-theory, all the excitations are in the upper-left $N'\times N'$ block up to gauge transformation. Because we are considering the free limit, we can cut out the $N'\times N'$ block and obtain a counterpart in the SU($N'$)-theory with the same energy.\footnote{Such a truncation would change the energy if the coupling constant is nonzero.} Note that, as long as $N'\times N'$ contains all the excitation, the size of $N’$ is arbitrary below $N$. The agreement of behaviors pointed out above means that, at $E-E_0\le\frac{N^{\prime 2}}{4}$, the entropy of the SU($N$)-theory can be reproduced just by considering such states. Namely, the excitations in the SU($N$)-theory are taking place \textit{only} in the $N'\times N'$ block up to gauge transformation. 
In addition, the excitations in the SU($N$)-theory cannot fit into the $N'\times N'$ block beyond the GWW point of the SU($N'$)-theory. It reasonably tells us that the SU($N$)-theory at $E-E_0=\frac{N^{\prime 2}}{4}$ is dominated by the SU($N'$)-partially-deconfined states and the deconfined sector should be identified with the GWW point of the SU($N'$)-theory. 

\begin{figure}[htbp]
\begin{center}
\rotatebox{-90}{
\scalebox{0.3}{
\includegraphics{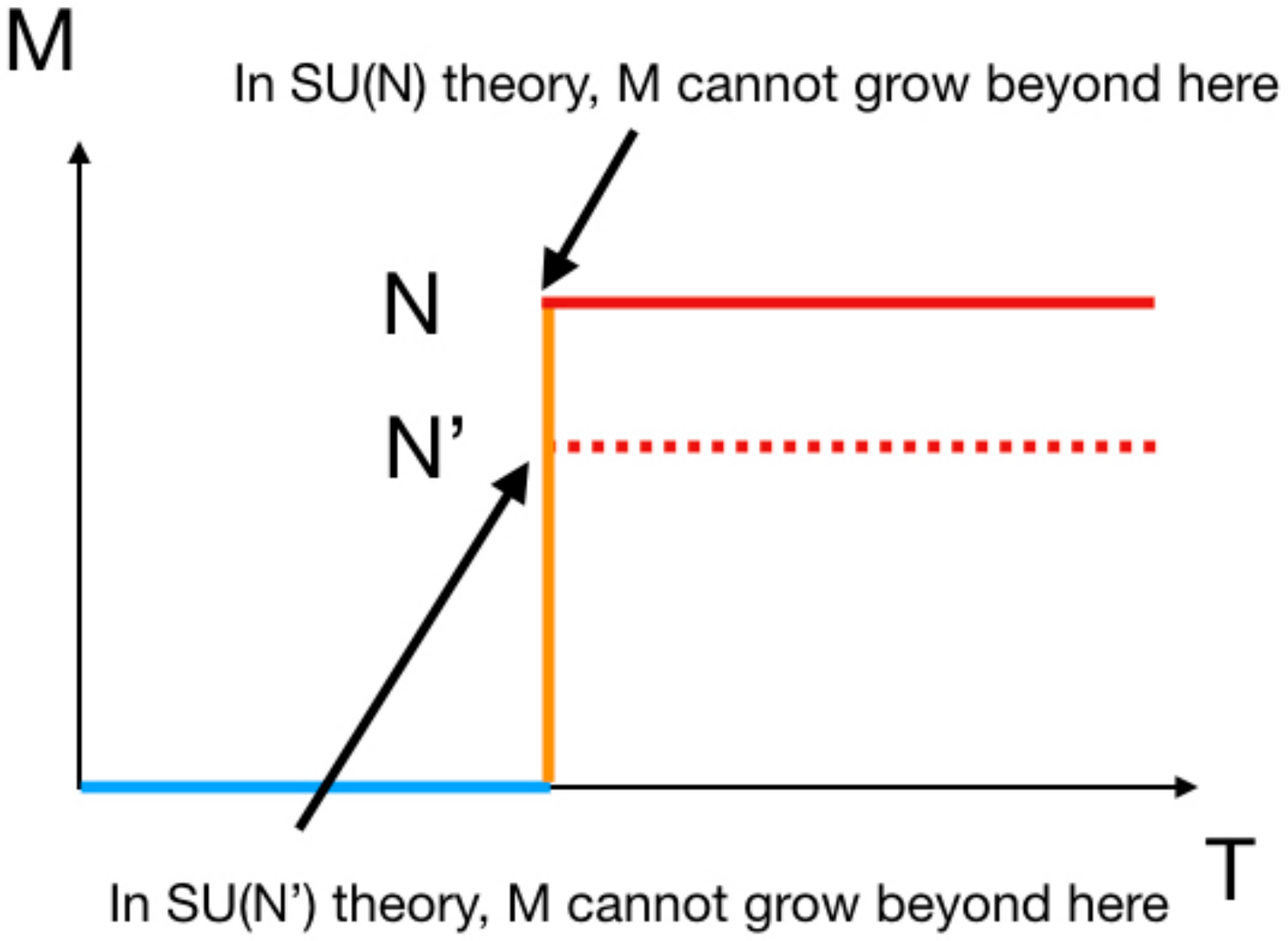}}}
\end{center}
\caption{
With the absence of the interaction, the deconfined sector of SU($N$)-theory and SU($N'$)-theory behave in the same way up to $M=N'$. 
From this, it follows that the SU($M$)-deconfined sector in the SU($N$)-theory corresponds to the GWW-transition point of the SU($M$) theory. 
These pictures appeared originally in Ref.~\cite{Hanada:2019rzv}.
}\label{fig:why_GWW_free_2}
\end{figure}

By switching the letter $N'$ to $M$, we obtain the precise picture of the SU($M$)-deconfined states in the free theory: \textit{SU($M$)-deconfined states are identified with the states at the GWW point of the SU($M$)-theory}. See also Fig.~\ref{fig:why_GWW_free_2}.   
The Polyakov loop $P$ in the SU($N$)-theory should be related to $M$ by $N^2P^2=\frac{M^2}{4}$, or equivalently, 
\begin{align}
P=\frac{M}{2N}\, . 
\end{align}
This is consistent with \eqref{eq:Gaussian_Polyakov_loop}: $M$ of the $N$ phases correspond to the GWW point, while the rest correspond to the confined phase (uniform distribution). A more precise understanding regarding this point is given in Sec.~\ref{sec:underlying_mechanism}. 

Step 3 is rather obvious from the above discussion. We can see the contributions from confined and deconfined sectors in $E$, $S$, and $\rho(\theta)$ as follows:
\begin{eqnarray}
E(T=T_{\rm c}, P=\frac{M}{2N},N)
=
\textcolor{blue}{\frac{D}{2}\cdot(N^2-M^2)}
+
\textcolor{red}{
\left(
\frac{D}{2}
+
\frac{1}{4}
\right)\cdot
M^2}\, ,  
\end{eqnarray}
\begin{eqnarray}
S(T=T_{\rm c}, P=\frac{M}{2N},N)
=
\textcolor{blue}{0\cdot (N^2-M^2)}
+
\textcolor{red}{\frac{\log D}{4}\cdot M^2}\, ,  
\end{eqnarray}
\begin{eqnarray}
\left.\rho(\theta)\right|_{T=T_{\rm c}, P=\frac{M}{2N}}
=
\textcolor{blue}{
\frac{1}{2\pi}
\cdot
\left(
1-\frac{M}{N}
\right) 
}
+
\textcolor{red}{
\frac{1}{2\pi}\left(
1+\cos\theta
\right)\cdot\frac{M}{N}
}\, . 
\end{eqnarray}
Blue and red are contributions from the confined and deconfined sectors, respectively. The SU($M$)-deconfined states dominate microcanonical thermodynamics because they explain the entropy precisely. 
%%%%%%%%%%%%
%%%%%%%%%%%%
\subsection{Underlying mechanism}\label{sec:underlying_mechanism}
\hspace{0.51cm}
%%%%%%%%%%%%
%%%%%%%%%%%%
In the previous subsection, we confirmed the formation of the SU($M$)-deconfined block by direct computations. Admittedly, such computations are not tractable for generic interacting theories. Furthermore, although we could confirm partial deconfinement, the physics principle behind it was not clear. 
In this subsection, we explain the underlying mechanism of partial deconfinement that generalizes to the interacting theories with generic matter content. 
The key is a deep connection between Bose-Einstein condensation (BEC) and color confinement~\cite{Hanada:2020uvt}.   
In retrospect, historically the first example of non-Abelian gauge theory was the system of $N$ identical bosons, which has S$_N$ permutation group as the gauge group. It was also the first example of the large-$N$ gauge theory because the thermodynamic limit $N\to\infty$ was studied. 
As discovered by Einstein, Bose-Einstein condensation is formed at low temperatures. As we will see, BEC and confinement are essentially the same phenomenon. 

Let us consider a system of $N$ indistinguishable bosons in the harmonic potential. 
The Hamiltonian is given by 
\begin{align}
\hat{H}
=
\frac{1}{2}\sum_{i=1}^N\left(
\hat{p}_{xi}^2
+
\hat{x}_{i}^2
+
\hat{p}_{yi}^2
+
\hat{y}_{i}^2
+
\hat{p}_{zi}^2
+
\hat{z}_{i}^2
\right).  
\end{align}
This Hamiltonian is invariant under the S$_N$ permutation group which exchanges particles, $i\to\sigma(i)$ ($\sigma\in\textrm{S}_N$). 
That $N$ bosons are ``indistinguishable" means that this S$_N$ symmetry is gauged.

It is convenient to use the Fock states to describe the system. 
We use $\vec{n}_i=(n_{x,i},n_{y,i},n_{z,i})$ to denote the excitation level of the $i$-th boson. 
Then, the Fock states are defined as
\begin{align}
\ket{\vec{n}_1,\vec{n}_2,\cdots,\vec{n}_N}
\equiv
\prod_{j=x,y,z}
\frac{\hat{a}_{1j}^{\dagger n_{1j}}}{\sqrt{n_{1j}!}}
\frac{\hat{a}_{2j}^{\dagger n_{2j}}}{\sqrt{n_{2j}!}}
\cdots
\frac{\hat{a}_{Nj}^{\dagger n_{Nj}}}{\sqrt{n_{Nj}!}}
\ket{0}.
\label{harmonic_oscillator_basis}
\end{align} 
These states span the extended Hilbert space $\mathcal{H}_{\textrm{ext}}$. 
The Fock vacuum $\ket{0}=\ket{\vec{0},\cdots,\vec{0}}$ is the ground state. 

The canonical partition function can be written in two ways, like in \eqref{eq:Z-H-inv-MM} and \eqref{eq:Z-H-ext-MM}. 
Here we use the latter. By replacing $G$ and integral with S$_N$ and sum, we obtain
\begin{align}
Z
=
\frac{1}{N!}
\sum_{\sigma\in\mathrm{S}_N}\textrm{Tr}_{\mathcal{H}_{\rm ext}}\left(
\hat{\sigma} e^{-\beta\hat{H}}
\right).   
\label{eq:partition-function-identical-bosons}
\end{align}
Here, $N!$ is the number of the elements of S$_N$, which corresponds to the volume of $G$. 
Trace over extended Hilbert space is the sum over Fock states, and hence
\begin{align}
Z
&=
\frac{1}{N!}
\sum_{\sigma\in\mathrm{S}_N}
\sum_{\vec{n}_1,\cdots,\vec{n}_N}
\langle\vec{n}_1,\cdots,\vec{n}_N|
\hat{\sigma} e^{-\beta\hat{H}}
|\vec{n}_1,\cdots,\vec{n}_N\rangle
\nonumber\\
&=
\frac{1}{N!}
\sum_{\vec{n}_1,\cdots,\vec{n}_N}
e^{-\beta\left(E_{\vec{n}_1}+\,\cdots\,+ E_{\vec{n}_N}\right)}
\sum_{\sigma\in\mathrm{S}_N}
\langle\vec{n}_1,\cdots,\vec{n}_N
|\vec{n}_{\sigma(1)},\cdots,\vec{n}_{\sigma(N)}\rangle
\nonumber\\
&=
\frac{1}{N!}
\sum_{\vec{n}_1,\cdots,\vec{n}_N}
e^{-\beta\left(E_{\vec{n}_1}+\,\cdots\,+ E_{\vec{n}_N}\right)}
\sum_{\sigma\in\mathrm{S}_N}
\delta_{\vec{n}_1\vec{n}_{\sigma(1)}}
\delta_{\vec{n}_2\vec{n}_{\sigma(2)}}
\cdots
\delta_{\vec{n}_N\vec{n}_{\sigma(N)}}
\nonumber\\
&\equiv
\frac{1}{N!}
\sum_{\vec{n}_1,\cdots,\vec{n}_N}
e^{-\beta\left(E_{\vec{n}_1}+\,\cdots\,+ E_{\vec{n}_N}\right)}
V_{\ket{\vec{n}_1,\cdots,\vec{n}_N}}. 
\end{align}
Here, 
\begin{align}
V_{\ket{\vec{n}_1,\cdots,\vec{n}_N}}
\equiv
\sum_{\sigma\in\mathrm{S}_N}
\delta_{\vec{n}_1\vec{n}_{\sigma(1)}}
\delta_{\vec{n}_2\vec{n}_{\sigma(2)}}
\cdots
\delta_{\vec{n}_N\vec{n}_{\sigma(N)}}
\end{align}
is the volume of the stabilizer of the state $\ket{\vec{n}_1,\cdots,\vec{n}_N}$ in the extended Hilbert space $\mathcal{H}_{\rm ext}$, i.e., 
\begin{align}
V_{\ket{\vec{n}_1,\cdots,\vec{n}_N}}
&=
\left(
\textrm{\#\ of}\ \sigma\in\textrm{S}_N\ \textrm{s.t.}\ \ket{\vec{n}_1,\cdots,\vec{n}_N}=\hat{\sigma}\ket{\vec{n}_1,\cdots,\vec{n}_N}
\right)
\nonumber\\
&=
\left(
\textrm{The\ volume\ of\ the\ stabilizer\ of\ }\ket{\vec{n}_1,\cdots,\vec{n}_N}\ \textrm{in}\ \mathcal{H}_{\rm ext}
\right)\, .
\end{align}

If all $N$ particles are in different states, 
only $\sigma=\textbf{1}$ gives rise to a nonzero contribution, and hence, the volume of the stabilizer is 1.  
On the other hand, if all $N$ particles are in the same state, all $\sigma$'s return the same nonzero contribution, 
and hence, the volume of the stabilizer is $N!$.
For a state of the form $\ket{\vec{n}_1,\cdots,\vec{n}_M,\vec{0},\cdots,\vec{0}}$, where $\vec{n}_1,\cdots,\vec{n}_M$ are generic excitations and most of them are different from each other, the BEC-sector consisting of $N-M$ particles in the ground state leads to the enhancement factor $(N-M)!$. 
BEC is formed, i.e., many particles fall into the ground state, due to such an enhancement factor. 

The same mechanism explains partial deconfinement. 
An SU($M$)-deconfined state $\ket{\Phi}$ is invariant under the action of $\textrm{SU}(N-M)\subset\textrm{SU}(N)$, and hence\footnote{
We use $\sim$ in \eqref{enhancement-for-partially-deconfined-state} because the deconfinement sector may have a small stabilizer.
}
\begin{align}
V_{\ket{\Phi}}
\sim
\textrm{vol}(\textrm{SU}(N-M))
\sim e^{(N-M)^2}. 
\label{enhancement-for-partially-deconfined-state}
\end{align}
The excitations are rolled up to form the $M\times M$ block structure because of this enhancement factor. 

Note that such an enhancement can exist for more complicated states that consist of the low-energy regime of the interacting theories. 
The explicit construction for the matrix model at nonzero coupling is given in Sec.~\ref{sec:interaction-and-wave-packet}. 

\begin{figure}[htbp]
  \begin{center}
  \includegraphics[width=0.8\textwidth]{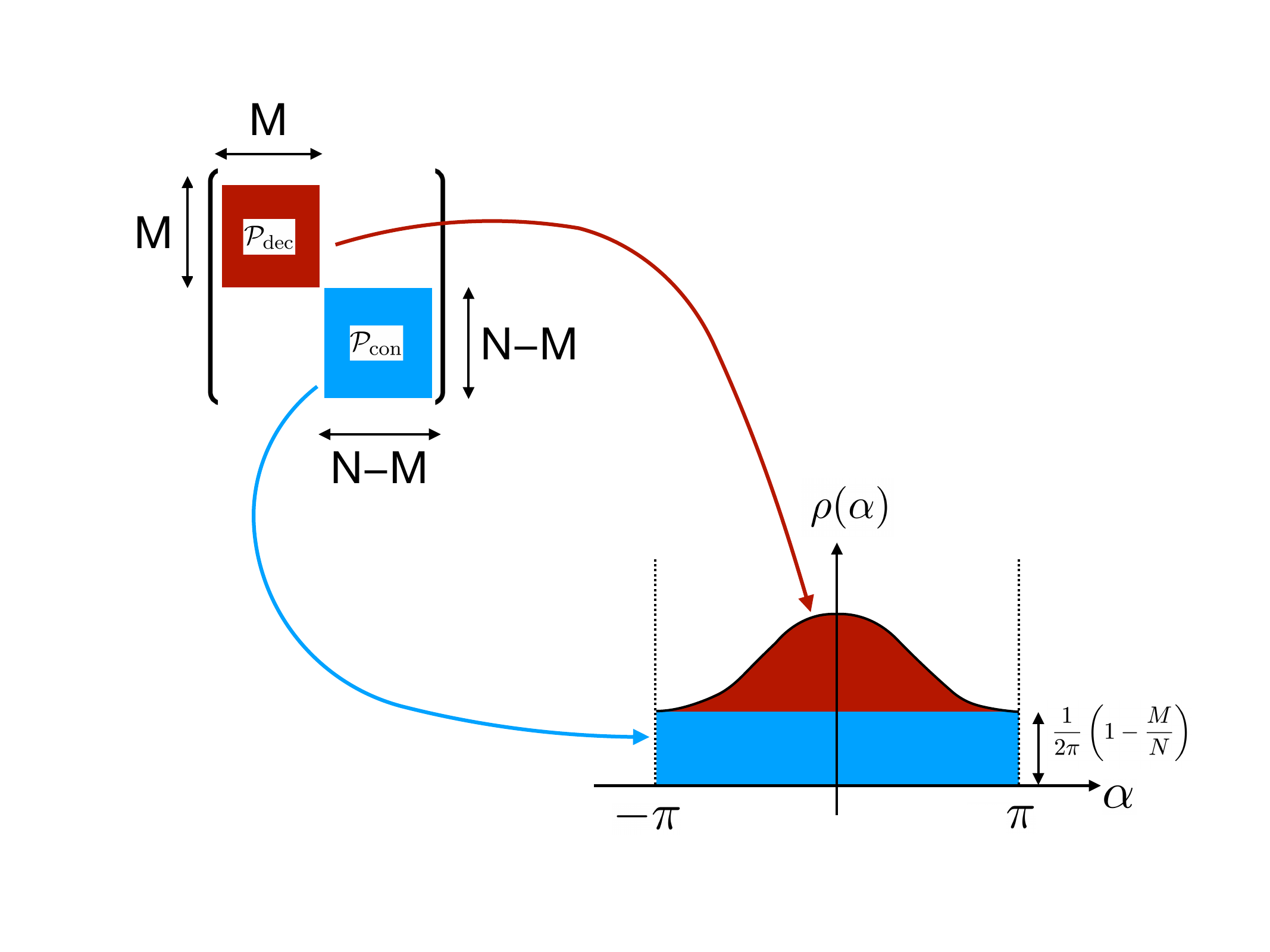}
  \end{center}
  \caption{Sketches of the distribution of the phases of the Polyakov line in the partially-deconfined phase. Constant offset comes from the confined Polyakov loop $\tr\mathcal{P}_{\rm con}$ while the non-uniform part comes from the deconfined Polyakov loop $\tr\mathcal{P}_{\rm dec}$. This picture is taken from Ref.~\cite{Gautam:2022exf}.
  }\label{fig:phase-distribution}
\end{figure}

As explained in Appendix~\ref{sec:Hamiltonian-to-Lagrangian}, $g\in G$ in \eqref{eq:Z-H-ext-MM} is the Polyakov line. (Specifically, this is the bare loop without renormalization. We will comment on the effect of renormalization in Sec.~\ref{sec:YM-continuum-limit}.)
Then, $\sigma\in\textrm{S}_N$ in \eqref{eq:partition-function-identical-bosons} can also be regarded as the Polyakov line. 
Let us use $\mathcal{P}$ to denote them. 
When BEC or confinement set in, typical states dominating thermodynamics are invariant under $\text{S}_{N-M}\subset\text{S}_{N}$ or $\textrm{SU}(N-M)\subset\textrm{SU}(N)$. 
Specifically, by choosing the SU($N-M$) to correspond to the lower-right $(N-M)\times (N-M)$-block, we can choose the SU($M$)-deconfined sector to be the upper-left $M\times M$-block as in Fig.~\ref{fig:matrix}. This choice of embedding of SU($N-M$) into SU($N$) fixes SU($N$) down to $\mathrm{SU}(M)\times\mathrm{SU}(N-M)$. The Polyakov line $\mathcal{P}$ takes the following form:
\begin{align}
    \mathcal{P}=
    \left(
    \begin{array}{cc}
    \mathcal{P}_{\rm dec} & 0\\
    0 & \mathcal{P}_{\rm con}
    \end{array}
    \right)\, .
    \label{eq:gauge-fixing-Polyakov}
\end{align}
Here, $\mathcal{P}_{\rm con}$ can be any element of SU($N-M$), and the generic phase distribution in this part is uniform in the limit of $N-M\to\infty$. The phases of $\mathcal{P}_{\rm dec}$ and $\mathcal{P}_{\rm con}$ are $\theta_1,\cdots,\theta_M$ and $\theta_{M+1},\cdots,\theta_N$, respectively. We can determine the distribution of the phases $\rho_{\rm dec}(\theta)=\rho_{\rm GWW}(\theta)$ and $\rho_{\rm con}(\theta)$. The latter is constant,
\begin{align}
    \rho_{\rm con}(\theta)=\frac{1}{2\pi}\, , 
\end{align}
while the former is not and its smallest value is zero. 
As we will discuss in Sec.~\ref{sec:QFT} and Sec.~\ref{sec:pol_dist_finite_N}, this constant distribution is tightly connected to the concept of \textit{Haar randomness}.
For the model under consideration, we can fix center symmetry in such a way that
\begin{align}
    \rho_{\rm GWW}(\pm\pi)=0\, . 
    \label{phase:dec_is_GWW}
\end{align}
The full distribution is 
\begin{align}
    \rho(\theta)
    &=
    \left(1-\frac{M}{N}\right)\cdot\rho_{\rm con}(\theta)
    +
    \frac{M}{N}\cdot\rho_{\rm GWW}(\theta)
    \nonumber\\
    &=
    \frac{1}{2\pi}\cdot\left(1-\frac{M}{N}\right)
    +
    \frac{M}{N}\cdot\rho_{\rm GWW}(\theta)\, . 
    \label{phase:full=con+dec}
\end{align}
See Fig.~\ref{fig:phase-distribution}. An important consequence is the relationship between the phase distribution and the size of the deconfined sector, which is  
\begin{eqnarray}
{\rm The\ constant\ offset}
=
\frac{1}{2\pi}\left(1-\frac{M}{N}\right). 
\label{Polyakov-ODLRO}
\end{eqnarray}
This relation holds for the system of indistinguishable bosons as well. The counterpart of $\mathcal{P}_{\rm con}$ can be any element of S$_{N-M}$, which leads to the same constant offset. Indeed, Feynman used a similar characterization for superfluid. The `Polyakov line' Feynman studied was the permutation matrix associated with the twisted boundary condition in the finite-temperature Euclidean path integral. If Feynman had known the Polyakov line and Gross-Witten-Wadia transition in 1953, he could have identified the onset of BEC with the GWW point!

%%%%%%%%%%%%
%%%%%%%%%%%%
\subsection{Partial deconfinement at nonzero coupling}\label{sec:interaction-and-wave-packet}
\hspace{0.51cm}
%%%%%%%%%%%%
%%%%%%%%%%%%
In free theories, some degrees of freedom can be excited without affecting others. We can split the Hilbert space into confined and deconfined sectors without caring about the interaction between them (because there is no interaction!) and perform simple analyses as we have done in Sec.~\ref{sec:Gaussian-Matrix-Model}. 
In this subsection, we explain how the characterization based on the unbroken symmetry in the extended Hilbert space and the relation \eqref{Polyakov-ODLRO} can be used even for the interacting theories~\cite{Hanada:2021ipb,Hanada:2021swb}. 

The key concept is the wave packet in the space in which the matrix components live. Let us first recall the wave packet in a one-particle system, and then generalize it to a gauged matrix model. We will formulate SU($M$)-deconfined wave packets which are invariant under SU($N-M$). Generic SU($M$)-deconfined states can be written as linear combinations of such wave packets.
%%%%%%%%%%%%
%%%%%%%%%%%%
\subsubsection*{Wave packet in one-particle system}
\hspace{0.51cm}
%%%%%%%%%%%%
%%%%%%%%%%%%
Let us consider a particle moving in a one-dimensional space. The Hamiltonian is 
\begin{align}
\hat{H}
=
\frac{\hat{p}^2}{2}
+
V(\hat{x}). 
\end{align}

Usually, energy eigenstates are far from our classical intuition. To connect classical and quantum descriptions, we use a \textit{wave packet} localized both in coordinate and momentum bases. We can obtain a low-energy wave packet $\ket{y,q}$ localized about $y$ in the coordinate basis and $q$ in the momentum basis by minimizing the energy $\bra{y,q}\hat{H}\ket{y,q}$ with constraints $\bra{y,q}\hat{x}\ket{y,q}=y$ and $\bra{y,q}\hat{p}\ket{y,q}=q$. When the size of the wave packet is negligible compared to the typical values of $q$ and $y$, such a wave packet can be interpreted as a classical particle at location $y$ moving with momentum $q$.

As a concrete example, let us consider the harmonic potential, $V(\hat{x})=\frac{\hat{x}^2}{2}$. The low-energy wave packet mentioned above is the coherent state
\begin{align}
\ket{y,q}
=
e^{i(q\hat{x}-y\hat{p})}
\ket{0,0}, 
\end{align}
where $\ket{0,0}$ is the same as the Fock vacuum. 
When $q=0$, it is merely a Gaussian wave packet in the coordinate basis, 
\begin{align}
\bra{x}\ket{y,q=0}
=
\frac{e^{-(x-y)^2/2}}{\pi^{1/4}}. 
\end{align}
In terms of the creation and annihilation operators $\hat{a}^\dagger=\frac{\hat{x}-i\hat{p}}{\sqrt{2}}$ and $\hat{a}=\frac{\hat{x}+i\hat{p}}{\sqrt{2}}$, the coherent state is expressed as
\begin{align}
\ket{y,q}
=
e^{\xi\hat{a}^\dagger-\xi^\ast\hat{a}}\ket{0}
=
e^{-\frac{1}{2}\xi^\ast\xi}
e^{\xi\hat{a}^\dagger}e^{-\xi^\ast\hat{a}}\ket{0}\, , 
\qquad
\xi
=
\frac{y+iq}{\sqrt{2}}\, , 
\end{align}
%%%
\begin{align}
e^{i(q\hat{x}-y\hat{p})}\ket{0}
=
e^{-\frac{1}{2}\xi^\ast\xi}
\sum_{n=0}^\infty\frac{\xi^n}{\sqrt{n!}}\ket{n}
\end{align}
and
\begin{align}
\hat{a}\ket{y,q}
=
\xi\ket{y,q}\, . 
\end{align}
Combined with $\hat{H}=\hat{a}^\dagger\hat{a}+\frac{1}{2}$, we obtain
\begin{align}
\bra{y,q}\hat{H}\ket{y,q}
=
\xi\xi^\ast+\frac{1}{2}
=
\frac{y^2+q^2+1}{2}\, . 
\end{align}
Time evolution is simply 
\begin{align}
e^{-i\hat{H}t}\ket{\xi}
&=
e^{-\frac{1}{2}\xi^\ast\xi}
\sum_{n=0}^\infty\frac{e^{-i(n+1/2)t}\xi^n}{\sqrt{n!}}\ket{n}
\nonumber\\
&=
e^{-it/2}
e^{-\frac{1}{2}\xi^\ast\xi}
\sum_{n=0}^\infty\frac{(e^{-it}\xi)^n}{\sqrt{n!}}\ket{n}\, .
\end{align}
Therefore, $\xi=\frac{y+iq}{\sqrt{2}}$ evolves as $e^{-it}\xi$, i.e., the center of the wave packet behaves as a classical harmonic oscillator. (Note that the overall phase factor $e^{-it/2}$ does not affect the interpretation.)
%%%%%%%%%%%%
%%%%%%%%%%%%
\subsubsection*{Wave packet in matrix space}
\begin{figure}[htbp]
  \begin{center}
   \includegraphics[width=80mm]{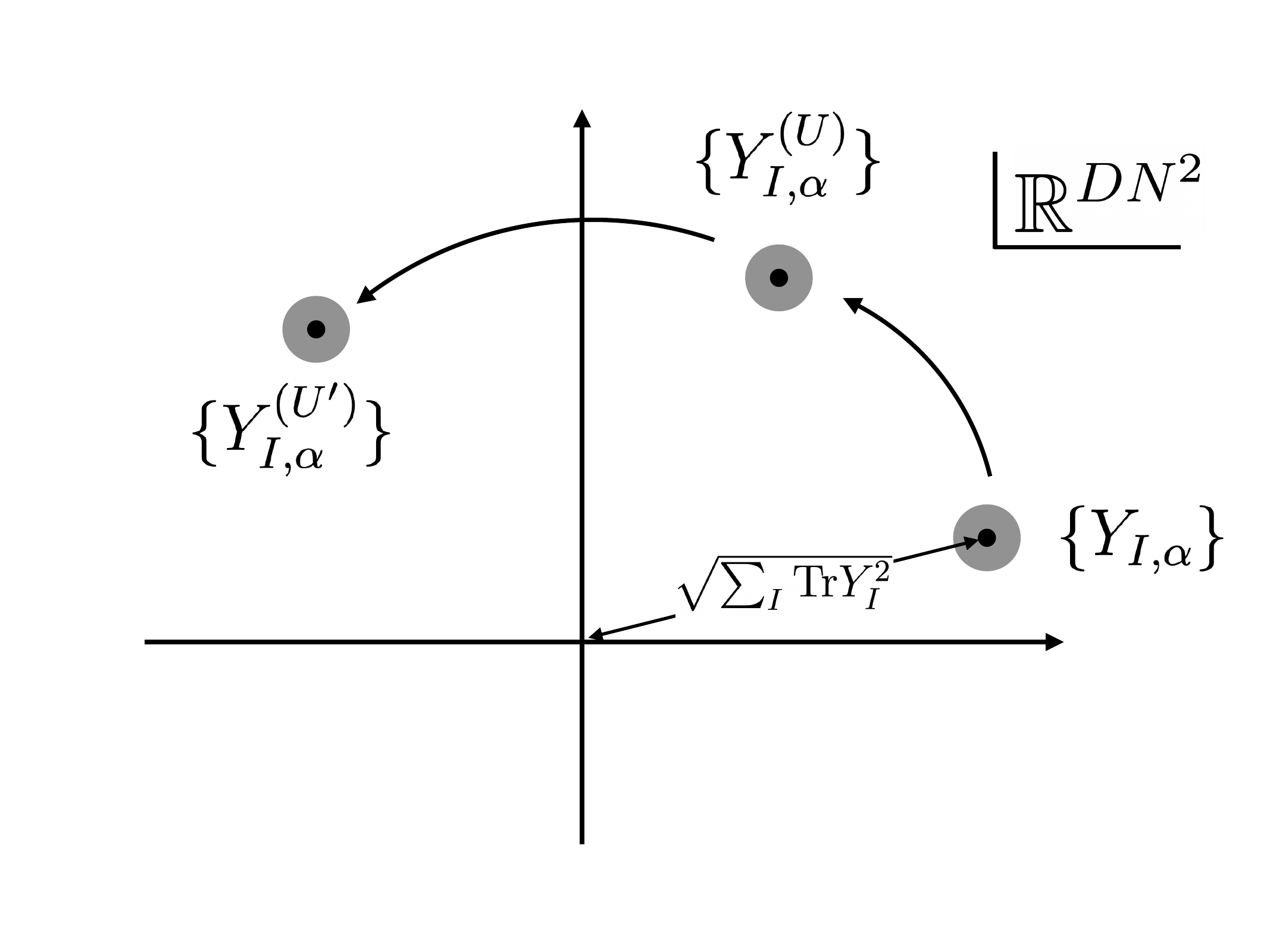}
  \end{center}
  \caption{
Wave packets in the coordinate space $\mathbb{R}^{DN^2}$. Each wave packet smoothly spreads in $\mathbb{R}^{DN^2}$. Along each of $DN^2$ dimensions, the size of the wave packet is at most of order $N^0$. 
Black points are the centers of wave packets ($Y$, $Y^{(U)}$ and $Y^{(U')}$), and gray disks are the wave packets (more precisely, the regions where the wave functions are not vanishingly small). 
Under the gauge transformation, the center of the wave packet moves as $Y\to Y^{(U)}=U^{-1}YU$, but the shape of the wave packet does not change. 
The center of the wave packet gives `slow modes' that describe the geometry consisting of D-branes and strings. 
This figure is a slight modification of a similar one in Ref.~\cite{Hanada:2021ipb}. 
  }\label{fig:wave-packet}
\end{figure}

\hspace{0.51cm}
%%%%%%%%%%%%
%%%%%%%%%%%%
Now we consider the extended Hilbert space of the matrix model. The only difference from the one-particle example is that the coordinate space and momentum space are $\mathbb{R}^{DN^2}$ rather than $\mathbb{R}$. 
We consider a wave packet around $Y\in\mathbb{R}^{DN^2}$ in the coordinate basis and $Q\in\mathbb{R}^{DN^2}$ in the momentum basis. 
A natural way to obtain such a low-energy wave packet is to find $\ket{Y,Q}$ that minimizes the energy
\begin{align}
\langle Y,Q|\hat{H}|Y,Q\rangle
\end{align}
satisfying the constraints
\begin{align}
\langle Y,Q|\hat{X}_I|Y,Q\rangle = Y_I, 
\qquad
\langle Y,Q|\hat{P}_I|Y,Q\rangle = Q_I. 
\end{align}
For the free theory, such a wave packet is the coherent state:
\begin{align}
\ket{Y,Q}
=
e^{i\textrm{Tr}(Q_I\hat{X}_I-Y_I\hat{P}_I)}
\ket{0,0}.  
\qquad
\textrm{(for free theory)}
\end{align}
The energy is (again, for the free theory)
\begin{align}
\bra{Y,Q}\hat{H}\ket{Y,Q}
=
\frac{1}{2}\left(\sum_I\textrm{Tr}(Y_I^2)+\sum_I\textrm{Tr}(Q_I^2)+DN^2\right). 
\end{align}
The ground state $\ket{Y=0,Q=0}$ is the Fock vacuum, which is SU($N$)-invariant. 
Coherent states are transformed as
\begin{align}
\ket{Y,Q}\to\ket{U^{-1}YU,U^{-1}QU}. 
\label{wave-packet-SUN-transf}
\end{align}
See Fig.~\ref{fig:wave-packet}. 
To see it by using equations, it is instructive to see a Gaussian wave packet in the coordinate basis
\begin{align}
\bra{X}\ket{Y,Q=0}
=
\frac{e^{-\sum_{I=1}^D\textrm{Tr}(X_I-Y_I)^2/2}}{\pi^{DN^2/4}}\, . 
\label{eq:Gaussian-wave-packet}
\end{align} 
From this, we can see that
\begin{align}
&
\ket{Y,Q=0}
=
\int d^{DN^2}X\ket{X}
\bra{X}\ket{Y,Q=0}
\nonumber\\
&\qquad
\to
\int d^{DN^2}X\ket{U^{-1}XU}
\frac{e^{-\sum_{I=1}^D\textrm{Tr}(X_I-Y_I)^2/2}}{\pi^{DN^2/4}}
\nonumber\\
&\qquad\quad
=
\int d^{DN^2}X\ket{X}
\frac{e^{-\sum_{I=1}^D\textrm{Tr}(UX_IU^{-1}-Y_I)^2/2}}{\pi^{DN^2/4}}
\nonumber\\
&\qquad\quad
=
\int d^{DN^2}X\ket{X}
\frac{e^{-\sum_{I=1}^D\textrm{Tr}(X_I-U^{-1}Y_IU)^2/2}}{\pi^{DN^2/4}}\, .
\end{align} 

Note that the transformation \eqref{wave-packet-SUN-transf} is not affected by the center-transformation $U\to U'=e^{i\delta}U$, where $\delta$ is any phase factor for U($N$) theory and $\mathbb{Z}_N$-phase factor $\frac{2\pi in}{N} (n=1,2,\cdots,N-1)$ for SU($N$) theory, respectively. When we study the distribution of Polyakov line phases, we can use the center transformation to make $P=\textrm{Tr}\mathcal{P}$ as close to real and positive as possible.\footnote{This is not the case for theories without center symmetry such as QCD; see Sec.~\ref{sec:GWW-in-QCD?}.} 
   
\begin{figure}[htbp]
  \begin{center}
   \includegraphics[width=80mm]{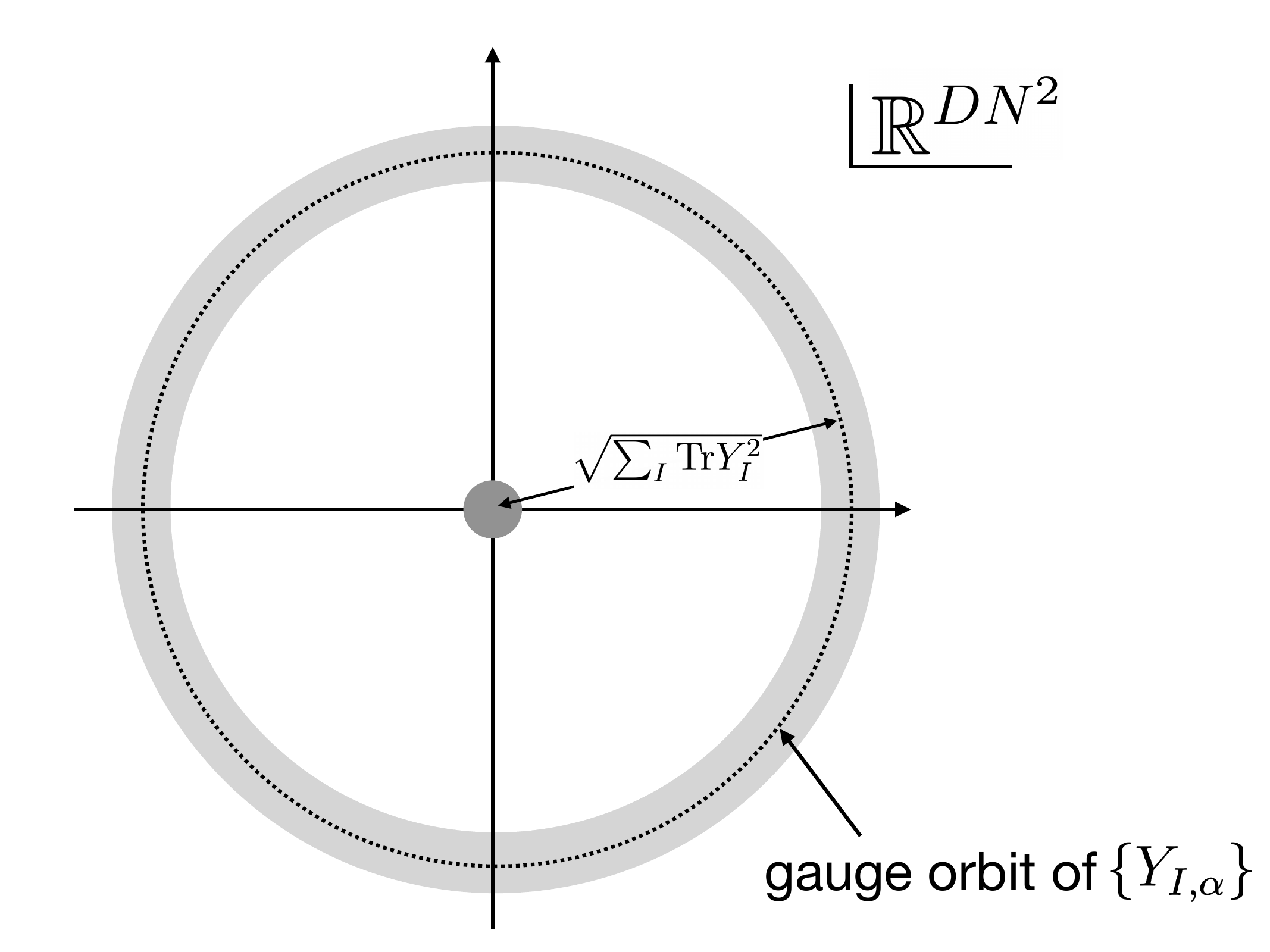}
  \end{center}
  \caption{
 The gauge-singlet that is obtained by acting the projection operator $\hat{\pi}$ on the wave packet $\ket{Y,Q}$. The wave function is localized about the gauge orbit of $\{Y_{I,\alpha}\}$ in $\mathbb{R}^{DN^2}$.  
This figure is a slight modification of the one in Ref.~\cite{Hanada:2021ipb}. 
  }\label{fig:wave-packet-gauge-inv}
\end{figure}

Because of the symmetry in the minimization problem, the same transformation law \eqref{wave-packet-SUN-transf} applies to the low-energy wave packets in the interacting theory, while the shape of the wave packet, such as the width along each direction, depends on the details of the interaction. States that transform to each other via gauge transformation should be identified.
The corresponding element in $\mathcal{H}_{\rm inv}$ can be obtained by acting the projector $\hat{\pi}$ defined in \eqref{eq:projector}, up to a normalization constant:
\begin{align}
\ket{Y,Q}
\to
\mathcal{N}^{-1/2}\hat{\pi}\ket{Y,Q}, 
\qquad
\mathcal{N}\equiv\bra{Y,Q}\hat{\pi}\ket{Y,Q}.
\label{Wave-packet-projection}
\end{align}
See Fig.~\ref{fig:wave-packet-gauge-inv}. A gauge singlet corresponds to a gauge orbit in $\mathcal{H}_{\rm ext}$. 
We do not expect that the expectation value of $\hat{H}$ changes significantly,\footnote{
This is not the case when localized wave packets are constructed from the coordinate eigenstates by taking linear combinations because the coordinate eigenstates are not localized in the momentum space, and the kinetic energy can be reduced significantly.} because the overlap between $\ket{Y,Q}$ and $\ket{U^{-1}YU,U^{-1}QU}$ is very small unless $U$ is very close to the stabilizer of $\ket{Y,Q}$. The normalization factor $\mathcal{N}$ is essentially the volume of the stabilizer of $\ket{Y,Q}$. If the stabilizer is large, then the enhancement factor in the partition function is large. 

Each wave packet smoothly spreads in the coordinate space $\mathbb{R}^{DN^2}$, and also in the momentum space $\mathbb{R}^{DN^2}$.  
Along each of $DN^2$ dimensions, the size is at most of order $N^0$. 
Two wave packets centered around $Y$ and $Y'$ in $\mathbb{R}^{DN^2}$ have vanishingly small overlap if the centers of wave packets are well separated, i.e., $\sqrt{{\rm Tr}(Y-Y')^2}\gg 1$.

%%%%%%%%%%%%
%%%%%%%%%%%%
\subsubsection*{Partial deconfinement at nonzero coupling}
\hspace{0.51cm}
%%%%%%%%%%%%
%%%%%%%%%%%%
The SU($M$)-deconfined states can be constructed as linear combinations of wave packets of the form 
\begin{align}
Y_I
=
\left(
\begin{array}{cc}
 \tilde{Y}_I& 0\\
0& 0
\end{array}
\right), 
\qquad
Q_I
=
\left(
\begin{array}{cc}
\tilde{Q}_I& 0\\
0& 0
\end{array}
\right). 
\label{partially-deconfined-state}
\end{align}
Here, $\tilde{Y}_I$ and $\tilde{Q}_I$ are $M\times M$ matrices. By construction, such wave packets are invariant under SU($N-M$)-transformations that do not touch the upper-right $M\times M$ block, i.e., 
\begin{align}
\left(
\begin{array}{cc}
 \textbf{1}_M& 0\\
0& \Omega^{-1}
\end{array}
\right)
\left(
\begin{array}{cc}
 \tilde{Y}_I& 0\\
0& 0
\end{array}
\right)
\left(
\begin{array}{cc}
 \textbf{1}_M& 0\\
0& \Omega
\end{array}
\right)
=
\left(
\begin{array}{cc}
 \tilde{Y}_I& 0\\
0& 0
\end{array}
\right)
\end{align}
and
\begin{align}
\left(
\begin{array}{cc}
 \textbf{1}_M& 0\\
0& \Omega^{-1}
\end{array}
\right)
\left(
\begin{array}{cc}
 \tilde{Q}_I& 0\\
0& 0
\end{array}
\right)
\left(
\begin{array}{cc}
 \textbf{1}_M& 0\\
0& \Omega
\end{array}
\right)
=
\left(
\begin{array}{cc}
 \tilde{Q}_I& 0\\
0& 0
\end{array}
\right)
\end{align}
hold for any $\Omega\in{\rm SU}(N-M)$. 
Let us write a generic linear combination of such states as
\begin{align}
|\Phi\rangle
=
\sum_{a=1}^K
c_a|Y_{[a]},Q_{[a]}\rangle\, , 
\qquad 
K=e^S\, , 
\qquad
|c_a|\sim\frac{1}{\sqrt{K}}\, . 
\label{eq:partially-deconfined-state}
\end{align}
Here, $S$ is the coarse-grained entropy that counts independent wave packets, which is $\sim M^2$ at weak coupling. 
By construction, such linear combinations are invariant under the same SU($N-M$)-transformations. 
Therefore, the mechanism explained in Sec.~\ref{sec:underlying_mechanism} is applicable even with the presence of the interactions.
The invariance under the $SU(N-M)$-transformation leads to the enhancement factor $\sim e^{(N-M)^2}$. 
The value of $M$ is determined in such a way that the product of this enhancement factor and $K\sim e^S$ is maximized. 

Such states are quite intriguing also from the D-brane point of view: the SU($M$)-deconfined states are interpreted through the holographic perspective by the situation that $M$ D-branes and open strings between them are excited and form an extended object. 
Other $N-M$ D-branes are sitting at the origin and no open strings are excited between them, or between those $N-M$ D-branes and the other $M$ D-branes. Note that all D-branes contribute to the exterior geometry. 
Such states can describe a small black hole, with no radiation emitted yet~\cite{Hanada:2021ipb,Gautam:2022akq,Hanada:2016pwv}. 
%%%%%%%%%%%%
%%%%%%%%%%%%
\subsection{Euclidean path integral and Monte Carlo simulation}\label{sec:Path-integral-gauge-fixing}
\hspace{0.51cm}
%%%%%%%%%%%%
%%%%%%%%%%%%
Euclidean path integral and Monte Carlo simulation are powerful and, at least at present, practically the only tools to study strongly-coupled nonperturbative dynamics of matrix models quantitatively. Therefore, it is important to understand how partial deconfinement can be observed in path-integral formalism. Specifically, we want a gauge fixing that makes the splitting of confined and deconfined sectors manifest as in Fig.~\ref{fig:matrix}. 

The relation between the Polyakov loop and partial deconfinement reviewed in Sec.~\ref{sec:underlying_mechanism} plays the central role for this purpose. If the splitting depicted in Fig.~\ref{fig:matrix} is realized, then the Polyakov loop and its phase distribution take the forms shown in Fig.~\ref{fig:phase-distribution} and Eqs.~\eqref{eq:gauge-fixing-Polyakov} --- \eqref{phase:full=con+dec}. Therefore, we take the static-diagonal gauge, i.e.,
\begin{align}
A_t(t)
=
T\times\textrm{diag}(\theta_1,\cdots,\theta_N)\, , 
\qquad
-\pi\le\theta_i<\pi\, , 
\end{align}
where $\theta_1,\cdots,\theta_N$ are the phases of the Polyakov line.
They fix the residual S$_N$ permutation symmetry down to $\mathrm{S}_M\times\mathrm{S}_{N-M}$ in such a way that $\theta_{M+1},\cdots,\theta_N$ are distributed uniformly and describe the constant offset in the density distribution function~\cite{Watanabe:2020ufk,Gautam:2022exf}. Associated with the static diagonal gauge, the Faddeev-Popov term of the following form is added~\cite{Furuuchi:2003sy,Kawahara:2006hs}:
\begin{align}
S_{\rm F.P.}
=
-\sum_{i<j}2\log
\left|
\sin\left(\frac{\theta_i-\theta_j}{2}\right)
\right|\, . 
\end{align}
Note that the Polyakov loop takes the same form everywhere on the temporal circle, and hence, the same separation of the form shown in Fig.~\ref{fig:matrix} is realized everywhere. 

This gauge fixing was applied to the matrix model~\cite{Watanabe:2020ufk} and partial deconfinement was demonstrated by using lattice Monte Carlo simulations. Ref.~\cite{Gautam:2022exf} studied the flux tube in the confined sector of the partially-deconfined saddle of Yang-Mills theory, as we will review in Sec.~\ref{sec:flux-tube}. 
%%%%%%%%%%%%
%%%%%%%%%%%%
\subsection{Introducing fermions}\label{sec:fermion-in-matrix-model}
\hspace{0.51cm}
%%%%%%%%%%%%
%%%%%%%%%%%%
In this subsection, we add fermions to the analysis. 
Let $\psi$ be a complex adjoint fermion as a concrete example. We regard $\hat{\psi}^\dagger$ and $\hat{\psi}$ as the creation and annihilation operators, respectively. 
The anticommutation relation is
\begin{align}
\{
\hat{\psi}_{\alpha,a},
\hat{\psi}^\dagger_{\beta,b}
\}
=
\delta_{\alpha\beta}\delta_{ab}
\end{align}
where $\alpha,\beta$ are spinor indices and $a,b$ are adjoint indices.

Let us start with the free theory, with the fermionic part of the Hamiltonian $\mu{\rm Tr}(\hat{\psi}^\dagger\hat{\psi})$, $\mu>0$. 
Then, the Fock vacuum $\ket{0}=\otimes_{\alpha,a}\ket{0}_{\alpha,a}$ characterized by $\hat{\psi}_{\alpha,a}\ket{0}=0$ is the ground state.
Note that the Fock vacuum is SU($N$)-invariant. The highest excited level $\otimes_{\alpha,a}\ket{1}_{\alpha,a}$ is also SU($N$)-invariant. 
Fermions are excited ($\hat{\psi}^\dagger$ can be multiplied) only in the deconfined sector.

For bosons, we used the wave-packet description for the generalization to finite coupling. The reason was that the coordinate eigenstates were not low-energy states. Ultimately, it was because the dimension of the Hilbert space for bosons is infinite. 
The situation is different for fermions, and it is not clear if we need to use such a thing like a `wave packet'. Furthermore, it is difficult to make sense of the `wave packet' of fermionic degrees of freedom. In fact, a coherent state for fermion is tricky because it is written as $\ket{\eta}=\ket{0}+\eta\ket{1}$ where $\eta$ is a Grassmann number, and hence, it is not a quantum state in the usual sense. 
Note also that Fock states $\ket{0}$ and $\ket{1}$ are already analogous to wave packets, in that they are low-energy Fock states.
(The Fock vacuum for boson is the Gaussian wave packet around the origin.)
 
Below, we use wave packets only for the bosonic sector. 
Let $\mathcal{H}_{\rm ext}=\mathcal{H}_{\rm ext, b}\otimes\mathcal{H}_{\rm ext, f}$. For each coordinate eigenstate $\ket{X}\in\mathcal{H}_{\rm ext, b}$, we can associate $\ket{\psi_X}\in\mathcal{H}_{\rm ext, f}$ and obtain a generic state in $\mathcal{H}_{\rm ext}$ as $\int dX f(X)\ket{X}\otimes\ket{\psi_X}$. Such a state is transformed as 
 \begin{align}
 \ket{\Phi}
 =
 \int dX f(X)\ket{X}\otimes\ket{\psi_X}
 \longrightarrow
 \hat{U}\ket{\Phi}
 &=
 \int dX f(X)\ket{U^{-1}XU}\otimes(\hat{U}\ket{\psi_X})
 \nonumber\\
 &=
 \int dX f(UXU^{-1})\ket{X}\otimes(\hat{U}\ket{\psi_{UXU^{-1}}})\, . 
 \end{align}
The stabilizer $G_{\ket{\Phi}}$ is 
\begin{align}
G_{\ket{\Phi}}
=
\{U\in{\rm SU}(N)|f(UXU^{-1})=f(X)\ {\rm and}\ \hat{U}\ket{\psi_{UXU^{-1}}}=\ket{\psi_X}\}\, . 
\label{eq:stabilizer-with-fermion}
\end{align} 
If we define a low-energy wave packet by minimizing the energy fixing the one-point function of $\hat{X}_I$ and $\hat{P}_I$ to $Y_I$ and $Q_I$, it is natural to expect that the stabilizer of $Y_I$ and $Q_I$ satisfy \eqref{eq:stabilizer-with-fermion} due to the symmetry of the problem. 
%%%%%%%%%%%%
%%%%%%%%%%%%
\subsubsection*{Analogy to BEC with fermionic degrees of freedom}
\hspace{0.51cm}
%%%%%%%%%%%%
%%%%%%%%%%%%
The analogy between confinement and BEC is not affected by the existence of fermions. 
Let us consider Fock space for fermionic oscillators, 
\begin{align}
\ket{\vec{n}_1,\vec{n}_2,\cdots,\vec{n}_N}
\equiv
\prod_{j}
\hat{c}_{1j}^{\dagger n_{1j}}
\hat{c}_{2j}^{\dagger n_{2j}}
\cdots
\hat{c}_{Nj}^{\dagger n_{Nj}}
\ket{0}.
\label{harmonic_oscillator_basis_fermion}
\end{align} 
Here, $j$ represents the spinor index, flavor index, etc. 
Each $n_{kj}$ is 0 or 1. 
Let us see how the state changes by permutation. 
If $\sigma\in\textrm{S}_N$ exchanges only 1 and 2 (i.e., $\sigma(1)=2, \sigma(2)=1, \sigma(j)=j$ for $j\ge 3$), then
\begin{align}
\hat{\sigma}\ket{\vec{n}_1,\vec{n}_2,\vec{n}_3,\cdots,\vec{n}_N}
&\equiv
\ket{\vec{n}_2,\vec{n}_1,\vec{n}_3,\cdots,\vec{n}_N}
\nonumber\\
&=
\prod_{j}
\hat{c}_{2j}^{\dagger n_{1j}}
\hat{c}_{1j}^{\dagger n_{2j}}
\cdots
\hat{c}_{Nj}^{\dagger n_{Nj}}
\ket{0}
\nonumber\\
&=
\prod_{j}
(-1)^{n_{1j}n_{2j}}
\hat{c}_{1j}^{\dagger n_{2j}}
\hat{c}_{2j}^{\dagger n_{1j}}
\cdots
\hat{c}_{Nj}^{\dagger n_{Nj}}
\ket{0}
\nonumber\\
&=
(-1)^{\sum_{j}n_{1j}n_{2j}}
\ket{\vec{n}_1,\vec{n}_2,\vec{n}_3,\cdots,\vec{n}_N}. 
\label{eq:fermion-permutation}
\end{align}
Here, we assumed that the Fock vacuum is permutation invariant. This assumption is consistent with \eqref{eq:fermion-permutation}, by setting all $\vec{n}_i$ to be $\vec{0}$.
If $\vec{n}_1=\vec{n}_2=\vec{m}$, 
\begin{align}
\hat{\sigma}\ket{\vec{m},\vec{m},\vec{n}_3,\cdots,\vec{n}_N}
=
(-1)^{\sum_{j}m_{j}}
\ket{\vec{m},\vec{m},\vec{n}_3,\cdots,\vec{n}_N}. 
\end{align}
Therefore, $\ket{\vec{m},\vec{m},\vec{n}_3,\cdots,\vec{n}_N}$ and $\hat{\sigma}\ket{\vec{m},\vec{m},\vec{n}_3,\cdots,\vec{n}_N}$ cancel out by the symmetrization if $\sum_{j}m_j$ is odd, but they survive if $\sum_{j}m_j$ is even. In this case, an even number of fermion creation operators are combined to form a bosonic excitation.

A similar argument holds for SU($N$). Suppose the fermions are in the fundamental representation, and consider the SU(2) subgroup acting on the first two elements as $\hat{c}^\dagger_{1j}\to \sum_{k=1}^2U_{1k}\hat{c}^\dagger_{kj}$ and $\hat{c}^\dagger_{2j}\to \sum_{k=1}^2U_{2k}\hat{c}^\dagger_{kj}$. Then, 
\begin{align}
\hat{c}_{1j}^\dagger
\hat{c}_{2j}^\dagger
\to
\left(\sum_{k=1}^2U_{1k}\hat{c}^\dagger_{kj}\right)
\left(\sum_{k=1}^2U_{2k}\hat{c}^\dagger_{kj}\right)
=
\hat{c}_{1j}^\dagger
\hat{c}_{2j}^\dagger
\left(U_{11}U_{22}-U_{12}U_{21}\right)
=
\hat{c}_{1j}^\dagger
\hat{c}_{2j}^\dagger. 
\end{align}
Because $m_j$ is 0 or 1, $\hat{c}_{1j}^{\dagger m_j}\hat{c}_{2j}^{\dagger m_j}$, and hence $\ket{\vec{m},\vec{m},\vec{n}_3,\cdots,\vec{n}_N}$, are invariant under SU(2). 
%%%%%%%%%%%%
%%%%%%%%%%%%
\section{Partial deconfinement in large-$N$ QFT}\label{sec:QFT}
\hspace{0.51cm}
%%%%%%%%%%%%
%%%%%%%%%%%%
Let us consider gauge theory living in nonzero spatial dimensions, i.e., QFT. 
Intuitively, the mechanism introduced for matrix models should apply locally. 
The precise meaning of this is discussed below. In short, the same argument can be used for \textit{slowly-varying} SU($N$) transformations. 

In Sec.~\ref{sec:YM}, we consider Yang-Mills theory. To make the discussion precise, we adopt lattice regularization and use the Kogut-Susskind Hamiltonian formulation. We start with studying the strong-coupling limit in Sec.~\ref{sec:strong-coupling-lattice}. 
The ground state is gauge invariant in $\mathcal{H}_{\rm ext}$, and the mechanism discussed for the case of the matrix model is applicable without modification. In Sec.~\ref{sec:YM-continuum-limit}, we consider the continuum limit that requires taking weak coupling. 
%In this case, the ground state 
As opposed to the previous case, the ground state at weak coupling is a localized wave packet which is \textit{not} gauge-invariant in $\mathcal{H}_{\rm ext}$. Still, essentially the same mechanism is applicable, by considering the slowly-varying SU($N$) transformations. 
The generalization to the large-$N$ QCD is provided in Sec.~\ref{sec:large-N-QCD}
and some physical consequences are discussed in Sec.~\ref{sec:physical-consequence}. 

Note that we consider the bare Polyakov loops, i.e., without renormalization, which corresponds to the group element in the projector from $\mathcal{H}_{\rm ext}$ to $\mathcal{H}_{\rm inv}$. We comment on renormalization at the end of Sec.~\ref{sec:YM-continuum-limit}. 
%%%%%%%%%%%%
%%%%%%%%%%%%
\subsection{Yang-Mills theory}\label{sec:YM}
\hspace{0.51cm}
%%%%%%%%%%%%
%%%%%%%%%%%%
The meaning of partial deconfinement is clearer if we look directly at quantum states by working in the operator formalism.  We describe the partially-deconfined phase by using the Hamiltonian formulation by Kogut and Susskind~\cite{Kogut:1974ag}. 
We consider the (3+1)-d Yang-Mills theory with the SU($N$) gauge group. 

The Kogut-Susskind Hamiltonian consists of the electric and magnetic terms, 
\begin{eqnarray}
\hat{H}
=
\hat{H}_{\rm E}
+
\hat{H}_{\rm B}.  
\end{eqnarray}
The electric part of the Hamiltonian is
\begin{eqnarray}
\hat{H}_{\rm E}
=
\frac{a^3}{2}
\sum_{\vec{n}}\sum_{\mu=1}^3\sum_{\alpha=1}^{N^2}\left(\hat{E}_{\mu,\vec{n}}^{\alpha}\right)^2. 
\end{eqnarray}
Here $a$ is the lattice spacing, and $\hat{E}_{\mu,\vec{n}}^{\alpha}$ is the electric field. 
The magnetic part is 
\begin{eqnarray}
\hat{H}_{\rm B}
=
-\frac{1}{2ag^2}\sum_{\vec{n}}\sum_{\mu<\nu}
\left(
\hat{U}_{\mu,\vec{n}}
\hat{U}_{\nu,\vec{n}+\hat{\mu}}
\hat{U}^\dagger_{\mu,\vec{n}+\hat{\nu}}
\hat{U}^\dagger_{\nu,\vec{n}}
+
\textrm{h.c.}
\right). 
\label{eq:KS-Hamiltonian_magnetic-part}
\end{eqnarray}
At weak coupling, the link variable $U_\mu$ is regarded as $U_\mu=e^{iagA_\mu}$. 
The commutation relations are 
\begin{eqnarray}
\left[
\hat{U},
\hat{U}
\right]
=
\left[
\hat{U},
\hat{U}^\dagger
\right]
=
\left[
\hat{U}^\dagger,
\hat{U}^\dagger
\right]
=
0.   
\end{eqnarray}
In addition, 
\begin{eqnarray}
\left[
\hat{E}_{\mu,\vec{n}}^\alpha,
\hat{U}_{\nu,\vec{n}'}
\right]
=
\delta_{\mu\nu}\delta_{\vec{n}\vec{n}'}\tau_\alpha\hat{U}_{\nu,\vec{n}'}, 
\qquad
\left[
\hat{E}_{\mu,\vec{n}}^\alpha,
\hat{U}^\dagger_{\nu,\vec{n}'}
\right]
=
-\delta_{\mu\nu}\delta_{\vec{n}\vec{n}'}\hat{U}_{\nu,\vec{n}'}^\dagger\tau_\alpha,   
\end{eqnarray}
\begin{eqnarray}
\left[
\hat{E}_{\mu,\vec{n}}^\alpha,
\hat{E}_{\nu,\vec{n}'}^\beta
\right]
=
-if^{\alpha\beta\gamma}\delta_{\mu\nu}\delta_{\vec{n}\vec{n}'}\hat{E}^\gamma_{\nu,\vec{n}'},  
\end{eqnarray}
where $f^{\alpha\beta\gamma}$ is the structure constant of SU($N$).~\footnote{  
We can choose the generators in such a way that
$\textrm{Tr}(\tau^\alpha\tau^\beta)=\delta^{\alpha\beta}$ and
$
\sum_{\alpha=1}^{N^2-1}\tau^{\alpha}_{pq}\tau^{\alpha}_{rs}
=
\delta_{ps}\delta_{qr}
-
\frac{\delta_{pq}\delta_{rs}}{N}
$
are satisfied. 
}

The operator $\hat{U}_{\mu,\vec{n}}$ is interpreted as the coordinate of the group manifold SU($N$) for the link variable on the site $\vec{n}$ in the $\mu$-direction. The operators $\hat{U}$ and $\hat{E}$ are defined on the extended Hilbert space $\mathcal{H}_{\rm ext}$ that contains gauge non-singlet modes. 
A convenient basis of $\mathcal{H}_{\rm ext}$ is the coordinate representation, 
\begin{eqnarray}
{\cal H}_{\rm ext}
=
\otimes_{\mu,\vec{n}}{\cal H}_{\mu,\vec{n}}
\sim
\otimes_{\mu,\vec{n}}
\left(
\oplus_{g\in{\rm SU}(N)}
|g\rangle_{\mu,\vec{n}}
\right), 
\end{eqnarray}
 where
 \begin{eqnarray}
\hat{U}_{\mu,\vec{n}}
|g\rangle_{\mu,\vec{n}}
=
g|g\rangle_{\mu,\vec{n}}
\qquad 
g\in{\rm SU}(N). 
\end{eqnarray}
More precisely, we will consider only the Hilbert space of square-integrable wave functions.

Let $\mathcal{G}=\prod_{\vec{n}}[{\rm SU}(N)]_{\vec{n}}$ be the group of all local gauge transformations. 
Gauge transformation by $\hat{\Omega}=\otimes_{\vec{n}}\hat{\Omega}_{\vec{n}}\in\mathcal{G}$ is defined by 
\begin{align}
\hat{\Omega}
\left(
\otimes_{\mu,\vec{n}}
|g\rangle_{\mu,\vec{n}}
\right)
=
\otimes_{\mu,\vec{n}}
|\Omega_{\vec{n}}^{-1}g\Omega_{\vec{n}+\hat{\mu}}\rangle_{\mu,\vec{n}}. 
\end{align}
%%%%%%%%%%%%
%%%%%%%%%%%%
\subsubsection{Strong coupling on lattice}\label{sec:strong-coupling-lattice}
\hspace{0.51cm}
%%%%%%%%%%%%
%%%%%%%%%%%%
In the `strong coupling limit'~\cite{Susskind:1979up}, we formally take the limit of $g^2\to\infty$ and drop the magnetic term~\eqref{eq:KS-Hamiltonian_magnetic-part}. 
Although such a limit contains lattice artifacts, however, it is qualitatively a good model of confinement/deconfinement transition. 
Partial deconfinement in this limit was studied in Refs.~\cite{Hanada:2021ksu,Gautam:2022exf}. 
In the strong coupling limit, $a^3$ is merely an overall factor, so we set $a=1$ for simplicity.%:
\begin{comment}
\begin{eqnarray}
\hat{H}_{\rm E}
=
\frac{1}{2}
\sum_{\vec{n}}\sum_{\mu=1}^3\sum_{\alpha=1}^{N^2}\left(\hat{E}_{\mu,\vec{n}}^{\alpha}\right)^2\, . 
\end{eqnarray}
\end{comment}

The ground state of this strong-coupling theory $|{\rm g.s.}\rangle$ satisfies $\hat{E}_{\mu,\vec{n}}^\alpha|{\rm g.s.}\rangle = 0$ for any $\alpha,\mu$, and $\vec{n}$. Hence, we introduce the notation $|E=0\rangle$ to denote the ground state. Because the electric field $\hat{E}_{\mu,\vec{n}}^\alpha$ is momentum on the group manifold, zero modes associated with the electric field is the constant on the group manifold, 
\begin{align}
|E=0\rangle
=
\otimes_{\mu,\vec{n}}
|E=0\rangle_{\mu,\vec{n}}\, ,
\qquad
|E=0\rangle_{\mu,\vec{n}}
=
\frac{1}{\sqrt{{\rm vol (SU}(N))}}\int_{{\rm SU}(N)} dg
|g\rangle_{\mu,\vec{n}}\, . 
\end{align}
Namely, the wave function $\langle g|E=0\rangle$ is constant. 
%Perhaps we should interpret it as the maximally-spread `wave packet'. 
The ground state is a well-localized wave packet in the weak-coupling limit, but as coupling grows the wave packet gradually spreads out and eventually becomes constant in the strong-coupling limit.  

Note that the ground state is gauge-invariant:
\begin{align}
\hat{\Omega}|E=0\rangle= |E=0\rangle. 
\end{align}

Let us consider low-energy gauge-invariant states consisting of a small number of closed strings with a total length $L\sim N^0$. Then, there are at most order-$N^0$ number of intersections. The interaction (splitting or joining) at each intersection is suppressed by $\frac{1}{N}$, and hence the interaction is negligible at large $N$. The energy of the system is simply $\frac{L}{2}$. Such states are in the confined phase. 
If we introduce a probe quark-antiquark pair connected by the open Wilson line, the energy increases linearly as $E_{q\bar{q}}=\frac{L}{2}$. This leads to the linear confinement potential.

In the deconfined phase~\cite{Polyakov:1978vu,Susskind:1979up}, long strings with lengths of order $N^2$ condense. There are many intersections, and thus the $1/N$-suppressed interactions from many intersections pile up and become non-negligible as a whole. Intuitively, if we introduce a short open string, it interacts with a condensed long string and forms a long open string, which allows us to separate quark and antiquark without making the string longer.

Finally, we consider partial deconfinement. 
Let $\hat{U}_{\rm dec}$ be the SU($M$)-subsector. The SU($M$)-deconfined states can be constructed by acting long traces of $\hat{U}_{\rm dec}$'s on $|E=0\rangle$~\cite{Hanada:2019czd}.
By using the Wilson loops restricted to the SU($M$)-subsector 
\begin{align}
\hat{W}_{{\rm dec},C}={\rm Tr}(\hat{U}_{{\rm dec};\mu,\vec{n}}\hat{U}_{{\rm dec};\nu,\vec{n}+\hat{\mu}}\cdots),
\end{align}
we can construct multi-string states 
\begin{align}
\hat{W}_{{\rm dec},C}\hat{W}_{{\rm dec},C'}\cdots|E=0\rangle,
\end{align}
and further take a linear combination of them.  Such states are SU($N-M$)-invariant, but not SU($N$)-invariant. If we want to %get 
obtain an SU($N$)-invariant state, we just act the projector $\hat{\pi}$. 
%%%%%%%%%%%%
%%%%%%%%%%%%
\subsubsection{Continuum limit (weak coupling on lattice)}\label{sec:YM-continuum-limit}
\hspace{0.51cm}
%%%%%%%%%%%%
%%%%%%%%%%%%
Now we discuss how the wave-packet picture generalizes to the continuum limit of lattice-regularized Yang-Mills, which was not explicitly discussed in the past.

To obtain Yang-Mills theory from Wilson's plaquette action, one usually considers fluctuation about $U_{\mu,\vec{x}}=\textbf{1}$ 
and correspondingly
$A_{\mu,\vec{x}}=\textbf{0}$. Strictly speaking, this is not a gauge-invariant statement, because this configuration transforms as $\textbf{1}\to\Omega_{\vec{n}}^{-1}\Omega_{\vec{n}+\hat{\mu}}$ by a gauge transformation. More precisely, one considers fluctuations around a pure-gauge configuration
\begin{align}
U_{\mu,\vec{n}}
=
\Omega_{\vec{n}}^{-1}\Omega_{\vec{n}+\hat{\mu}}\, . 
\label{pure-gauge}
\end{align}
Because both the plaquette action and the continuum action are SU($N$)-invariant, this ambiguity is not a problem at all. 
In terms of Hilbert space, however, this nature of gauge transformation leads to a puzzling conceptual issue, because such a state is \textit{not} invariant under local SU($N$) transformation.
More generally, no localized wave packet is SU($N$)-invariant because the center of a wave packet, which is a unitary matrix, cannot be SU($N$)-invariant. Then, how can the mechanism of partial deconfinement introduced in Sec.~\ref{sec:underlying_mechanism}, which is based on the invariance of the wave packet under a subgroup of SU($N$), work? 

To resolve this issue, let us consider the ground state first. The ground state should be the wave packet localized around a pure-gauge configuration \eqref{pure-gauge}. Such a state is invariant under the `global' SU($N$) transformation by $\Omega'_{\vec{n}}\equiv\Omega_{\vec{n}}^{-1}V\Omega_{\vec{n}}$, where $V$ is independent of $\vec{n}$:
\begin{align}
\Omega_{\vec{n}}^{\prime -1}
(\Omega_{\vec{n}}^{-1}
\Omega_{\vec{n}+\hat{\mu}})
\Omega^{\prime}_{\vec{n}+\hat{\mu}}
=\Omega_{\vec{n}}^{-1}\Omega_{\vec{n}+\hat{\mu}}\, . 
\end{align}
The invariance under this `global' SU($N$) transformation can lead to the enhancement factor discussed in Sec.~\ref{sec:underlying_mechanism}. Furthermore, essentially the same enhancement factor can appear if $V$ depending on $\vec{n}$ is sufficiently slowly varying. Namely, for $\Omega'_{\vec{n}}\equiv\Omega_{\vec{n}}^{-1}V_{\vec{n}}\Omega_{\vec{n}}$, 
\begin{align}
\Omega_{\vec{n}}^{\prime -1}(\Omega_{\vec{n}}^{-1}\Omega_{\vec{n}+\hat{\mu}})\Omega^{\prime}_{\vec{n}+\hat{\mu}}
=
\Omega_{\vec{n}}^{-1}
V_{\vec{n}}^{-1}
V_{\vec{n}+\hat{\mu}}
\Omega_{\vec{n}+\hat{\mu}}\, , 
\end{align}
and hence, if $V_{\vec{n}}^{-1}V_{\vec{n}+\hat{\mu}}$ is sufficiently close to $\textbf{1}$ we can still have a significant enhancement factor. 

Next, we consider the excited states. Because we are interested in the continuum limit, we write generic link variables as
\begin{align}
U_{\mu,\vec{n}}
=
\Omega_{\vec{n}}^{-1}
e^{iaA_{\mu,\vec{n}}}
\Omega_{\vec{n}+\hat{\mu}} 
\end{align}
and assume that $aA_{\mu,\vec{n}}$ converges to zero in the continuum limit ($a\to 0$). 
By expanding the lattice theory with respect to this $A_{\mu,\vec{n}}$, we would obtain the continuum theory. 
Suppose that we could take $A_{\mu,\vec{n}}$ to be invariant under the adjoint action of $\textrm{SU}(N-M)\subset\textrm{SU}(N)$, i.e., $V^{-1}A_{\mu,\vec{n}}V=A_{\mu,\vec{n}}$ for $V\in\textrm{SU}(N-M)\subset\textrm{SU}(N)$. Then, such a state acquires the enhancement factor under the `global' SU($N-M$) transformation by $\Omega'_{\vec{n}}=\Omega_{\vec{n}}^{-1}V\Omega_{\vec{n}}$. 
Now we let $V=e^{i\Lambda}$ depend on $\vec{n}$, but we assume it changes slowly. 
%(We explain the precise meaning of `slowly' below.) 
Then, by ignoring the higher-order terms with respect to the lattice spacing $a$, the gauge transformation is 
\begin{align}
A_{\mu,\vec{n}}
\to
A_{\mu,\vec{n}}
+
i[A_{\mu,\vec{n}},\Lambda_{\vec{n}}]
+
\frac{\Lambda_{\vec{n}+\hat{\mu}}-\Lambda_{\vec{n}}}{a}. 
\end{align}
By assumption, $[A_{\mu,\vec{n}},\Lambda_{\vec{n}}]$ is zero. 
The last term corresponds to the derivative $\partial\Lambda$ in the continuum limit. 
If this term is much smaller than the size of the wave packet (specifically if the shift of the center of wave packet from the origin, $\sqrt{\textrm{Tr}(\partial\Lambda)^2}$, is much smaller than the spread of the wave packet along each direction of $\mathbb{R}^{3(N^2-1)}$), then a sufficiently large overlap exists between the wave packet before and after the gauge transformation. 
This gives a substantial enhancement factor compared to a generic wave packet localized around a nonzero value of $A_\mu$. 
Note that the requirement of small $\partial\Lambda$ removes the potential ultraviolet divergence of the contribution to free energy from the enhancement factor.

%\textcolor{red}{
%In the case of lattice regularization, the link variable $U_{\mu,\vec{x}}$ and continuum gauge field $A_{\mu,\vec{x}}$ are related by $U_{\mu,\vec{x}}=e^{iag_{\rm YM}A_{\mu,\vec{x}}}$. 
%The coupling constant $g_{\rm YM}$ is of order $a^0$ (or logarithm of $a$), and $A_{\mu,\vec{x}}$ is of order $a^0$, and hence, the size of the wave packet is roughly of order $a^1$. Therefore, $V_{\vec{n}}^{-1}V_{\vec{n}+\hat{\mu}}=e^{iaX_{\vec{n},\mu}}$, where $X_{\vec{n},\mu}$ is a random matrix with $O(1)$ entries, are allowed.  
%}

When partial deconfinement takes place, the plaquette $U_{\square,\mu,\nu,\vec{n}}$ would be centered around
\begin{align}
\left(
\begin{array}{cc}
U'_{\square,\mu,\nu,\vec{n}} & \textbf{0} \\
\textbf{0} & \textbf{1}
\end{array}
\right)
\end{align}
up to gauge transformation, and 
$F_{\mu\nu,\vec{n}}$ would be centered around
\begin{align}
\left(
\begin{array}{cc}
F'_{\mu\nu,\vec{n}} & \textbf{0} \\
\textbf{0} & \textbf{0}
\end{array}
\right).
\end{align}
Such $V=e^{i\Lambda}$ that is slowly varying and leaves this block structure invariant gives the enhancement factor. 

For the confined vacuum which is genuinely gauge-invariant, such a $V$ can be arbitrary, modulo the slowly-varying nature. 
%Therefore, 
This fact tells us that \textit{the distribution of Polyakov line phases must be Haar-random.}
We emphasize that this is a stronger condition than the unbroken center symmetry, and it can be generalized to QCD, even at finite $N$. 
%%%%%%%%%%%%
%%%%%%%%%%%%
\subsubsection*{Comments on renormalization}
\hspace{0.51cm}
%%%%%%%%%%%%
%%%%%%%%%%%%
The discussion above relates the bare Polyakov line and symmetry of wave function on each spatial link. To obtain a reasonable physical picture that is not sensitive to UV cutoff, we implicitly assumed that the lattice spacing is not small. When we take the continuum limit, we need to take renormalization into account. 

Let us consider the expectation value of the Polyakov loop. In the continuum limit $a\to 0$ with fixed temperature, the expectation value of the bare Polyakov loop in any nontrivial representation vanishes even in the deconfined phase and the bare Polyakov line becomes Haar random. A multiplicative renormalization factor is needed to keep the expectation value finite~\cite{Gupta:2007ax,Mykkanen:2012ri}. 

This is not surprising from the symmetry point of view because it should be hard to distinguish the wave function on each link from the ground state when we zoom in to a very short distance scale. Nontrivial properties at IR can appear because a small deviation at each link accumulates into a significant difference due to the growing number of links. 

To see the properties at IR, it would be better to use an appropriately renormalized Polyakov line. Corresponding to each renormalization scheme of the Polyakov loop, the renormalization scheme of the Polyakov line can be obtained via the character expansion. Another possible option is to use a smeared gauge field and define a smeared version of the Polyakov loop~\cite{Datta:2015bzm,Petreczky:2015yta}.

%%%%%%%%%%n factor is needed to %%
%%%%%%%%%%%%
\subsection{Large-$N$ QCD}\label{sec:large-N-QCD}
\hspace{0.51cm}
%%%%%%%%%%%%
%%%%%%%%%%%%
Let us consider SU($N$) QCD with $N_{\rm f}$ flavors of fundamental quarks. The only difference from Yang-Mills theory is the existence of fermions in the fundamental representation. 
Therefore, the mechanism discussed above is applicable to QCD.

\begin{figure}[htbp]
\begin{center}
\scalebox{0.3}{
\includegraphics{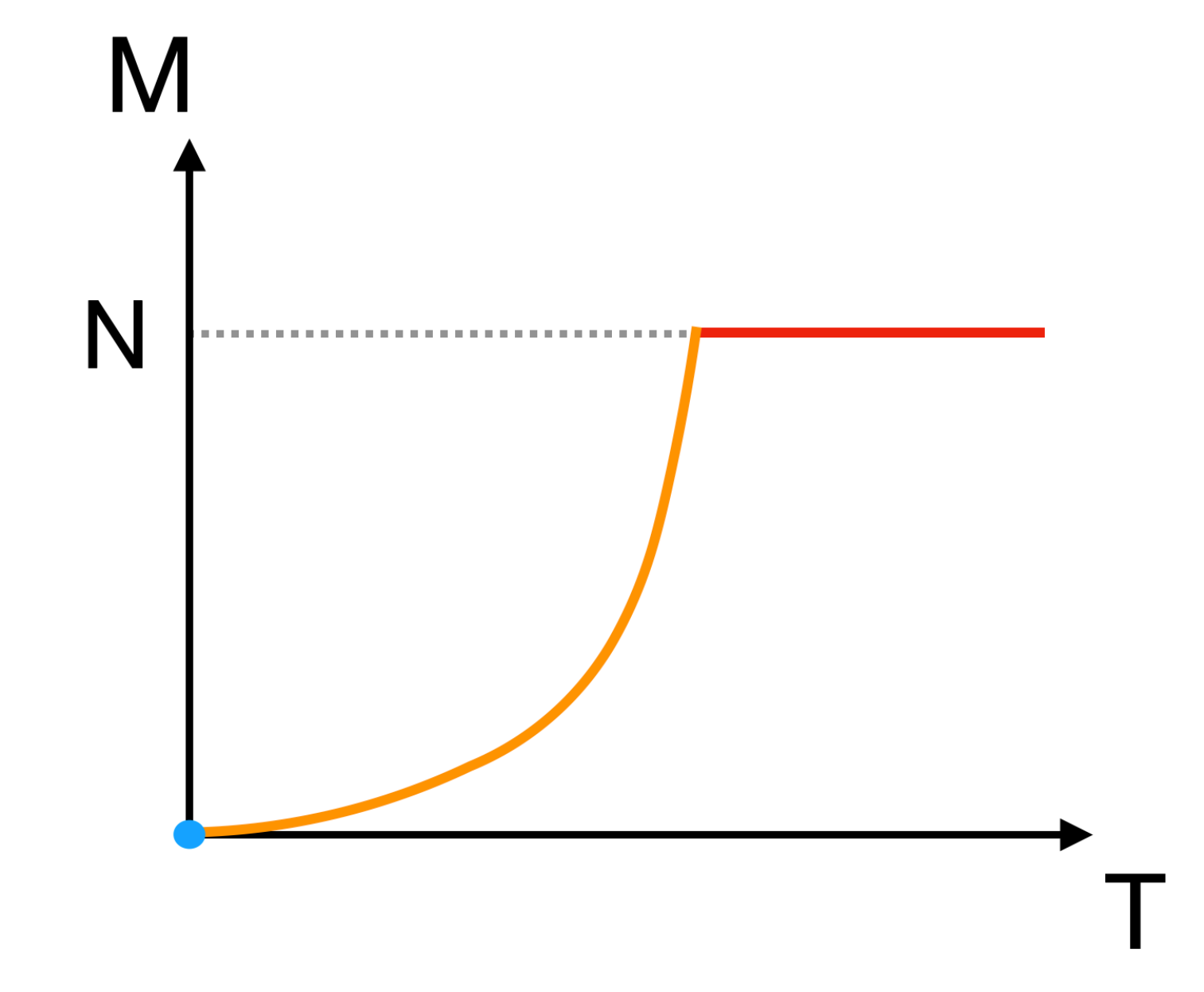}}
\end{center}
\caption{
A schematic picture of the temperature dependence of the size of the deconfined sector $M$ (see Fig.~\ref{fig:QCD-deconf-pattern} for the definition) in weakly-coupled QCD on S$^3$~\cite{Hanada:2019czd}. Complete confinement is realized only at zero temperature. 
}\label{fig:QCD-deconf-temperature}
\end{figure}

We assume that quark chemical potential is zero so that the fermion determinant is real and positive. Then, due to the Vafa-Witten theorem~\cite{Vafa:1983tf}, the vector part of the flavor symmetry should not be broken. Therefore, we expect the deconfinement pattern depicted in Fig.~\ref{fig:QCD-deconf-pattern}. This was confirmed for the weakly-coupled QCD on three-sphere, by re-interpreting the results obtained by Schnitzer~\cite{Schnitzer:2004qt}. The temperature dependence is depicted schematically in Fig.~\ref{fig:QCD-deconf-temperature}. For details, see~\cite{Hanada:2019kue}.  

%%%%%%%%%%%%
%%%%%%%%%%%%
\subsection{Some physical consequences}\label{sec:physical-consequence}
\hspace{0.51cm}
%%%%%%%%%%%%
%%%%%%%%%%%%

%%%%%%%%%%%%
%%%%%%%%%%%%
\subsubsection{Flux tube in confined sector}\label{sec:flux-tube}
\hspace{0.51cm}
%%%%%%%%%%%%
%%%%%%%%%%%%
In pure Yang-Mills theory, the formation of the flux tube is a good characterization of confinement. When partial deconfinement takes place, we naturally expect the existence of the flux tube in the confined sector. 

This idea was tested for the case of large-$N$ strong-coupling limit of lattice gauge theory~\cite{Gautam:2022exf}. To separate confined and deconfined sectors, the gauge-fixing condition discussed in Sec.~\ref{sec:Path-integral-gauge-fixing} was used. Then, the Polyakov loop in those sectors can be defined by
\begin{align}
P_{\rm dec} = \frac{1}{M}\sum_{j=1}^M e^{i\theta_j}\, , 
\qquad
P_{\rm con} = \frac{1}{N-M}\sum_{j=M+1}^N e^{i\theta_j}\, .  
\end{align}
See also Fig.~\ref{fig:phase-distribution}. 
String tension can be obtained from the two-point function of Polyakov loops. Modulo mild assumptions (including that partial deconfinement is taking place), a few exact results can be obtained and they were reproduced by lattice simulations. For details, see Ref.~\cite{Gautam:2022exf}. 
%%%%%%%%%%%%
%%%%%%%%%%%%
\subsubsection{Chiral symmetry breaking}
\hspace{0.51cm}
%%%%%%%%%%%%
%%%%%%%%%%%%
There is a famous argument based on 't Hooft anomaly that relates chiral symmetry breaking and confinement~\cite{tHooft:1979rat}. At high energy, free quarks carry some 't Hooft anomaly. If confinement takes place, free quarks are gone (because of the flux tube discussed in Sec.~\ref{sec:flux-tube}), and hence some other degrees of freedom must reproduce the same anomaly. If chiral symmetry is broken, pions can carry the correct amount of the 't Hooft anomaly. 

Nothing forbids us from applying the same idea at the intermediate energy scale where partial deconfinement is taking place. In the confined sector, free quarks are gone, and hence, it is natural to expect that the chiral symmetry is broken so that pions in the confined sector can account for the anomaly~\cite{Hanada:2019kue}. If this is the case, the chiral-symmetry breaking should set in at or above the GWW point. Given the absence of other characteristic energy scales, there is no strong motivation to believe that the chiral-symmetry breaking and the GWW point do not coincide, and hence, it is natural to conjecture that the chiral symmetry breaks at the GWW transition point. This conjecture was confirmed for the case of large-$N$ strongly-coupled lattice gauge theory by lattice simulations~\cite{Hanada:2021ksu}.
%%%%%%%%%%%%
%%%%%%%%%%%%
\subsubsection{CP symmetry breaking at $\theta=\pi$}\label{sec:CP-symmetry-breaking}
\hspace{0.51cm}
%%%%%%%%%%%%
%%%%%%%%%%%%
Yang-Mills theory admits the $\theta$-term that is defined by the second term in the following Lagrangian density:
\begin{align}
\mathcal{L}
=
\frac{1}{4g^2}\textrm{Tr}(F_{\mu\nu}F^{\mu\nu})
-
\frac{i\theta}{8\pi^2}\textrm{Tr}(F_{\mu\nu}\tilde{F}^{\mu\nu})\, . 
\end{align}
The theory possesses the CP symmetry at $\theta=0$ and $\theta=\pi$. 
Interestingly, at $\theta=\pi$, the CP symmetry breaks spontaneously in the confined phase~\cite{Gaiotto:2017yup}. This motivates us to speculate that, similarly to the chiral symmetry breaking, CP symmetry breaking coincides with the GWW transition. 

The CP-symmetry breaking at $\theta=\pi$ can be studied systematically for the case of mass-deformed $\mathcal{N}=1$ SYM, by focusing on the weakly-coupled region~\cite{Chen:2020syd}. One can explicitly check that, in the large-$N$ limit, the CP symmetry breaks at the GWW point~\cite{Hanada:2021ksu}. Furthermore, one can also investigate finite-$N$ theories. At $N\ge 3$, we can find the unstable saddle that can be identified with the partially-deconfined phase and stable saddles that correspond to the completely-deconfined and completely-confined phases. The CP-symmetry breaking takes place precisely at the border between the partially-deconfined and completely-deconfined saddles.  
%%%%%%%%%%%%
%%%%%%%%%%%%
\section{Partial deconfinement at finite $N$}\label{sec:Finite-N}
\hspace{0.51cm}
%%%%%%%%%%%%
%%%%%%%%%%%%
Crisp characterization of phase transitions associated with partial deconfinement in the large-$N$ limit was given by using the distribution of Polyakov line phases: the departure from the constant distribution is the transition from the completely-confined phase to partially-deconfined phase, and the GWW transition is the transition from the partially-deconfined phase to completely-deconfined phase. These characterizations rely on the specific features of the large-$N$ limit.

In this section, we discuss how partial deconfinement could be observed in finite-$N$ theories. On top of that, we will discuss the application to SU(3) QCD in Sec.~\ref{sec:SU(3)-QCD}. We will consider two possibilities: to find the counterparts of constant distribution and GWW transition at finite $N$ from the Polyakov loop (Sec.~\ref{sec:pol_dist_finite_N}) or to introduce different characterizations that do not require large $N$ based on global symmetry and instanton (Sec.~\ref{sec:global_symmetry_finite_N}). 
In Sec.~\ref{sec:SU(3)-QCD}, we will examine these two perspectives in analyses of SU(3) gauge theory with dynamical fermions.
%We can compare them to check the consistency of our analyses of the SU(3) theory in Sec.~\ref{sec:SU(3)-QCD}. 
%%%%%%%%%%%%%%%%%%
\subsection{Distribution of Polyakov line phases at finite $N$}\label{sec:pol_dist_finite_N}
%%%%%%%%%%%%%%%%%%
Symmetry in extended Hilbert space and the amount of gauge redundancy are well-defined notions both at large $N$ and at finite $N$. 
The Polyakov loop $P_{\vec{x}}$ defined at each point $\vec{x}$ provides us with a convenient way to estimate them. The set of $P_{\vec{x}}$ specifies an element in a big group $\mathcal{G}=\prod_{\vec{x}}[{\rm SU}(N)]_{\vec{x}}$. We can diagonalize them and get $N$ eigenvalues at each point.
%%%%%%%%%%%%%%%%%%
\subsubsection*{Complete confinement and Haar-random distribution}
%%%%%%%%%%%%%%%%%%
For theories with a center symmetry such as SU($N$) pure Yang-Mills theory, an unbroken center symmetry provides a simple way to detect the complete confinement. However, the characterization by center symmetry does not apply to QCD, because such a symmetry does not exist. A much better characterization is based on the \textit{genuine gauge invariance} of the confined vacuum~\cite{Hanada:2020uvt}, i.e., the gauge invariance in the extended Hilbert space. 
As discussed in Sec.~\ref{sec:QFT}, it follows that \textit{the distribution of Polyakov line phases must be Haar-random}. The Haar-random distribution was constant in the large-$N$ limit. As we discuss in Appendix~\ref{sec:Haar-random-distribution}, the Haar-random distribution is given by 
\begin{align}
\rho_{\rm Haar}(\theta)=\frac{1}{2\pi}\left(1-(-1)^N\cdot\frac{2}{N}\cos(N\theta)\right)
\end{align}
for SU($N$). Deviation from this Haar-random distribution can be a good criterion to observe the transition from the completely-confined phase to the partially-deconfined phase, even at finite $N$. 
At large $N$, it is indeed true for the theories we have investigated.

Let us consider the Fourier expansion of the phase distribution, 
\begin{align}
\rho_{\rm Polyakov}(\theta)
&=
\frac{1}{2\pi}
+
\frac{1}{2\pi}
\sum_{n>0}
\left(
\tilde{\rho}_ne^{-in\theta}
+
\tilde{\rho}_{-n}e^{in\theta}
\right)
\nonumber\\
&=
\frac{1}{2\pi}
+
\frac{1}{2\pi}
\sum_{n>0}
2\tilde{\rho}_n\cos(n\theta)\, . 
\label{eq:Pol-dist-Fourier}
\end{align}
Here, we assumed $\tilde{\rho}_n=\tilde{\rho}_{-n}\in\mathbb{R}$ for simplicity, which is indeed the case for QCD at zero chemical potential. 
The coefficient $\tilde{\rho}_n$ is the expectation values of the $n$-wound Polyakov loop $u_n\equiv\frac{1}{N}\textrm{Tr}(\mathcal{P}^n)$: 
\begin{align}
\tilde{\rho}_n
=
\langle u_n\rangle
=
\frac{1}{N}
\left\langle
\textrm{Tr}(\mathcal{P}^n)\right\rangle\, . 
\end{align}
For the Haar-random distribution, these coefficients are
\begin{align}
\tilde{\rho}_{{\rm Haar},n}
=
\left\{
\begin{array}{cc}
\frac{(-1)^N}{N} & (n=\pm N)\\
0 & (n\neq\pm N)\, . 
\end{array}
\right.
\end{align}
Therefore, $\langle\textrm{Tr}\mathcal{P}^n\rangle$ is nonzero only for $n=\pm N$. This is consistent with the existence of baryon as a bound state of $N$ quarks.

One may not find the appearance of the Haar-random distribution too surprising in the strict zero-temperature limit, because the Polyakov loop in the path-integral formalism is the product of infinitely many unitary link variables. However, that the Haar-randomness should survive even at finite temperature modulo small corrections due to hadron gas is a highly nontrivial statement.\footnote{
The contribution from hadron gas is $1/N$-suppressed, and hence, it is hard to see as long as the 't Hooft expansion is valid. However, if we consider the parameter region where the 't Hooft $1/N$ expansion is not good, we may see a large correction to the Haar-random distribution. An example is the weakly-coupled regime of mass-deformed $\mathcal{N}=1$ SYM discussed in Sec.~\ref{sec:CP-symmetry-breaking}, where the phases are frozen at the multiples of $\frac{2\pi}{N}$. In this particular setup, the phase transition takes place in the weakly-coupled regime or, equivalently, at rather high temperatures. 
}  
%%%%%%%%%%%%%%%%%%
\subsubsection*{Finite-$N$ counterpart of GWW}
%%%%%%%%%%%%%%%%%%
One may also imagine that the onset of the gap in the phase distribution, which is, by definition, the GWW point in the large-$N$ limit, is somehow related to the transition between complete deconfinement and partial confinement. However, as we will see in Sec.~\ref{sec:SU(3)-QCD}, $N=3$ does not seem to be large enough to admit such a naive interpretation. 

A better criterion comes from the behavior of the multiply-wound Polyakov loops. At large $N$, the phase distribution $\rho(\theta)$ in the completely-deconfined phase is ``gapped'', i.e., $\rho(\theta)$ is zero in a certain finite interval. This requires that the expansion \eqref{eq:Pol-dist-Fourier} does not terminate at finite order.\footnote{
If a given function is identically zero at a finite interval, its derivatives at arbitrary order are also identically zero. It gives infinitely many constraints to the Fourier coefficients that cannot be solved if the expansion terminates at a finite order.  
} Therefore, multiply-wound loops with arbitrarily large winding numbers can have a nonzero expectation value. 
In particular, in the infinite-temperature limit, the distribution is the delta function and all loops contribute equally.
Intuitively, a state consisting of any number of probe quarks can be excited rather easily, which is a natural consequence of deconfinement. 

The absence of the gap in the distribution does not immediately guarantee the absence or suppression of the loops with large wingding numbers. Still, in the partially-confined phase of the large-$N$ theories we studied analytically or numerically in the past, we did not observe significant contributions from such loops. It would be a natural feature given that the deconfined sector is not sufficiently excited; when more energy is added to the system, it is used to make the deconfined sector larger (to increase $M$), and hence, the excitations are diluted across the deconfined degrees of freedom. Because this intuition does not require the large-$N$ limit, we conjecture that the condensation of multiply-wound Polyakov loops with large winding numbers can be regarded as a finite-$N$ counterpart of the GWW point.  

Equivalently, we can consider the characters $\chi_r(\mathcal{P})$ corresponding to irreducible representations $r$. The character $\chi_{r}$ is the Polyakov loop in the representation $r$. (The usual Polyakov loop is for the fundamental representation.) 
That the expectation value $\langle\chi_r\rangle$ is nonzero would mean that the excitation in the representation $r$ can be generated with finite free energy. 
%That the expectation value $\langle\chi_r\rangle$ is nonzero would mean that the excitation in the representation $r$ can be excited without costing free energy. 
In Sec.~\ref{sec:GWW-in-QCD?}, we will examine lattice QCD configurations and find that the fundamental representation deconfines earlier than other representations. 
%%%%%%%%%%%%%%%%%%
\subsection{Characterization via global symmetry and instanton condensation}\label{sec:global_symmetry_finite_N}
%%%%%%%%%%%%%%%%%%
Another natural characterization of partial confinement at finite $N$ is to make use of global symmetry~\cite{Hanada:2021ksu}. As discussed in Sec.~\ref{sec:physical-consequence}, the onset of partial deconfinement (i.e., the transition between complete deconfinement and partial deconfinement) can be associated with the breaking of global symmetry, such as chiral symmetry, at least in the concrete examples studied in Ref.~\cite{Hanada:2021ksu}. Breaking of global symmetry is a well-defined concept even at finite $N$, and hence, it is natural to adopt it to detect partial deconfinement at finite $N$.

When the confinement/deconfinement transition is of first order, the maximum of free energy provides us with a natural counterpart of the partially-deconfined phase that can be generalized to finite $N$. In the plot of the free energy as a function of temperature, the cusp can be regarded as the transition between the completely-deconfined phase and the partially-deconfined phase (Fig.~\ref{fig:three=patterns}, center-left).\footnote{
This identification may not always be valid. Specifically, superconformal index with imaginary potential can give a counter-example~\cite{Choi:2021lbk}: at large $N$, the cusp does not coincide with the GWW point. Whether a similar disagreement can be seen in a non-supersymmetric setup is currently not clear. We thank Jack Holden and Seok Kim for the discussions.
} 
In Ref.~\cite{Hanada:2021ksu}, such a cusp was studied for mass-deformed $\mathcal{N}=1$ super Yang-Mills on $\mathbb{R}^3\times\mathrm{S}^1$ with periodic boundary condition. When $\theta$-angle is $\pi$, the CP breaking takes place at this point even at finite $N$. This provides us with a further motivation to use the breaking of global symmetry to detect partial confinement.

Chiral symmetry is a well-defined concept only for massless quarks, while quarks in the real-world QCD are massive. Although one can still enjoy the chiral symmetry of a massless probe quark, its physical meaning is not so obvious.
A natural alternative that could capture partial deconfinement regardless of the mass of quarks is instanton condensation. 
Since the instantons are intimately connected with the chiral symmetry, it is natural to expect the behavior of instantons to change across the GWW transition.
This is indeed the case for the large-$N$ two-dimensional lattice gauge theory (i.e., the original Gross-Witten-Wadia model)~\cite{Buividovich:2015oju}. 
In Sec.~\ref{sec:SU(3)-QCD}, we will use instanton condensation as a probe for the finite-$N$ analog of the GWW transition and compare it to the analyses based on the Polyakov line. 
%%%%%%%%%%%%
%%%%%%%%%%%%
\section{Partial deconfinement in real-world QCD ($N=3$)
}\label{sec:SU(3)-QCD}
\hspace{0.51cm}
%%%%%%%%%%%%
%%%%%%%%%%%%
As discussed in the previous section, it is natural to expect that partial deconfinement takes place in the real-world QCD, with SU(3) gauge group and infinite volume, unless the intermediate-energy region is spatially inhomogeneous. 
In this section, we will address how we can detect partial deconfinement, adopting the ideas provided in Sec.~\ref{sec:Finite-N}. 

For this purpose, we use lattice QCD configurations generated by WHOT-QCD collaboration~\cite{Umeda:2012er}.
In the collaboration, $O(a)$-improved Wilson quark action~\cite{Sheikholeslami:1985ij} coupled with the RG-improved Iwasaki gauge action~\cite{Iwasaki:1983iya} was used to simulate $N_{\rm f}=2+1$ QCD, with lattice spacing is $a\simeq 0.07$ fm, up- and down-quark mass heavier than physical mass (i.e., $\frac{m_\pi}{m_\rho}\simeq 0.63$ is larger than the real-world value $\frac{m_\pi}{m_\rho}\simeq 0.18$) and approximately physical strange-quark mass. 
Although up and down quarks are heavier than in the real world, they are light enough so that there is no first-order phase transition, similar to the case of physical quark mass. 
The scale was set in such a way that the $\rho$-meson mass coincides with the physical value, i.e., $m_\rho=775.45$ MeV. 

Configurations used in our study are listed in Table~\ref{table:WHOT}. The lattice size is $N_t\times 32^3$, where the size along the imaginary-time direction is $N_t=4,6,\cdots,16$.
The third column of the table shows the phase (complete deconfinement, partial deconfinement, or complete confinement) determined based on the following criteria:
\begin{itemize}
    \item
    \textbf{Polyakov loop.}
    Departure from Haar-random distribution indicates partial deconfinement. Condensation of the loops in the higher representations indicates complete deconfinement.
    
    \item 
    \textbf{Instantons.}
    Instanton condensation indicates partial deconfinement or complete confinement. The absence of the instanton condensation indicates complete deconfinement. 
    
\end{itemize}
The details will be discussed below.    

\begin{table}[hbtp]
  \centering
  \begin{tabular}{|c|c|c|}
    \hline
    Lattice size  & Temperature & Phase\\
    \hline 
    $4\times 32^3$   &  697 MeV & \textcolor{red}{CD}\\
    $6\times 32^3$   &  464 MeV & \textcolor{red}{CD}\\
    $8\times 32^3$   &  348 MeV & \textcolor{orange}{PD} or \textcolor{red}{CD}\\
    $10\times 32^3$  &  279 MeV & \textcolor{orange}{PD} \\
    $12\times 32^3$  &  232 MeV & \textcolor{orange}{PD}\\
    $14\times 32^3$  &  199 MeV & \textcolor{orange}{PD}\\
    $16\times 32^3$  &  174 MeV & \textcolor{blue}{CC} or \textcolor{orange}{PD}\\  
    \hline
  \end{tabular}
    \caption{WHOT-QCD configurations of
    $N_{\rm f}=2+1$ QCD~\cite{Umeda:2012er}. Lattice spacing $a\simeq 0.07$ fm, with up and down quarks heavier than physical mass (i.e., $\frac{m_\pi}{m_\rho}\simeq 0.63$ is larger than the real-world value $\frac{m_\pi}{m_\rho}\simeq 0.18$) and approximately physical strange-quark mass. The third column is our estimate of the phase discussed in the main text. 
    \textcolor{red}{CD}, \textcolor{orange}{PD}, and \textcolor{blue}{CC} denote 
    \textcolor{red}{Complete Deconfinement}, 
    \textcolor{orange}{Partial Deconfinement}, and \textcolor{blue}{Complete Confinement}. 
}\label{table:WHOT}
\end{table}
%%%%%%%%%%%%%%%%%%%%%%
%%%%%%%%%%%%%%%%%%%%%%
\subsection{Polyakov line phases vs. partial deconfinement}\label{sec:GWW-in-QCD?}
%%%%%%%%%%%%%%%%%%%%%%
%%%%%%%%%%%%%%%%%%%%%%
The Polyakov loop $P_{\vec{x}}$ can be defined at each spatial point $\vec{x}$. They characterize an element in the group of local transformations $\mathcal{G}=\prod_{\vec{x}}[{\rm SU}(N)]_{\vec{x}}$. 
%that describes the local gauge transformation. 
We can diagonalize them at each point and obtain $3n_{\rm space}$ eigenvalues, where $n_{\rm space}$ is the number of spatial points on the lattice ($n_{\rm space}=32^3=32768$ for the configurations used below). 
In the limit of $n_{\rm space}\to\infty$, we can define continuous phase distribution. When the physical volume goes to infinity, a phase transition can possibly take place, even if $N$ is finite.

In this section, we consider bare, rather than renormalized, Polyakov loops.
Note also that we show only the real part of $\tilde{\rho}_n$'s because, as we explained, they are real from the theoretical point of view. We confirmed that the imaginary part is consistent with zero.

%%%%%%%%%%%%%%%%%%%%%%%%%%%%%%%%%
%%%%%%%%%%%%%%%%%%%%%%%%%%%%%%%%%
\subsubsection*{No emergent center symmetry at high temperatures}
%%%%%%%%%%%%%%%%%%%%%%%%%%%%%%%%%
%%%%%%%%%%%%%%%%%%%%%%%%%%%%%%%%%
At asymptotically high temperatures, we expect that typical states are not invariant under any element of $\mathcal{G}$ but the identity.
The phase distribution should be peaked at $\theta=0$. 

In the context of the Euclidean path integral with imaginary time, people often say `fermions decouple at high temperatures'. This is because fermions have the anti-periodic boundary condition and hence have parametrically large Kaluza-Klein momentum (or Matsubara frequency) with respect to the imaginary time direction which is a circle with circumference $\frac{1}{T}$. Because of this, one may think the high-temperature limit reduces to pure Yang-Mills and the $\mathbb{Z}_3$-center symmetry emerges. Although this intuition correctly explains the dominance of gluonic contributions to many observables, fermions still contribute to the Polyakov loop because they are transformed by $\mathcal{G}$. At high temperature, the fermionic part of typical quantum states is a generic superposition of the ground state and excited states in the Fock basis, i.e., $\otimes_{\vec{n},\alpha,c,f}\left(c^{(0)}_{\vec{n},\alpha,c,f}\ket{0}_{\vec{n},\alpha,c,f}+c^{(1)}_{\vec{n},\alpha,c,f}\ket{1}_{\vec{n},\alpha,c,f}\right)$, where $|c^{(0)}_{\vec{n},\alpha,c,f}|^2\sim |c^{(1)}_{\vec{n},\alpha,c,f}|^2\sim\frac{1}{2}$. 
The indices $\vec{n},\alpha,c$, and $f$ represent lattice site, spinor, color, and flavor. Because fermions are in the fundamental representation, the nontrivial elements of $\mathbb{Z}_3$-transformation act on them as multiplication of phase $e^{\pm\frac{2\pi i}{3}}$. The Fock vacuum does not change, $\ket{0}_{\vec{n},\alpha,c,f}\to\ket{0}_{\vec{n},\alpha,c,f}$, but the excited state transforms as $\ket{1}_{\vec{n},\alpha,c,f}\to e^{\pm\frac{2\pi i}{3}}\ket{1}_{\vec{n},\alpha,c,f}$. This leads to a substantial suppression factor in the partition function, and hence, the $\mathbb{Z}_3$-center symmetry does not emerge.
Therefore, there is no ambiguity regarding the phase distribution analogous to the one in pure Yang-Mills associated with the spontaneous breaking of the center symmetry. 

The situation is different in the 't Hooft large-$N$ limit ($N_{\rm f}$ fixed and $N\to\infty$), where such an effect is negligible compared to the contributions from the gluon sector due to the $1/N$-suppression. The ordering of large-$N$ and high-temperature limits can matter, because the `decoupling' of fermions takes place for different reasons.
%%%%%%%%%%%%%%%%%%%%%%%%%%%%%%%%%
%%%%%%%%%%%%%%%%%%%%%%%%%%%%%%%%%
\subsubsection*{Haar-random distribution at low temperatures}
%%%%%%%%%%%%%%%%%%%%%%%%%%%%%%%%%
%%%%%%%%%%%%%%%%%%%%%%%%%%%%%%%%%
Next, let us consider low temperatures. We expect that the ground state is an element in the extended Hilbert space which is invariant under the slowly-varying part of $\mathcal{G}$. In particular, the phase distribution coincides with that of Haar-random SU(3). As we show in Appendix~\ref{sec:Haar-random-distribution}, the Haar-random distribution is 
\begin{align}
\rho_{\rm Haar}(\theta)=\frac{1}{2\pi}\left(1+\frac{2}{3}\cos(3\theta)\right)\, . 
\end{align}
Therefore, $\langle\textrm{Tr}\mathcal{P}^n\rangle$ is nonzero for $n=\pm 3$, that is consistent with the existence of baryon.\footnote{
One may wonder why other multiples of 3 cannot have nonzero expectation values. A more precise statement is that only $\textrm{Tr}\mathcal{P}^3$ couples to SU(3)-singlet and $\textrm{Tr}\mathcal{P}^6$, $\textrm{Tr}\mathcal{P}^9$, etc, do not couple to two-, three-, or more-baryon states. See Appendix~\ref{sec:winding-loop-vs-character} for details. 
} 

\begin{figure}[htbp]
\begin{center}
\scalebox{0.4}{
\includegraphics{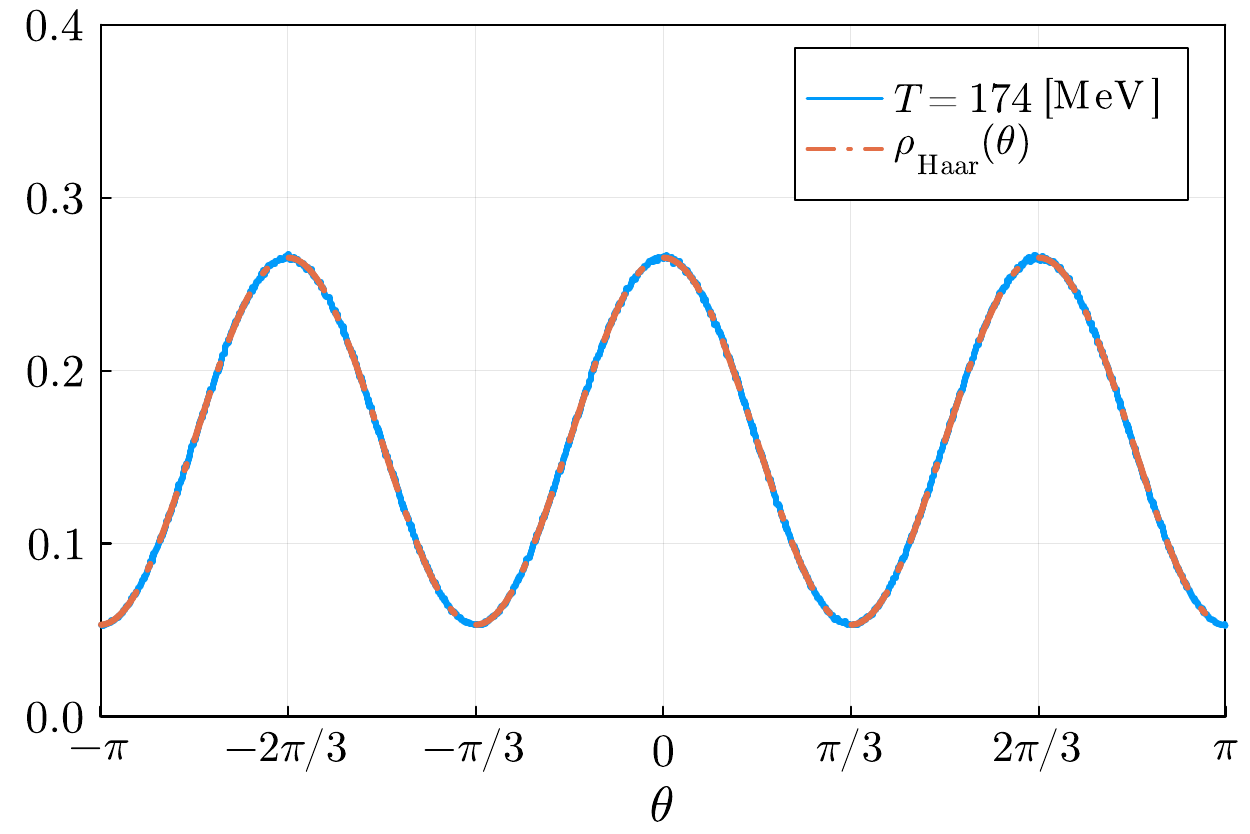}}
\end{center}
\caption{Haar-random distribution vs. Polyakov line phase distribution at 174 MeV, obtained from the WHOT-QCD configurations. The lattice size is $16\times 32^3$ and 599 configurations were used. 
These histograms are drawn with 992 bins.
}\label{fig:Pol-vs-Haar-lowT}
\end{figure}

We emphasize that the center symmetry does not play any role in our argument. We can use the Polyakov loop to describe confinement and deconfinement even for theories without center symmetry including QCD. The vanishing expectation value of the Polyakov loop in QCD at zero temperature is a consequence of the Haar-randomness. In Fig.~\ref{fig:Pol-vs-Haar-lowT}, we compared the Haar-random distribution and the phase distribution obtained from the WHOT-QCD collaboration's configurations at 174 MeV. We can see a good agreement.  

%%%%%%%%%%%%%%%%%%%%%%%%%%%%%%%%
%%%%%%%%%%%%%%%%%%%%%%%%%%%%%%%%
\subsubsection*{Deviation from Haar-random distribution and Hagedorn point}
%%%%%%%%%%%%%%%%%%%%%%%%%%%%%%%%
%%%%%%%%%%%%%%%%%%%%%%%%%%%%%%%%
The Hagedorn point, i.e., the transition between completely-confined and partially-confined phases, can be seen as the departure from the Haar-random distribution. In fact, that the phase distribution becomes Haar random is already somewhat nontrivial. 

\begin{figure}[htbp]
\begin{center}
\scalebox{0.33}{
\includegraphics{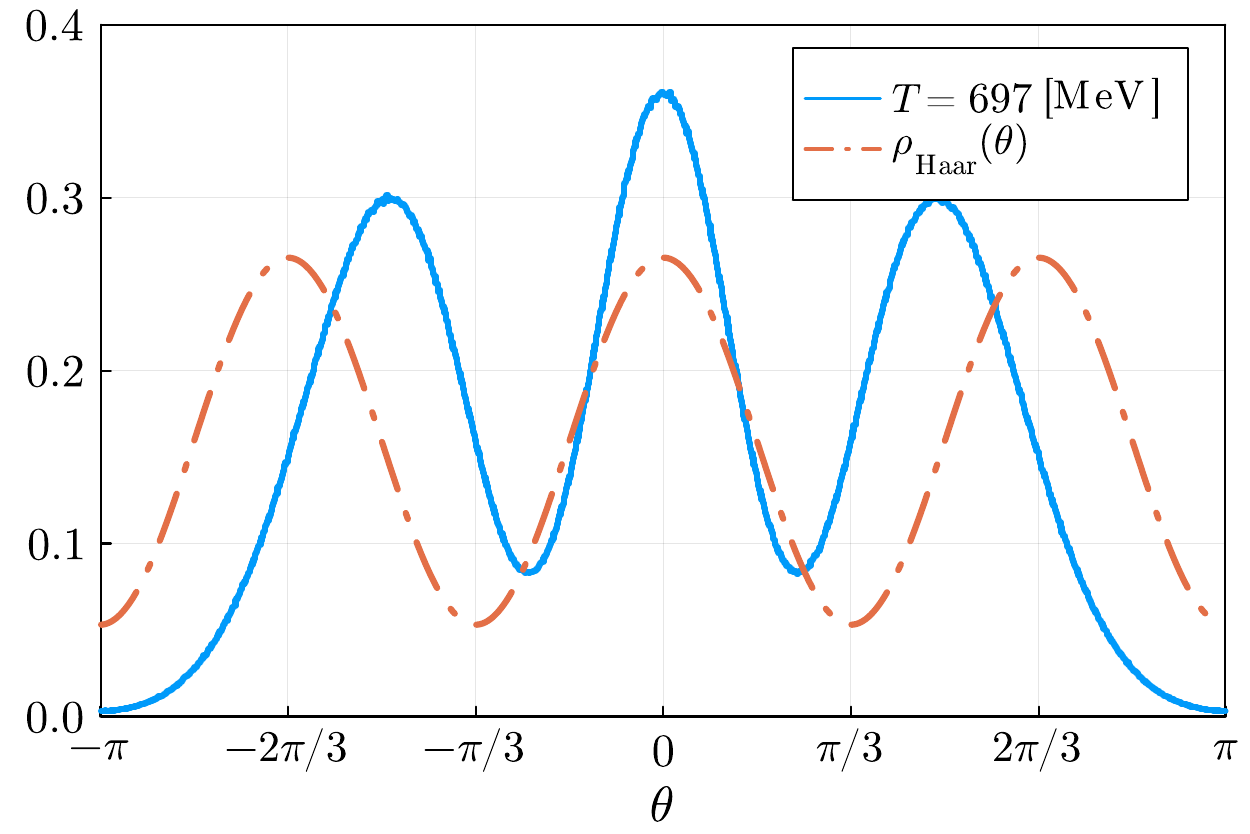}}
\scalebox{0.33}{
\includegraphics{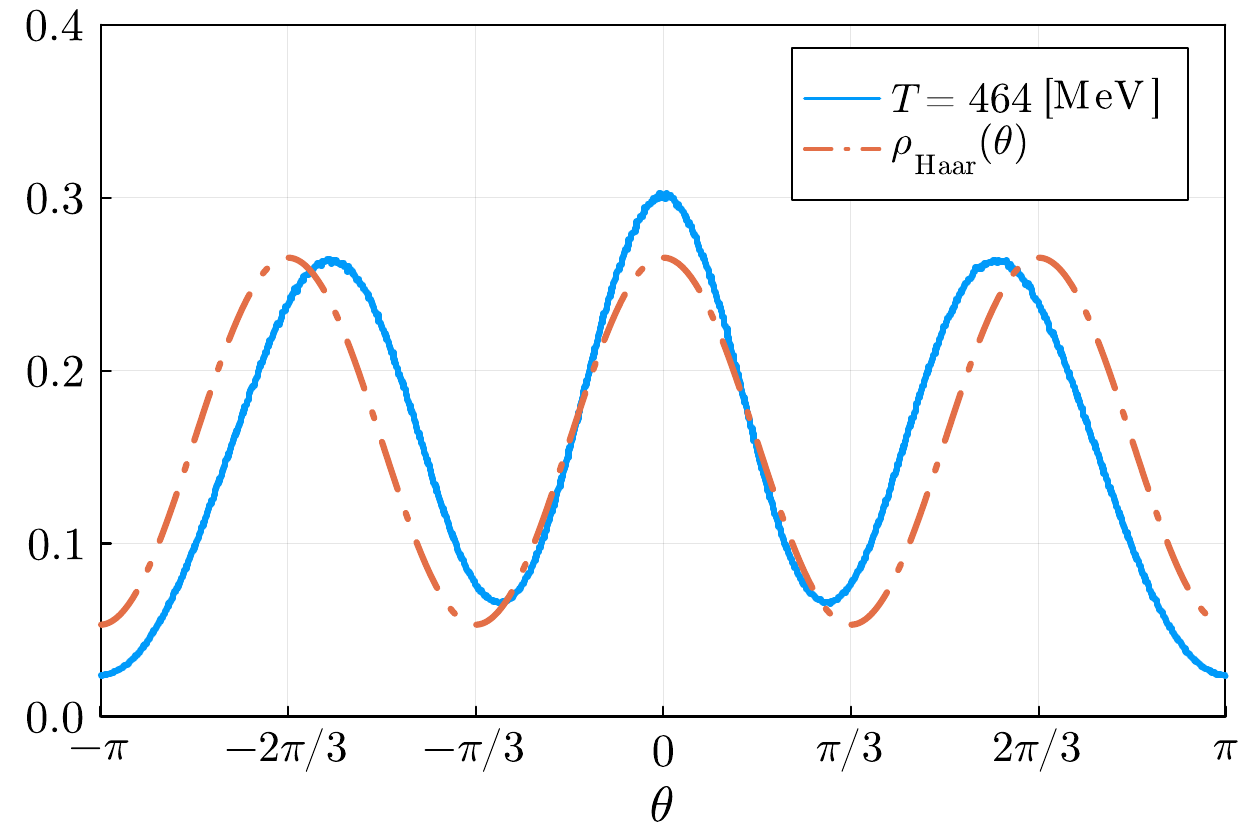}}
\scalebox{0.33}{
\includegraphics{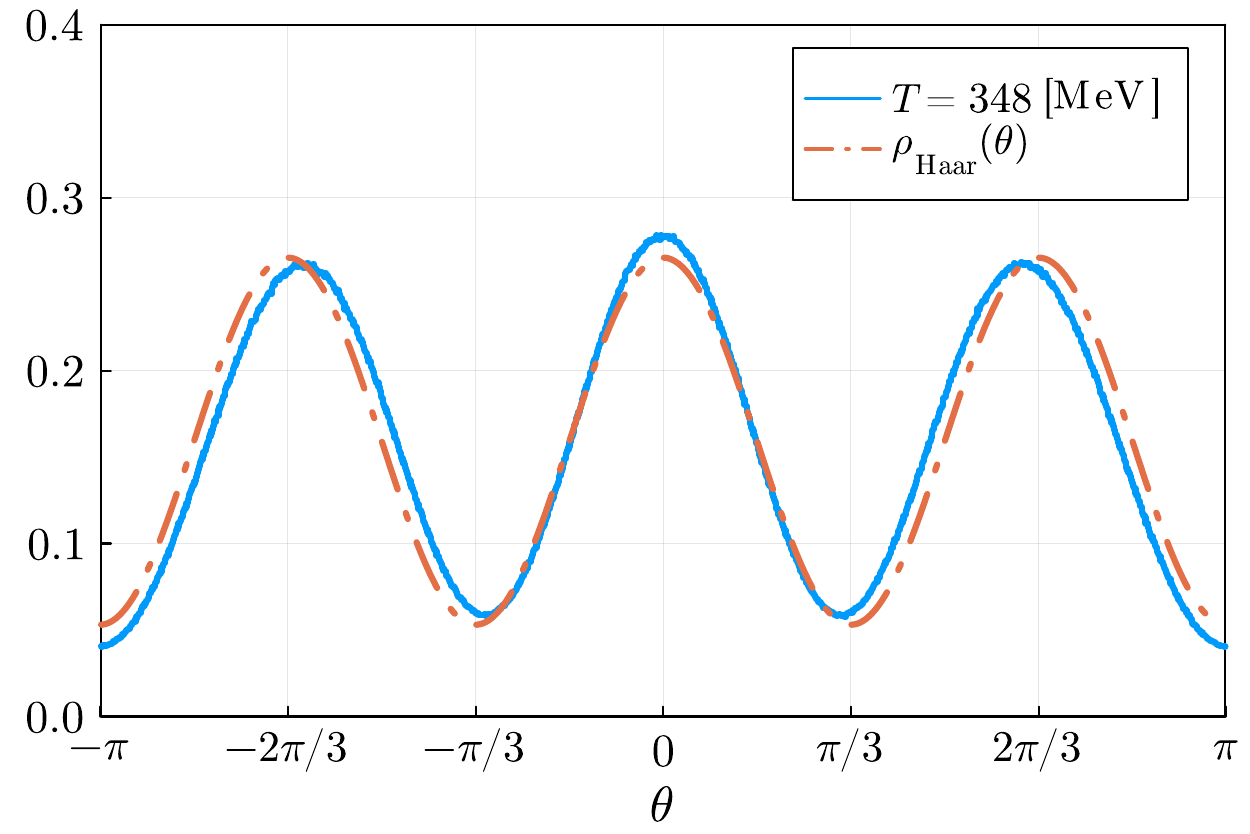}}
\scalebox{0.33}{
\includegraphics{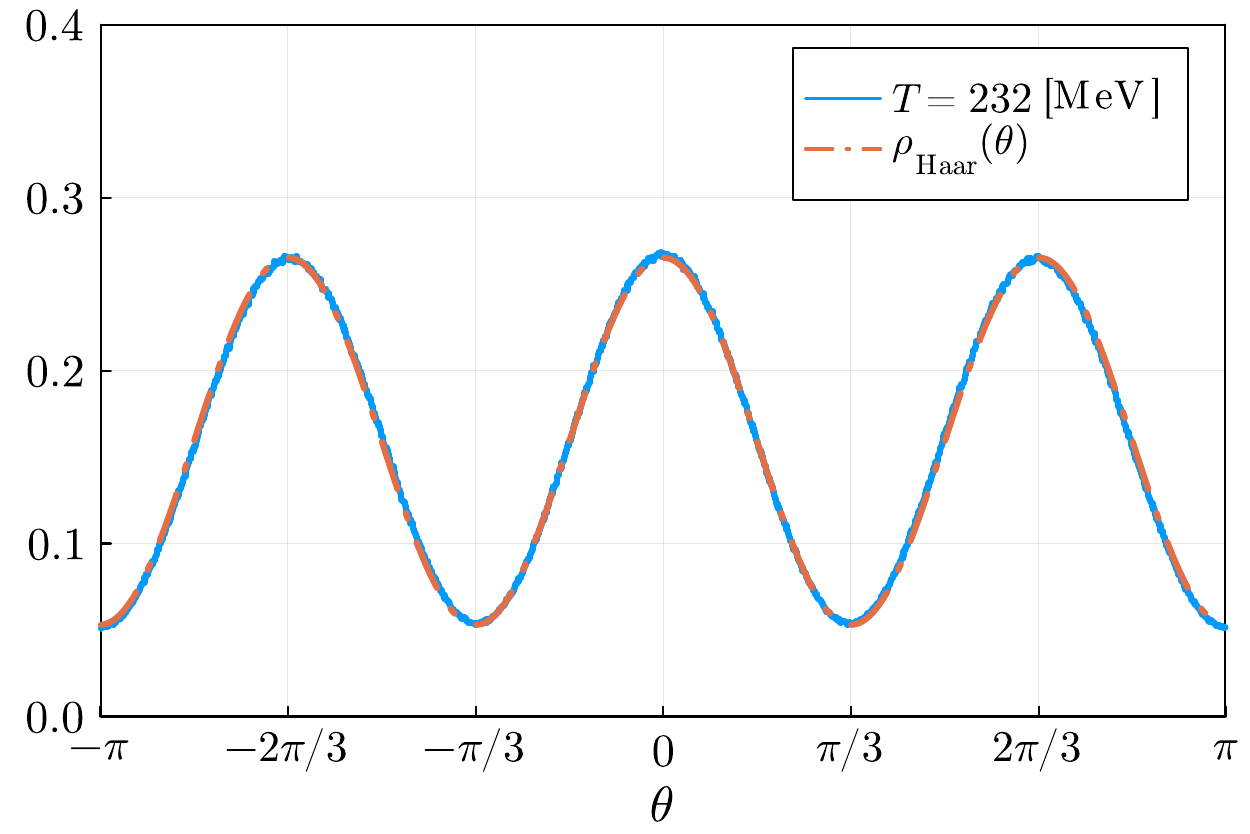}}
\end{center}
\caption{Distributions of Polyakov line phases, $N_t\times 32^3$ lattice, obtained from the WHOT-QCD configurations. The lattice size is $N_t\times 32^3$ ($N_t=4,6,8,12$ and correspondingly $T =$ 697, 464, 348, and 232 MeV.) and 599 configurations were used. Although the agreement with Haar-random distribution seems to be good at $T=232$ MeV, more careful investigation shows the small deviation and hence the onset of partial deconfinement; see Fig.~\ref{fig:multiply-wound}, Fig.~\ref{fig:T-vs-character}, and main text.
See also Fig.~\ref{fig:Pol-vs-Haar-lowT} for $T = $ 174 MeV.
}\label{fig:WHOT-phase}
\end{figure}

\begin{figure}[htbp]
\begin{center}
\scalebox{0.35}{
\includegraphics{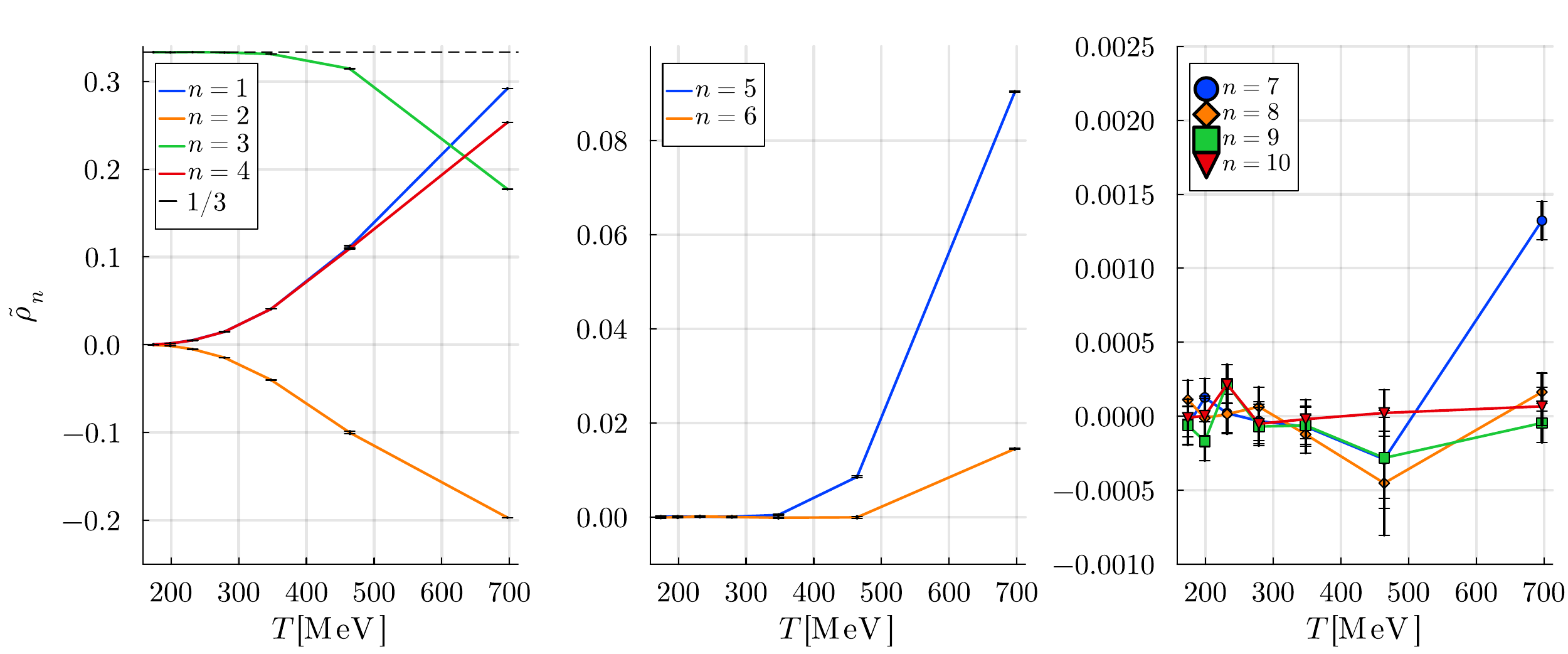}}
\scalebox{0.35}{
\includegraphics{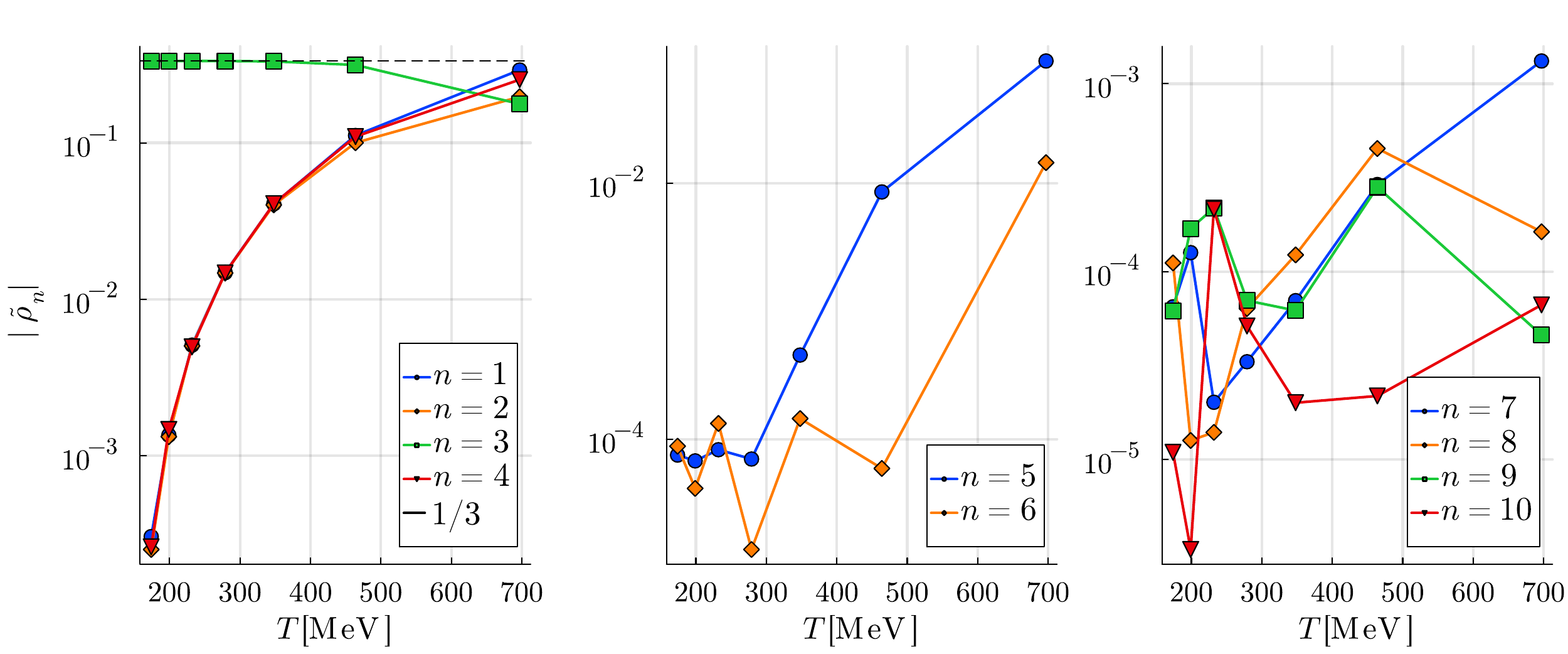}}
\scalebox{0.33}{
\includegraphics{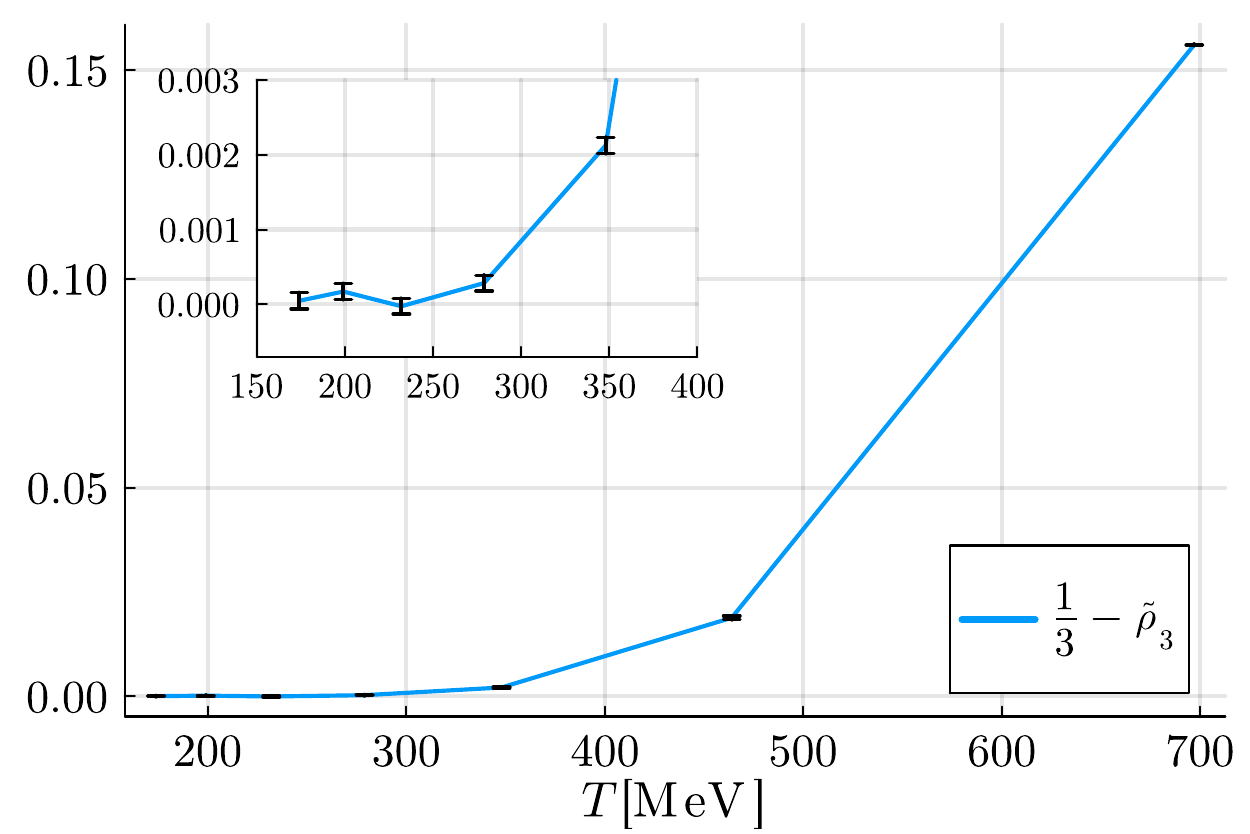}}
\scalebox{0.33}{
\includegraphics{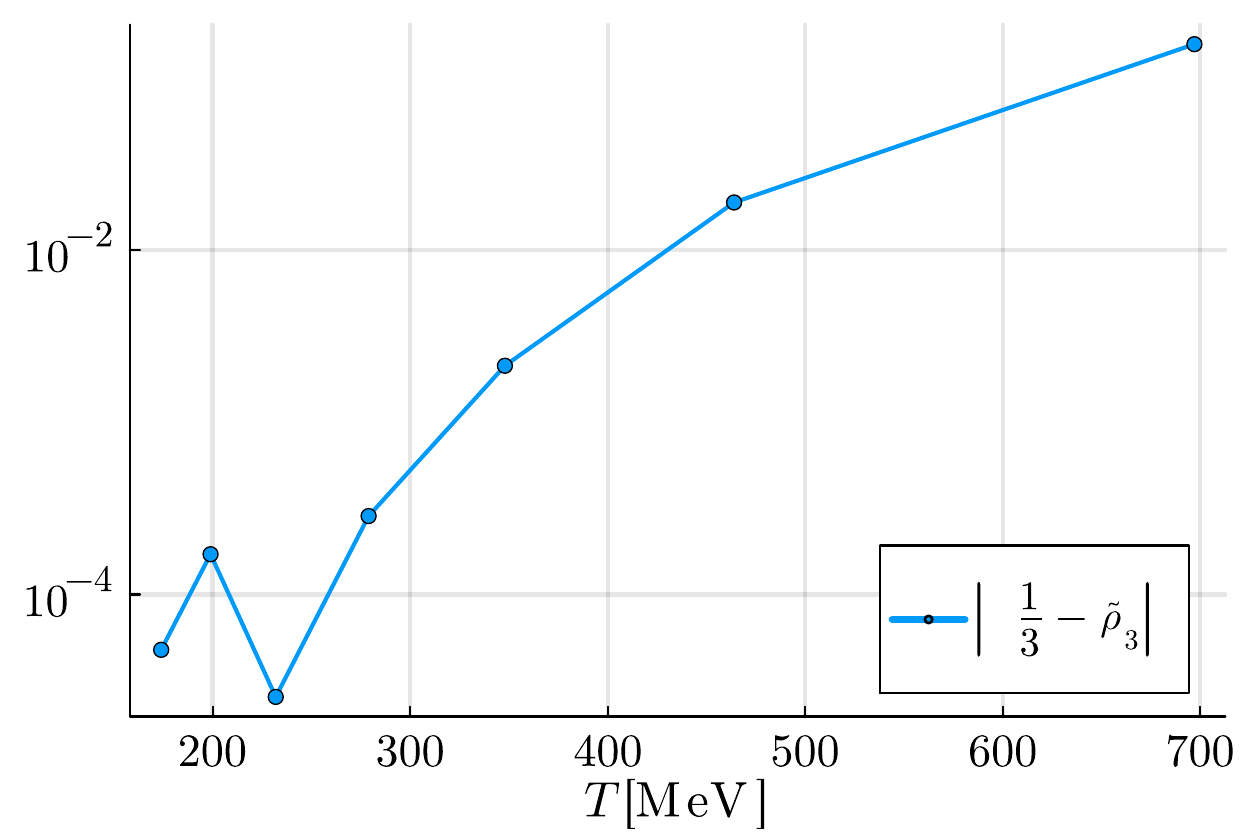}}
\end{center}
\caption{Expectation values of multiply-wound Polyakov loops $\tilde{\rho}_n=\langle\frac{1}{N}{\rm Tr}\mathcal{P}^n\rangle$ obtained from WHOT-QCD configurations. The lattice size is $N_t\times 32^3$ ($N_t=4,\cdots, 16$ and correspondingly, $T = 174, \cdots, 697$ MeV) and 599 configurations were used.
These histograms are drawn with 992 bins.
[Top] $\tilde{\rho}_n$ vs. temperature. 
[Middle] $|\tilde{\rho}_n|$ vs. temperature in the log scale.
[Bottom] $\frac{1}{3}-\tilde{\rho}_3$ vs. temperature and $\left|\frac{1}{3}-\tilde{\rho}_3\right|$ vs. temperature in the log scale. The error bars are not shown in the log plots.
}\label{fig:multiply-wound}
\end{figure}

In Fig.~\ref{fig:WHOT-phase}, we plot the distribution of Polyakov line phases obtained from the configurations created by the WHOT-QCD collaboration. 
We can see larger deviations from the Haar-random distribution at higher temperatures.  
To see the deviations quantitatively, we plot the Fourier coefficients $\tilde{\rho}_n$ in Fig.~\ref{fig:multiply-wound}. The Haar-random distribution specifies $\tilde{\rho}_3=\frac{1}{3}$, $\tilde{\rho}_n=0$ ($n=1,2,4,5,\cdots$). 
We can see the departure from Haar-random distribution around the lowest temperature, $T=174$ MeV, in our configuration set. 
This implies that  $T=174$ MeV is the completely-confined or partially-deconfined phase, and $T\ge 199$ MeV should be partially deconfined.

This `transition' may well be a crossover. We could presume that the Polyakov loops have small nonzero values that are exponential to the size of the representation. We cannot exclude this possibility with our numerical precision, and this is indeed the case for large-$N$ QCD on a small three-sphere~\cite{Hanada:2019kue}. Intuitively, a small deviation from Haar-randomness could be understood as contributions from sparse hadron gas. 
We could also anticipate nonzero expectation value for one-point functions of Polyakov loops in certain irreducible representations that are not forbidden by a symmetry (e.g., a center-neutral sector in pure Yang-Mills and any representation in QCD). Those quantities could behave as $e^{-m_{r}/T}$, where $m_{r}$ is the lightest excitation that couples to Polyakov loop in representation $r$.
If there is a mass gap, the correction to Haar randomness is parametrically small at low temperatures. 

%%%%%%%%%%%%%%%%%%%%%%%%%%%%%%%%
%%%%%%%%%%%%%%%%%%%%%%%%%%%%%%%%
\subsubsection*{Condensation of Multiply-wound loops and characters}
%%%%%%%%%%%%%%%%%%%%%%%%%%%%%%%%
%%%%%%%%%%%%%%%%%%%%%%%%%%%%%%%%
At $N=3$, the Haar-random distribution is rather far from the large-$N$ limit (i.e., the constant distribution). Therefore, it is hard to make sense of the `constant offset' that was the key to connecting the GWW point (onset of the gap) and partial deconfinement/complete deconfinement transition in the large-$N$ limit. In Fig.~\ref{fig:WHOT-phase}, even though we can observe the onset of the gap at $697$ MeV, this temperature is unnaturally high compared to other scales in QCD. 

Instead, we pay attention to the behavior of multiply-wound Polyakov loops. As we can see from Fig.~\ref{fig:multiply-wound}, the loops with winding number $n=1,2,4$ start to contribute at around 174 MeV,  while those with $n=5,6$ set in at around 348 MeV, and $n=3$ starts to decrease around there. The latter is likely to signal the transition from partial deconfinement to complete deconfinement.

An interesting feature is that $\tilde{\rho}_1$, $\tilde{\rho}_2$, and $\tilde{\rho}_4$ take the same magnitude at $174$ MeV $\lesssim T\lesssim$ $348$ MeV. This means only the fundamental representation is deconfined in this temperature range, as we will see shortly. 

To make the physical interpretation clearer, we also calculate the characters $\chi_r(\mathcal{P})$ corresponding to several irreducible representations. The character $\chi_{r}$ is the Polyakov loop in the representation $r$. (The usual Polyakov loop is for the fundamental representation. See Appendix~\ref{sec:character} for some properties of the characters.) For SU(3), we have
\begin{align}
\chi_{\rm fund.}
=
\sum_{j=1}^3
e^{i\theta_j}
\end{align}
for the fundamental representation, 
\begin{align}
\chi_{\rm adj.}
=
2
+
\sum_{j\neq k}
e^{i(\theta_j-\theta_k)}
\end{align}
for the adjoint representation, 
\begin{equation}
\chi_{\rm 2\mathchar`-sym.}
    =
    \sum_{j=1}^3
e^{2i\theta_j}
    +
    \sum_{j=1}^3
e^{-i\theta_j}\, 
\end{equation}
for the rank-2 symmetric representation, and
\begin{align}
\chi_{\rm 3\mathchar`-sym.}
=
1
+
\sum_{j=1}^3
e^{3i\theta_j}
+
\sum_{j\neq k}
e^{i(\theta_j-\theta_k)}
\end{align}
for the rank-3 symmetric representation. Notice when $\langle\chi_r\rangle$ is nonzero, the excitation in representation $r$ is deconfined. The characters are related to the multiply-wound loops $u_n\equiv\frac{1}{N}\textrm{Tr}(\mathcal{P}^n)$ as
\begin{align}
    3 u_1 &= \chi_{\rm fund.}\, ,
    \\
    3 u_2 
    &= 
    \chi_{\rm 2\mathchar`-sym.} - \left(\chi_{\rm fund.}\right)^\ast\, ,
    \\
    3 u_3
    &=\chi_{\rm 3\mathchar`-sym.} - \chi_{\rm adj.}+1\, ,
\end{align}
and so on.

\begin{figure}[htbp]
\begin{center}
\scalebox{0.33}{
\includegraphics{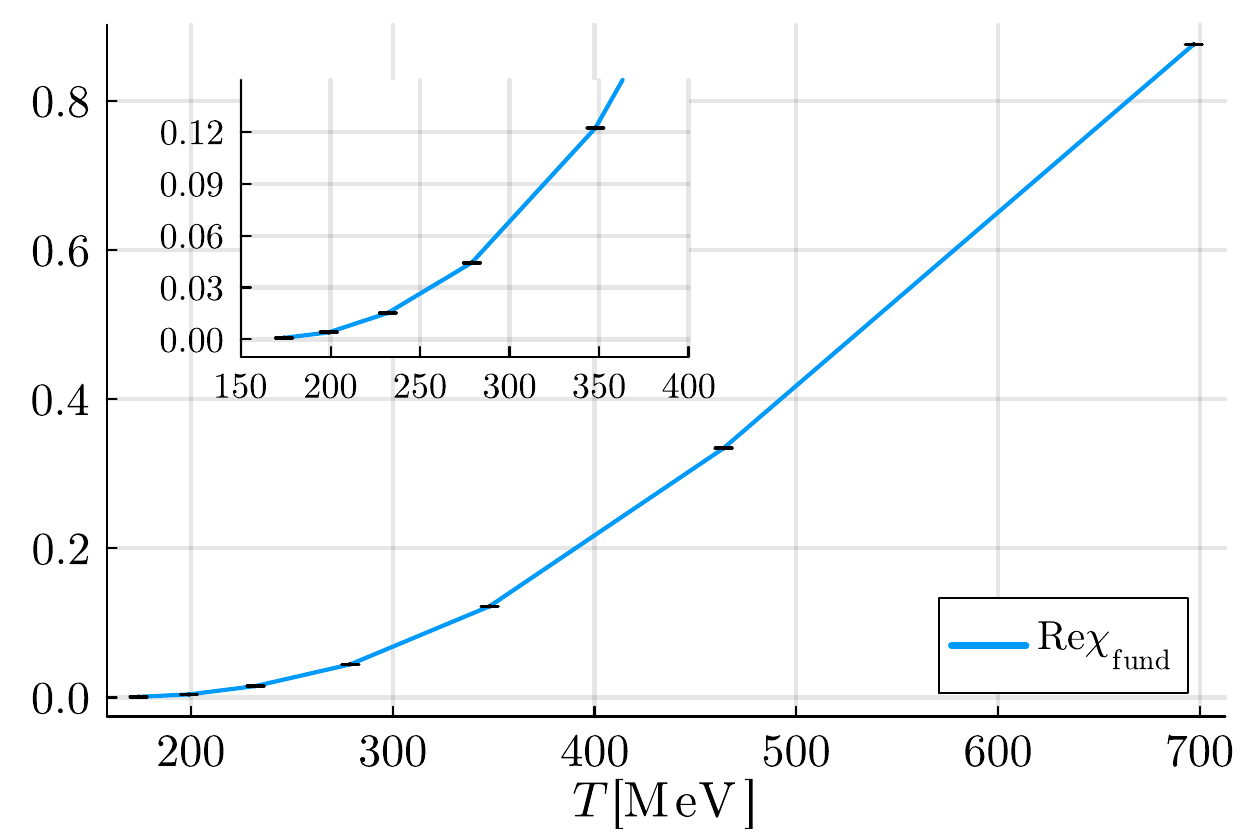}}
%\scalebox{0.33}{
%\includegraphics{Plot_char_(1,1)_Re_temp_0001-0599_v2.pdf}}
\scalebox{0.33}{
\includegraphics{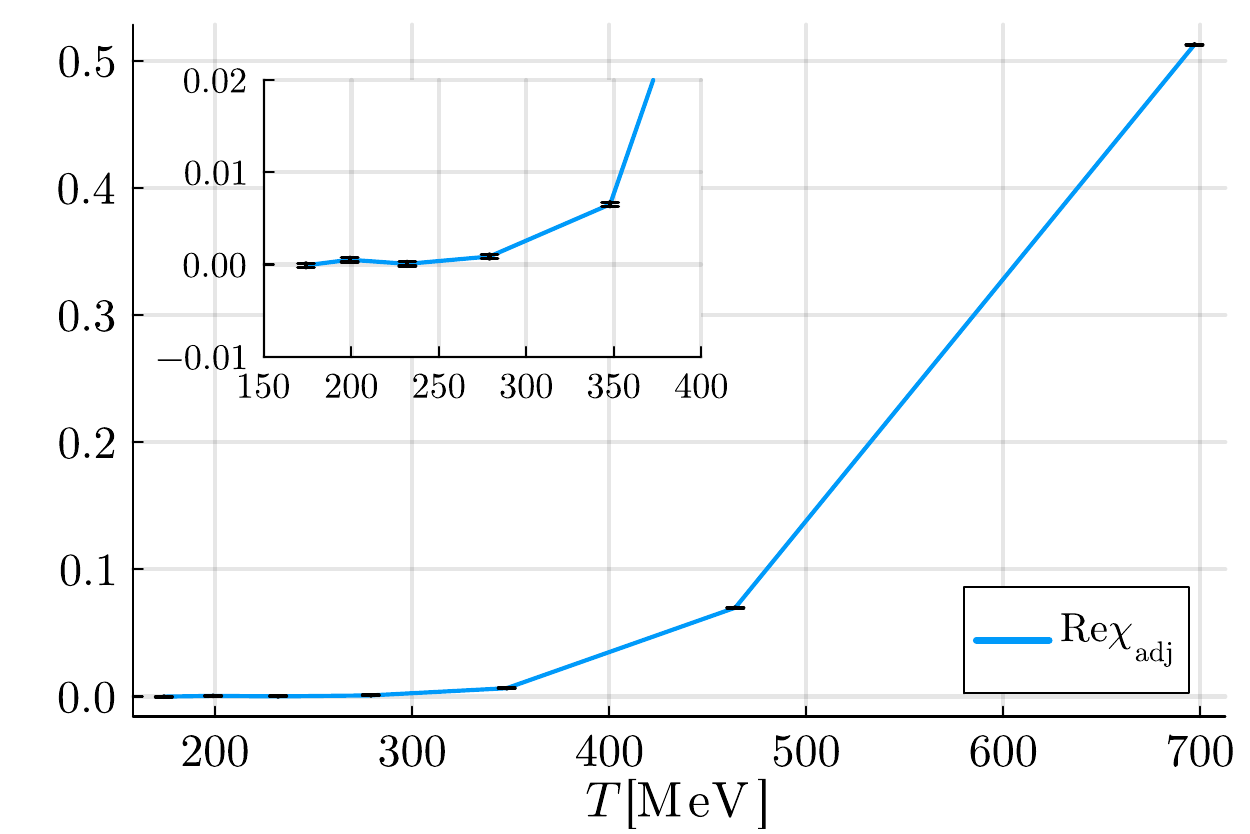}}
\scalebox{0.33}{
\includegraphics{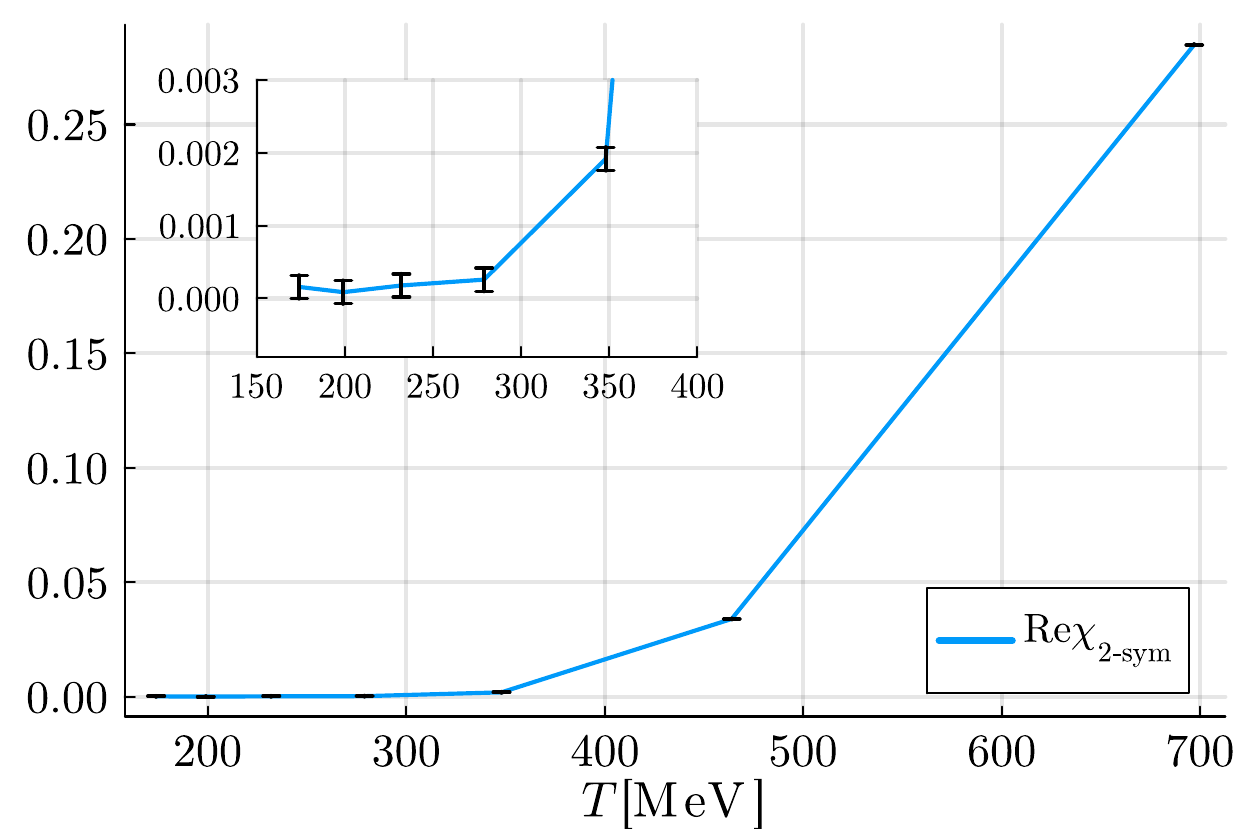}}
\scalebox{0.33}{
\includegraphics{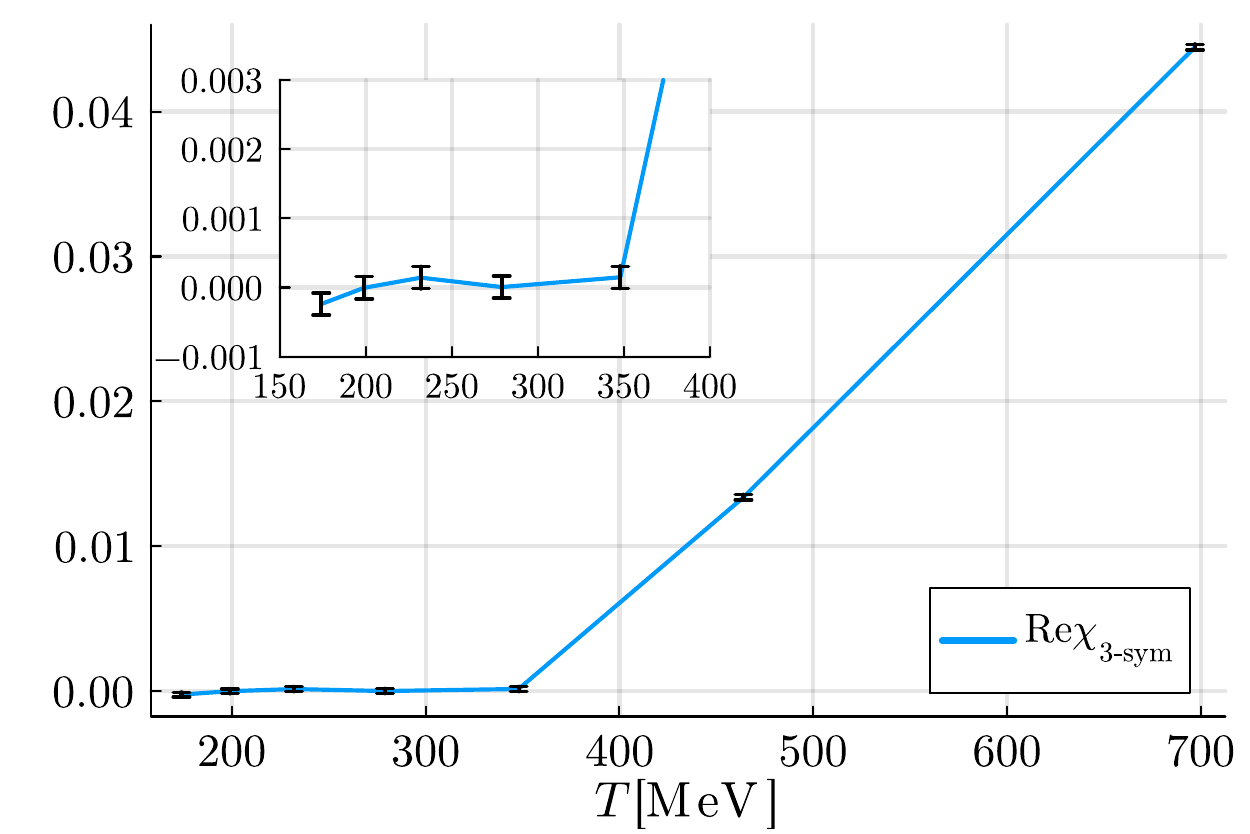}}
\scalebox{0.4}{
\includegraphics{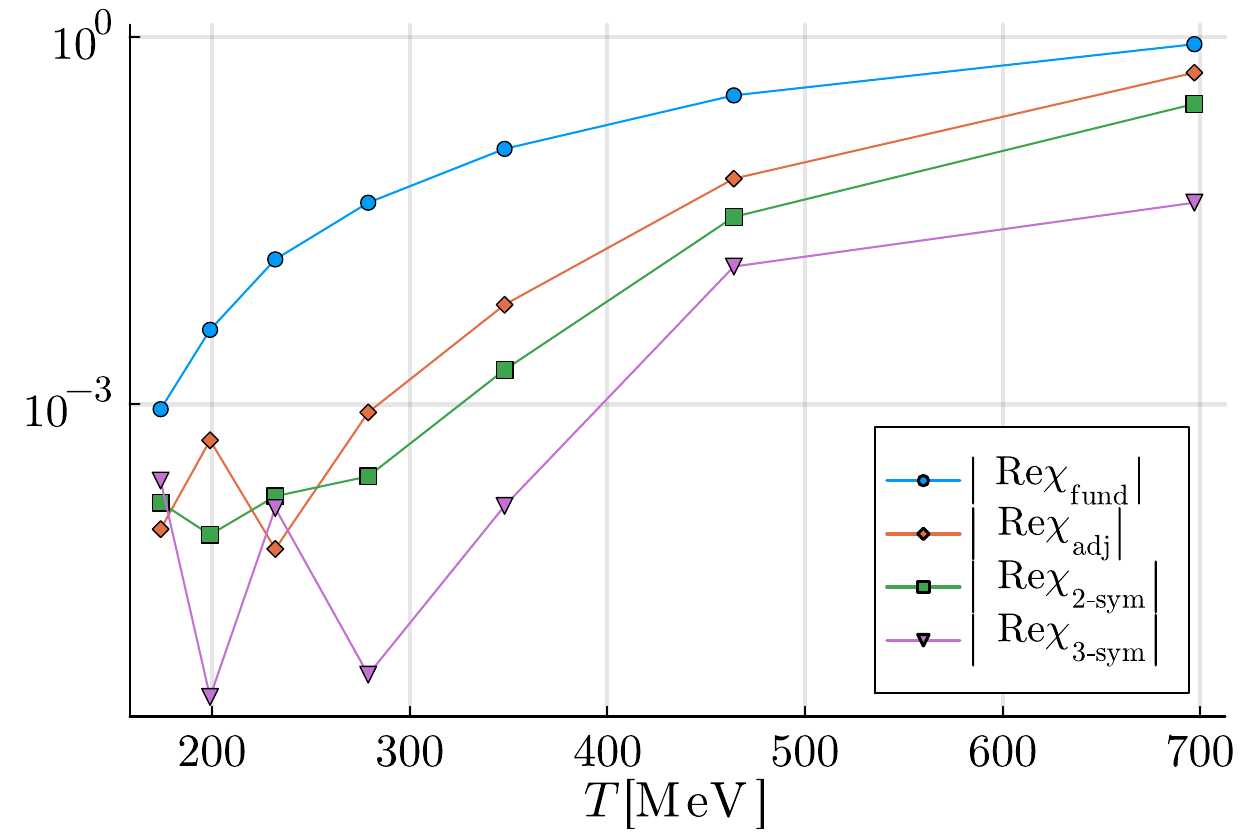}}
\end{center}
\caption{
The expectation values of characters vs. temperature for the fundamental, adjoint, rank-2 symmetric, and rank-3 symmetric representations, obtained from WHOT-QCD configurations. 
The bottom shows all at once in the log scale.
Note that the expectation values are real. 
}\label{fig:T-vs-character}
\end{figure}

In Fig.~\ref{fig:T-vs-character}, $\langle\chi_{\rm fund.}\rangle$, $\langle\chi_{\rm adj.}\rangle$, $\langle\chi_{\rm 2\mathchar`-sym.}\rangle$ and $\langle\chi_{\rm 3\mathchar`-sym.}\rangle$ are plotted. 
We can see that $\langle\chi_{\rm fund.}\rangle$ departs from zero at around $174~{\rm MeV}$. We can also see $\langle\chi_{\rm adj.}\rangle$, $\langle\chi_{\rm 2\mathchar`-sym.}\rangle$ and $\langle\chi_{\rm 3\mathchar`-sym.}\rangle$ depart from zero at higher temperatures ($T\gtrsim 348$ MeV). It would be natural to interpret that the onset of nonzero values of $\langle\chi_{\rm adj.}\rangle$, $\langle\chi_{\rm 2\mathchar`-sym.}\rangle$ and $\langle\chi_{\rm 3\mathchar`-sym.}\rangle$ is associated with the transition from partial deconfinement to complete deconfinement.

In summary, we estimate that $174$ MeV $\lesssim T\lesssim$ $348$ MeV is partially deconfined. 
%($T=174$ MeV may be completely confined, and $T=348$ MeV may be completely deconfined.)

The reason $\tilde{\rho}_1$, $\tilde{\rho}_2$, and $\tilde{\rho}_4$ take the same value up to sign at $174$ MeV $\lesssim T\lesssim$ $348$ MeV is that only the fundamental representation is deconfined. Indeed, from the equations above, we can see that $\tilde{\rho}_1=\langle u_1\rangle=\frac{1}{3}\langle\chi_{\rm fund.}\rangle$, and $\tilde{\rho}_2=\langle u_2\rangle=\frac{1}{3}\langle\chi_{\rm 2\mathchar`-sym.}\rangle-\frac{1}{3}\langle\chi_{\rm fund.}\rangle^\ast=-\frac{1}{3}\langle\chi_{\rm fund.}\rangle$, and hence, $\tilde{\rho}_1=-\tilde{\rho}_2$ in this temperature range. The agreement with $\tilde{\rho}_4$ can be shown by using the relations in Appendix~\ref{sec:character}, assuming only the fundamental representation is deconfined.

Our numerical observation does not exclude the possibility that Polyakov loops in large representations are very small but not exactly zero in the partially-deconfined phase. If this is the case, a natural possibility would be that the suppression is exponential with respect to the size of the representation in the partially-deconfined phase and power in the completely-deconfined phase. 

%%%%%%%%%%%%%%%%%%%%%%
%%%%%%%%%%%%%%%%%%%%%%
\subsection{Instanton condensation}\label{sec:instanton-condensation}
\begin{figure}[hbtp]
\scalebox{0.55}{
\includegraphics{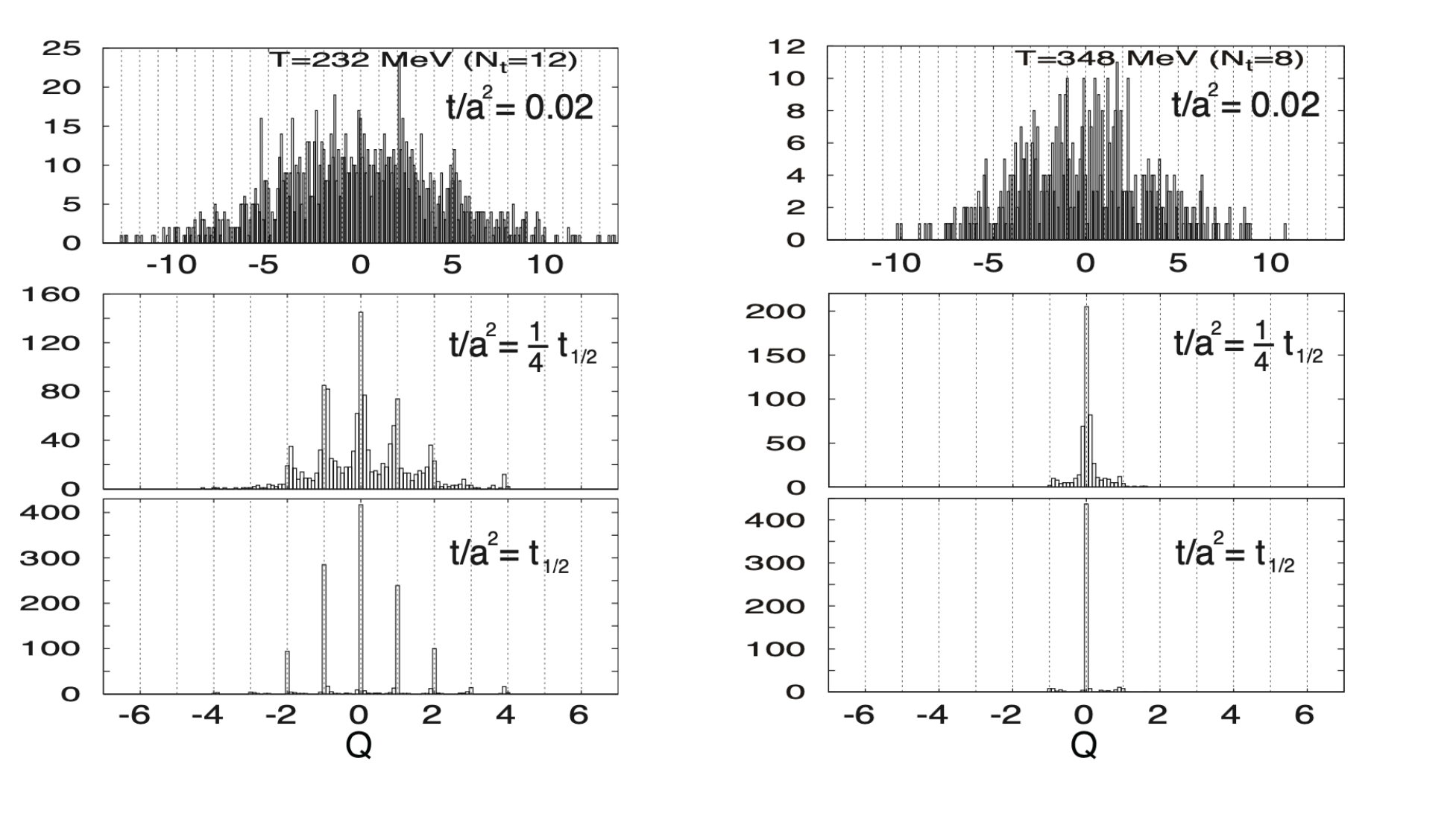}}
\caption{Histogram of the topological charge taken from Ref.~\cite{Taniguchi:2016tjc}. Each lattice configuration gives an integer value after smearing by the gradient flow. The parameter $t$ is the flow time. After sufficient smearing, charges peak at integer values (the third row).  At $T=232$ MeV, peaks at nonzero values signal the condensation of instantons. Flow time $t_{1/2}$ is defined in Ref.~\cite{Taniguchi:2016tjc} and chosen in such a way that unphysical effects from too much smearing can be avoided. 
}\label{fig:topological-charge-WHOT}
\end{figure}

\begin{figure}[hbtp]
\begin{minipage}{8cm}
\centering
\scalebox{0.55}{
\includegraphics{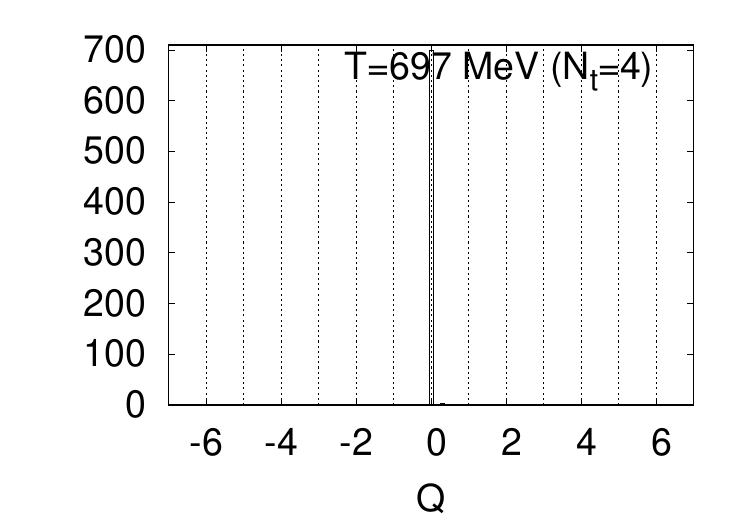}}
\end{minipage}
\begin{minipage}{8cm}
\centering
\scalebox{0.55}{
\includegraphics{
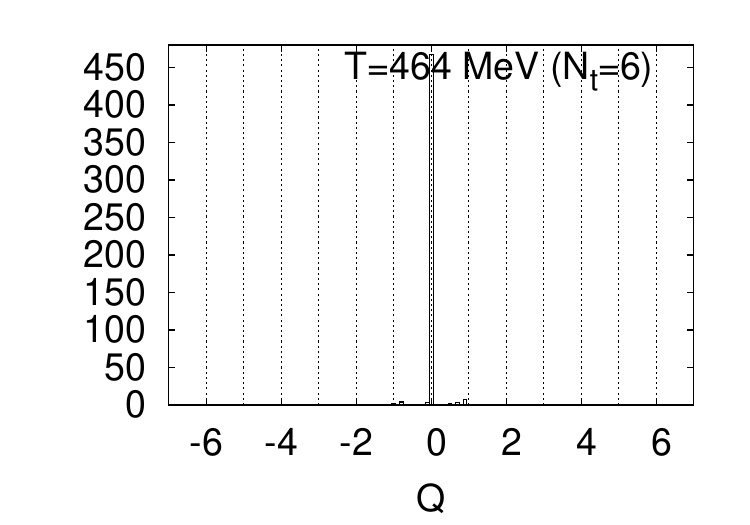}}
\end{minipage}
\begin{minipage}{8cm}
\centering
\scalebox{0.55}{
\includegraphics{
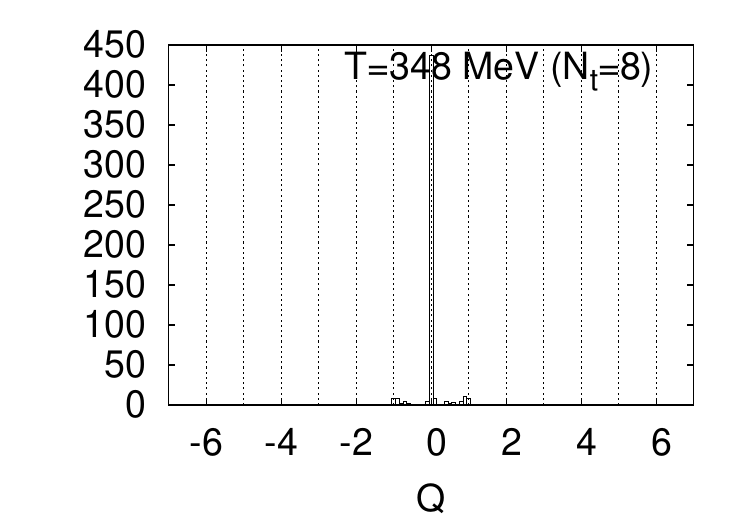
}}
\end{minipage}
\begin{minipage}{8cm}
\centering
\scalebox{0.55}{
\includegraphics{
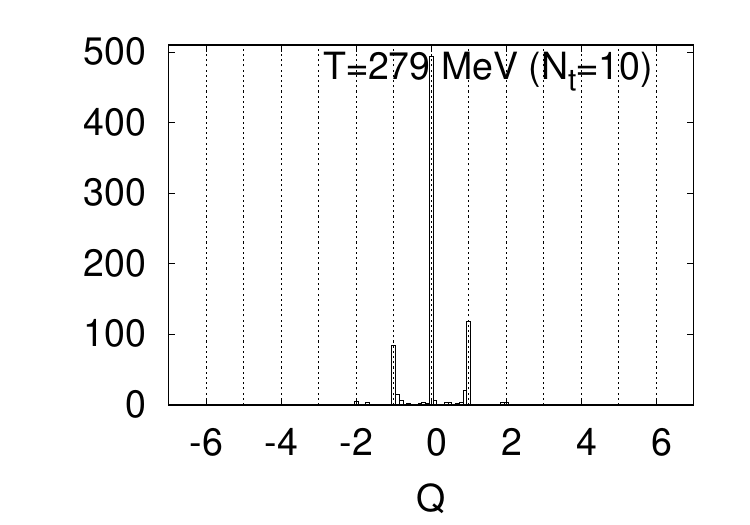}}
\end{minipage}
\begin{minipage}{8cm}
\centering
\scalebox{0.55}{
\includegraphics{
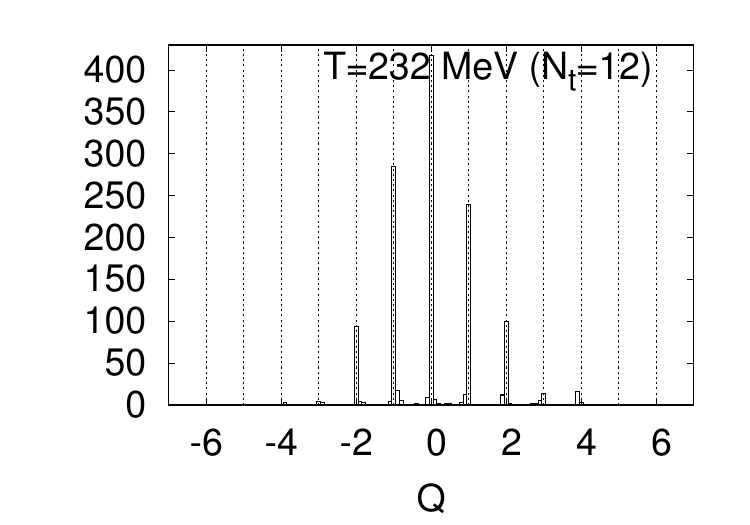
}}
\end{minipage}
\begin{minipage}{8cm}
\centering
\scalebox{0.56}{
\includegraphics{
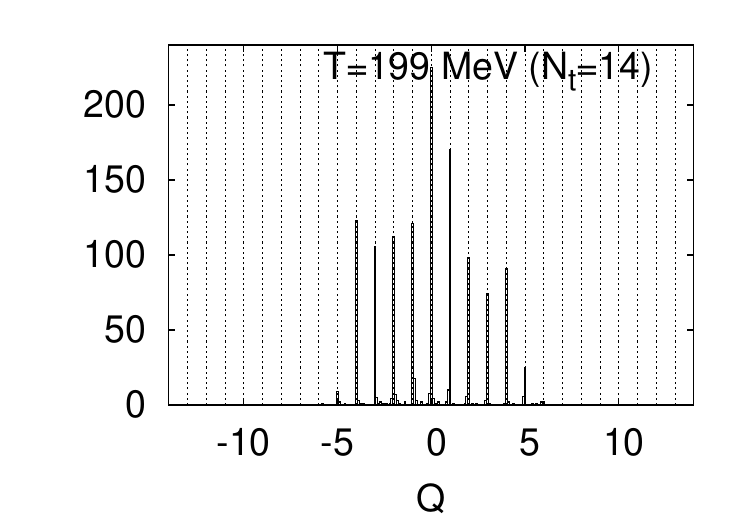}}
\end{minipage}
\begin{minipage}{8cm}
\centering
\scalebox{0.56}{
\includegraphics{
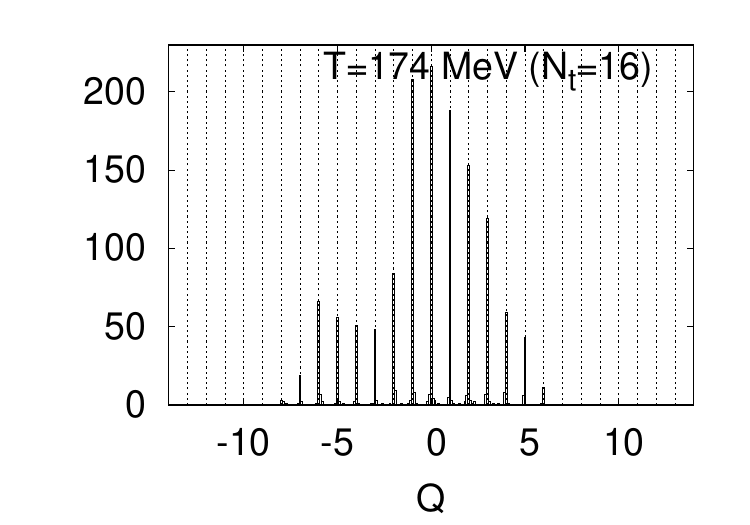}}
\end{minipage}
\begin{minipage}{8.7cm}
\centering
\scalebox{0.53}{
\includegraphics{
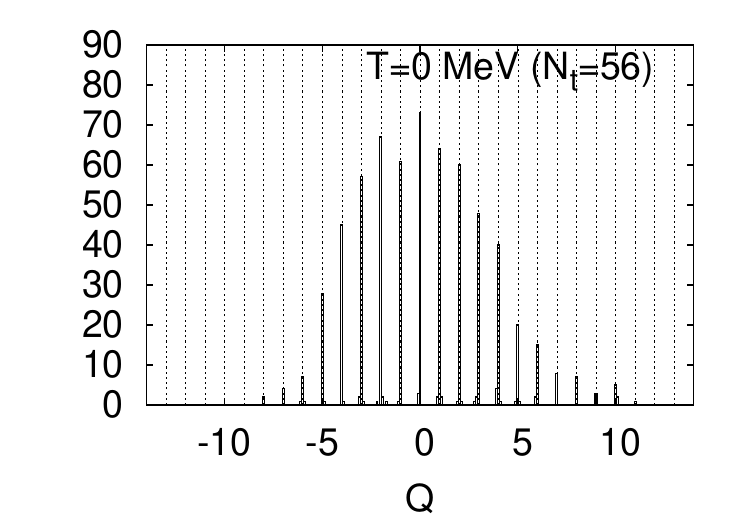
}}
\end{minipage}
\caption{Histogram of the topological charge provided by the authors of Ref.~\cite{Taniguchi:2016tjc}. Flow time $t_{1/2}$ is defined in Ref.~\cite{Taniguchi:2016tjc} and chosen in such a way that unphysical effects from too much smearing can be avoided. 
}\label{fig:topological-charge-WHOT-all-temperature}
\end{figure}

%%%%%%%%%%%%%%%%%%%%%%
%%%%%%%%%%%%%%%%%%%%%%
As discussed in Sec.~\ref{sec:global_symmetry_finite_N}, we expect that the instanton condensation is associated with the partial deconfinement. To detect the instanton condensation, we use the topological charge of each lattice configuration computed by the WHOT-QCD collaboration~\cite{Taniguchi:2016tjc}. Since lattice configurations are sensitive to the ultraviolet cutoff, they must be treated by the smearing with, for example, the gradient flow~\cite{Luscher:2010iy}. After smearing, each configuration returns an integer value, or more precisely, the histogram of the topological charge has peaks at integer values as shown in Fig.~\ref{fig:topological-charge-WHOT}.
As we can see from Fig.~\ref{fig:topological-charge-WHOT-all-temperature}, at $T\le 279$ MeV, clear multiple peaks appear at charge including the ones at $Q\neq 0$ that signal the instanton condensation. 
(Because lattice volume is finite, the peaks gradually become lower and eventually disappear.)  We can see wider distributions at lower temperatures. Peaks at $Q\neq 0$ disappear $T\gtrsim 348$ MeV. Therefore, we estimate that $T\lesssim$ $348$ MeV is the partially- or completely-confined phase. Remarkably, we have obtained the same estimate from the Polyakov loops and instantons!

%%%%%%%%%%%%%%%%%%%%%%
%%%%%%%%%%%%%%%%%%%%%%
\section{Conclusion and discussion}
%%%%%%%%%%%%%%%%%%%%%%
%%%%%%%%%%%%%%%%%%%%%%
In this paper, we described how the general mechanism of partial deconfinement~\cite{Hanada:2020uvt,Hanada:2021ipb,Hanada:2021swb} works for QCD and clarified the meaning of the Polyakov loop that does not rely on center symmetry. We studied the large-$N$ limit first and then provided a reasonable picture at finite $N$ combining inputs from the large-$N$ limit and lattice QCD simulation (WHOT-QCD collaborations' configurations). The distribution of the phases of the Polyakov line was a convenient tool to detect three phases of QCD at finite temperature: complete confinement, partial deconfinement, and complete deconfinement. 
Both at large $N$ and finite $N$, Haar-random distribution can give a good criterion for the complete confinement. In the WHOT-QCD configuration set, a remarkable agreement with the Haar-random distribution was observed at 174 MeV. 
In the large-$N$ limit, the partially-deconfined and completely-deconfined phases are separated by the GWW transition. We suggested that a natural counterpart of the GWW transition at finite $N$ is the condensation of the multiply-wound Polyakov loops with large winding numbers, and correspondingly Polyakov loops in large representations. In the set of WHOT-QCD configurations, only the Polyakov loop in the fundamental representation is condensed at $174~\textrm{MeV}\lesssim T\lesssim 348~\textrm{MeV}$, while other loops are condensed at $T\gtrsim  348~\textrm{MeV}$. Hence, we conjectured that the temperature range $174~\textrm{MeV}\lesssim T\lesssim 348~\textrm{MeV}$ is partially deconfined. As a nontrivial check, we monitored the topological charge. We argued that instanton condensation should be observed in partially-deconfined and completely-confined phases and confirmed, at $T\lesssim 348~\textrm{MeV}$, the nonzero topological charge, which indicates the instanton condensation. 
We also comment that the possibility of an intermediate phase in the region $T_\mathrm{c} \le T \lesssim 3T_\mathrm{c}$ has been discussed from various perspectives (see, e.g.,~\cite{Asakawa:1995zu,Glozman:2022zpy,Cohen:2023hbq}).

There are various directions to be explored. First and foremost, it is desirable to study other lattice configurations and check consistency with our results based on WHOT-QCD collaboration's configurations. Note that partial deconfinement is a generic property of gauge theories that does not rely on detail of gauge group or matter content. Therefore, it is significant to investigate various theories and confirm the consistency of our scenario. The computation of Polyakov loops would be straightforward. (At this moment, the authors are not aware of the study of the Polyakov loops in larger representations except for pure SU($N$) Yang-Mills theory for $N=2,3,4,5$ and $6$~\cite{Gupta:2007ax,Mykkanen:2012ri} in the context of Casimir scaling in the deconfined phase.\footnote{
Ref.~\cite{Gupta:2007ax} reported the nonzero value of the Polyakov loop in the adjoint representation in the confined phase of pure SU(3) Yang-Mills theory on a coarse lattice. It is interesting to see if the nonzero value survives in the continuum limit. }) When one considers the case a quark is massless, it is reasonable to anticipate that the chiral transition takes place at (the finite-$N$ analog of) the GWW point. 
In the current analysis, we focused on the one-point function of the Polyakov loops. It is interesting to consider whether one can also study other observables from the point of view of the Haar-randomness and the deviation from it.
See, Ref.~\cite{Bergner:2023rpw}, for development along this line, deriving the Casimir scaling of Polyakov loops in various representations from this point of view.
Presumably, the issue of renormalization is closely related to the spatial correlation of the Polyakov loops.

An obstacle to applying partial deconfinement to finite $N$ had been the meaning of the size of the deconfined sector $M$ ($0\le M \le N$), as shown in Fig.~\ref{fig:matrix}. Even in the large-$N$ limit, the size $M$ is not literally an integer. It could have an uncertainty of order $N^0$ which is negligible at large $N$. Admittedly, such an ambiguity makes the use of ``$M$'' very subtle at finite $N$, say, $N=3$. To circumvent this issue, we avoided the use of $M$ and relied on characters (the Polyakov loops in various representations).
The use of the character $\chi_r$ also has the advantage of being manifestly gauge invariant. Although in large $N$-theory using the size of the deconfined sector $M$ is shown to have gauge invariant meaning, as discussed in this paper, it may be worthwhile to revisit the analysis of partial deconfinement in the large-$N$ theory using the character expansion. Character expansion played an important role in the large-$N$ theory, for example in Refs.~\cite{Gross:1993hu, Gross:1993yt}. A recent work~\cite{Berenstein:2023srv} employs character expansion to study the deconfinement transition in large-$N$ theory.

Once we understand the nature of phase transitions in real-world QCD better, it would be fascinating if we could find experimental or cosmological signals. 
Recent studies of the large-$N$ perspective of the confinement/deconfinement phase transition in the context of cosmology can be seen, for instance, in Refs.~\cite{Yamada:2022imq,Fujikura:2023fbi}.

Last but not least, application to quantum gravity is important. Indeed, the original motivation for partial deconfinement~\cite{Hanada:2016pwv} was to study black hole geometry via holography. Specifically, the partially-deconfined phase is proposed as the counterpart of the small black hole state that appears in the context of the AdS/CFT correspondence~\cite{Maldacena:1997re,Gubser:1998bc,Witten:1998qj,Aharony:1999ti}.
It is interesting and worthwhile to investigate the finite-$N$ prospects of the large-$N$ (partial) deconfinement in the context of holography since $1/N$ corrections should play crucial roles in black hole physics as the quantum effects on the gravity side.
Hopefully, we can gain some intuition into quantum gravitational phenomena such as black hole evaporation from the QFT side.

%%%%%%%%%%%%%%%%%%%%%%
%%%%%%%%%%%%%%%%%%%%%%
\begin{center}
\Large{\textbf{Acknowledgement}}
\end{center}
%%%%%%%%%%%%%%%%%%%%%%
%%%%%%%%%%%%%%%%%%%%%%
Many of the new results in this paper were obtained in collaboration with Hiroki Ohata and Hidehiko Shimada. The letter version is Ref.~\cite{Hanada:2023krw}.
We would like to thank the members of the WHOT-QCD collaboration, including Shinji Ejiri, Kazuyuki Kanaya, Masakiyo Kitazawa, and Takashi Umeda, for providing us with their lattice configurations and many plots and having stimulating discussions with us. 
The analysis of topological charge in Ref.~\cite{Taniguchi:2016tjc}, which was crucial in Sec.~\ref{sec:instanton-condensation}, was led by Yusuke Taniguchi who passed away in 2022. Kazuyuki Kanaya collected the data and plots created by Yusuke Taniguchi for us. We deeply thank Yusuke Taniguchi and Kazuyuki Kanaya. We also thank Sinya Aoki, Hidenori Fukaya, Vaibhav Gautam, Yui Hayashi, Jack Holden, Seok Kim, Tamas Kovacs, Atsushi Nakamura, Marco Panero, Robert Pisarski, Enrico Rinaldi, Yuya Tanizaki, and Jacobus Verbaarschot for discussions and comments. Shohei Ohtani's unbelievable performance gave us the energy to survive the hot and humid summer of 2023. This work was supported by the Japan Lattice Data Grid (JLDG) constructed over the SINET5 of NII. M.~H. thanks for the STFC grants ST/R003599/1 and ST/X000656/1.
H.~W. is supported by the Japan Society for the Promotion of Science (JSPS) KAKENHI Grant number 22H01218.

\appendix

%%%%%%%%%%%%
%%%%%%%%%%%%
\section{Canonical ensemble and microcanonical ensemble}\label{sec:canonical-vs-microcanonical}
\hspace{0.51cm}
%%%%%%%%%%%%
%%%%%%%%%%%%

Let us consider an isolated system in which the energy is conserved. Let $\Omega(E)dE$ be the number of microscopic states in the energy range between $E$ and $E+dE$. The entropy $S(E)$ is defined by 
\begin{align}
\Omega(E)=e^{S(E)}\, . 
\end{align}

In the microcanonical ensemble, the energy is fixed in a narrow range $[E, E+dE]$, and all states in this energy range are allowed with equal probability. 
Many microscopic states correspond to the same macroscopic state. More precisely, many microscopic states give almost the same values as macroscopic observables. A macroscopic state can be determined by seeing the entropy as a function of macroscopic observables and maximizing it. 
The size of the deconfined sector $M$ is determined in the same manner, by seeing the entropy as a function of $E$ and $M$ and then maximizing it with respect to $M$. When there is more than one maxima, the one with the largest entropy is dominant. 

Temperature in the microcanonical ensemble is defined by 
\begin{align}
T
=
\left(
\frac{dS}{dE}
\right)^{-1}\, . 
\end{align}
Note that temperature is a function of energy: $T=T(E)$. The specific heat is defined by $C=\frac{dE}{dT}$. It can be positive or negative; we will discuss more regarding this point shortly.

In the canonical ensemble, the system is attached to a heat bath with temperature$T$, and the energy is allowed to change. 
Temperature is the controlling parameter, and energy is determined as a function of temperature: $E=E(T)$. The partition function is given by 
\begin{align}
Z(T)
=
\int dE
\Omega(E)e^{-E/T}
=
\int dE
e^{-F(E,T)/T}\, , 
\end{align}
where $e^{-E/T}$ is the Boltzmann factor and 
\begin{align}
F(E,T)
=
E-TS(E)
\label{def:canonical-partition-function}
\end{align}
is the free energy. The energy $E(T)$ is determined by minimizing the free energy. When there are multiple minima, the one with the smallest free energy is dominant. 

Let us see the meaning of minimum and maximum of the free energy more precisely. 
By taking the derivative of free energy with respect to the energy, we obtain 
\begin{align}
\left.
\frac{\partial F}{\partial E}\right|_{T_{\rm canonical}}
=
1-T_{\rm canonical}\frac{dS}{dE}
=
1-\frac{{T_{\rm canonical}}}{{T_{\rm microcanonical}}}\, . 
\end{align}
Therefore, at the minimum and maximum of the free energy, canonical temperature and microcanonical temperature coincide.  
Due to this property, the same energy-versus-temperature plot (like Fig.~\ref{fig:three=patterns}) is obtained from the canonical ensemble and microcanonical ensemble. The same holds for other quantities including the size of the deconfined sector. 

By taking the second derivative, we obtain
\begin{align}
\left.
\frac{\partial^2F}{\partial E^2}\right|_{T_{\rm canonical}}
=
\frac{{T_{\rm canonical}}}{{T_{\rm microcanonical}}^2}
\frac{dT_{\rm microcanonical}}{dE}
=
\frac{{T_{\rm canonical}}}{{T_{\rm microcanonical}}^2}\cdot C^{-1}, 
\end{align}
where $C$ is the specific heat in the microcanonical ensemble. Therefore,
\begin{align}
\textrm{Positive specific heat}
&\leftrightarrow
\textrm{Free energy minimum}
\nonumber\\
\textrm{Negative specific heat}
&\leftrightarrow
\textrm{Free energy maximum}
\nonumber
\end{align}
From this, we see that the phase in the microcanonical ensemble with negative specific heat is not dominant in the canonical ensemble. Whether a phase with negative specific heat is stable or not in the microcanonical ensemble depends on the details of the system, as we discuss next. 
%%%%%%%%%%%%
%%%%%%%%%%%%
\section{Phase coexistence in color space vs. phase coexistence in coordinate space}
\hspace{0.51cm}
%%%%%%%%%%%%
%%%%%%%%%%%%
Let us consider a first-order transition between water (liquid water) and ice (solid water) at 1 atm (Fig.~\ref{fig:water}). Supercooled water below 0\textcelsius\ and superheated ice above 0\textcelsius are not stable, and they turn to a mixture of water and ice just by a small perturbation as shown by arrows in Fig.~\ref{fig:water}. (Energy is fixed because the change takes place quickly.) Due to the local nature of the interaction, the temperature of the mixture is 0\textcelsius.  

Gauge theories admit two-phase-coexistence in the color space as well. Color space is nonlocal in the sense that the interactions are all-to-all. Furthermore, details of the splitting into two phases are different for different representations, as depicted in Fig.~\ref{fig:QCD-deconf-pattern} for the case of adjoint and fundamental representations. Therefore, the temperature is not necessarily fixed, as depicted in Fig.~\ref{fig:three=patterns}. 

When the transition is of the first order, splitting in color space or in the coordinate space can happen depending on the details of the model. For matrix models, for example, only the splitting in color space is allowed because the `coordinate space' is, by definition, just a point. Another important case without splitting in the coordinate space is the four-dimensional maximally supersymmetric Yang-Mills theory on the three-sphere. In such cases, the partially-deconfined phase is stable in a microcanonical ensemble. Pure Yang-Mills theory admits splitting in the coordinate space if the volume is sufficiently large. In such a case, the partially-deconfined phase is not stable even in a microcanonical ensemble.
%%%%%%%%%%%%
%%%%%%%%%%%%
\section{Operator formalism and Euclidean path integral}\label{sec:Hamiltonian-to-Lagrangian}
\hspace{0.51cm}
%%%%%%%%%%%%
%%%%%%%%%%%%
In this appendix, we show how the partition function \eqref{eq:Z-H-ext-MM} can be rewritten into the form of the path integral. 
This section is based on the appendix~A of \cite{Rinaldi:2021jbg}. Although the authors are not aware of an older reference that pointed out the connection, they think this was known by many people. Presumably, however, the fact that $g$ in \eqref{eq:Z-H-ext-MM} is the Polyakov line was not appreciated. 

As a concrete example, we consider the bosonic Hermitian matrix model. Generalizations to other cases are straightforward. 
The expression \eqref{eq:Z-H-ext-MM} can be rewritten as
\begin{align}
Z(T)
&=
\frac{1}{[{\rm vol}(G)]^K}
\int\left(\prod_{k=1}^KdU_{(k)}\right)
{\rm Tr}_{{\cal H}_{\rm ext}}
\Bigl(
\hat{U}_{(K)}
e^{-\frac{H(\hat{P},\hat{X})}{TK}}
\hat{U}_{(K-1)}^{-1}\hat{U}_{(K-1)}
\nonumber\\
&
\quad
e^{-\frac{H(\hat{P},\hat{X})}{TK}}
\hat{U}_{(K-2)}^{-1}\hat{U}_{(K-2)}
\cdots
\hat{U}_{(1)}^{-1}\hat{U}_{(1)}
e^{-\frac{H(\hat{P},\hat{X})}{TK}}
\Bigl)
\nonumber\\
&=
\frac{1}{[{\rm vol}(G)]^K}
\int\left(\prod_{k=1}^KdU_{(k)}\right)
\int\left(\prod_{k=1}^KdX_{(k)}\right)
\nonumber\\
&
\qquad
\langle X_{(K)}|
\hat{U}_{(K)}
e^{-\frac{H(\hat{P},\hat{X})}{TK}}
\hat{U}_{(K-1)}^{-1}
|X_{(K-1)}\rangle
\nonumber\\
&
\qquad
\times
\langle X_{(K-1)}|
\hat{U}_{(K-1)}
e^{-\frac{H(\hat{P},\hat{X})}{TK}}
\hat{U}_{(K-2)}^{-1}
|X_{(K-2)}\rangle
\nonumber\\
&
\qquad
\times
\cdots
\times
\langle X_{(1)}|
\hat{U}_{(1)}
e^{-\frac{H(\hat{P},\hat{X})}{TK}}
|X_{(K)}\rangle \ .
\end{align}
At the first line, we simply inserted $\hat{U}^{-1}\hat{U}=\hat{1}$ at many places.
$\hat{U}_{(K)}$ corresponds to $\hat{g}$.
At the next line, we inserted $\int dX_{(k)}|X_{(k)}\rangle\langle X_{(k)}|=\hat{1}$ at many places.
Each term in the product can be rewritten as 
\begin{align}
&
\langle X_{(k)}|
\hat{U}_{(k)}
e^{-\frac{H(\hat{P},\hat{X})}{TK}}
\hat{U}_{(k-1)}^{-1}
|X_{(k-1)}\rangle
\nonumber\\
&\quad=
\langle U_{(k)}X_{(k)}U_{(k)}^{-1}|
e^{-\frac{H(\hat{P},\hat{X})}{TK}}
|U_{(k-1)}X_{(k-1)}U_{(k-1)}^{-1}\rangle
\nonumber\\
&\quad=
\int dP
\langle U_{(k)}X_{(k)}U_{(k)}^{-1}|
e^{-\frac{H(\hat{P},\hat{X})}{TK}}
|P\rangle\langle P|
U_{(k-1)}X_{(k-1)}U_{(k-1)}^{-1}\rangle
\nonumber\\
&\quad=
\int dP
e^{i{\rm Tr}[P(U_{(k)}X_{(k)}U_{(k)}^{-1}-U_{(k-1)}X_{(k-1)}U_{(k-1)}^{-1})]}
e^{-H(P,U_{(k)}X_{(k)}U_{(k)}^{-1})/(TK)}
\nonumber\\
&\quad=
e^{-KT{\rm Tr}[(U_{(k)}X_{(k)}U_{(k)}^{-1}-U_{(k-1)}X_{(k-1)}U_{(k-1)}^{-1})^2]}
e^{-V(U_{(k)}X_{(k)}U_{(k)}^{-1})/(TK)}
\nonumber\\
&\quad\simeq
e^{-L[D_t(U_{(k)}X_{(k)}U_{(k)}^{-1}),(U_{(k)}X_{(k)}U_{(k)}^{-1})]/(TK)}
\nonumber\\
&\quad=
e^{-L[D_tX_{(k)},X_{(k)}]/(TK)} \ , 
\end{align}
where $L[D_tX,X]$ is the Lagrangian with the Euclidean signature.
We used
\begin{align}
U_{(k-1)}^{-1}U_{(k)}\equiv e^{iA_{(k)}/(KT)}
\end{align}
and
\begin{align}
X_{(k)}-(U_{(k-1)}U_{(k)}^{-1})^{-1}X_{(k-1)}(U_{(k-1)}U_{(k)}^{-1})
\simeq
\frac{D_tX_{(k)}}{KT}\, .
\end{align}
In the limit of $K\to\infty$, we obtain
\begin{align}
Z(T) &= \int [dA] [dX]e^{-\int dt L[D_tX,X]} \ .
\end{align}
%The temporal direction is compactified with the circumference $\frac{1}{T}$, and the periodic boundary condition is imposed to $X_I$ and $A_t$.
Therefore,
\begin{align}
g &= U_K
\nonumber\\
&=
U_K(U_{K-1}^{-1}U_{K-1})(U_{K-2}^{-1}U_{K-2})\cdots(U_{1}^{-1}U_{1})
\nonumber\\
&=
(U_KU_{K-1}^{-1})(U_{K-1}U_{K-2}^{-1})\cdots(U_{2}U_{1}^{-1})U_1
\nonumber\\
&=
{\rm P}e^{i\int_0^{1/T} dt A_t}
\end{align}
is the Polyakov line.
Here ${\rm P}$ stands for the path-ordered product.
%%%%%%%%%%%%%%%%%%%%%%%%%%%%%%%%
%%%%%%%%%%%%%%%%%%%%%%%%%%%%%%%%
\subsection*{Equivalence between \eqref{eq:Z-H-inv-MM} and \eqref{eq:Z-H-ext-MM}}
%%%%%%%%%%%%%%%%%%%%%%%%%%%%%%%%
%%%%%%%%%%%%%%%%%%%%%%%%%%%%%%%%
Let us also see why \eqref{eq:Z-H-inv-MM} and \eqref{eq:Z-H-ext-MM} are equivalent. As a simple example, let us start with the case of $N$ indistinguishable bosons. The gauge group $G$ is the symmetric group $S_{\rm N}$. From a Fock state $\ket{\phi}\equiv\ket{\vec{n}_1,\cdots,\vec{n}_N}\in \mathcal{H}_{\rm ext}$, we can obtain $\ket{\phi}_{\rm inv}\equiv\mathcal{N}_{\ket{\phi}}^{-1/2}\hat{\pi}\ket{\phi}\in \mathcal{H}_{\rm inv}$, where the normalization factor is
\begin{align}
\mathcal{N}_{\ket{\phi}}
\equiv
\bra{\phi}\hat{\pi}\ket{\phi}
=
\frac{{\rm Vol}G_{\ket{\phi}}}{{\rm Vol}G}
=
\frac{{\rm Vol}G_{\ket{\phi}}}{N!}\, 
\end{align}
where $G_{\ket{\phi}}$ is the stabilizer of the state $\ket{\phi}$, i.e., $G_{\ket{\phi}}=\{\sigma\in G\ |\ \ket{\phi}=\hat{\sigma}\ket{\phi}\}$. Because this is a finite group, the `volume' ${\rm Vol}G_{\ket{\phi}}$ is the number of the elements ($V_{\ket{\vec{n}_1,\cdots,\vec{n}_N}}$ in Sec.~\ref{sec:underlying_mechanism}). The number of states on the gauge orbit $\{\hat{\sigma}|\ket{\phi}\ |\ \sigma\in G\}$ is $\mathcal{N}_{\ket{\phi}}^{-1}$. If we naively took trace over the extended Hilbert space, we would count gauge-equivalent states $\mathcal{N}_{\ket{\phi}}^{-1}$ times. To relate the trace over the extended Hilbert $\textrm{Tr}_{\mathcal{H}_{\rm ext}}$ to the trace over the gauge-invariant Hilbert $\textrm{Tr}_{\mathcal{H}_{\rm inv}}$, we must remove such an over counting. Therefore, 
\begin{align}
    \sum_{\ket{\phi}_{\rm inv}\in\mathcal{H}_{\rm inv}} 
   \bra{\phi}_{\rm inv}
    e^{-\hat{H}/T}
    \ket{\phi}_{\rm inv}
    &=
    \sum_{\ket{\phi}\in\mathcal{H}_{\rm ext}}
    \mathcal{N}_{\ket{\phi}}
    \bra{\phi}_{\rm inv}
    e^{-\hat{H}/T}
    \ket{\phi}_{\rm inv}
    \nonumber\\
    &=
    \sum_{\ket{\phi}\in\mathcal{H}_{\rm ext}}
    \bra{\phi}\hat{\pi}
    e^{-\hat{H}/T}
    \hat{\pi}\ket{\phi}
    \nonumber\\
    &=
    \sum_{\ket{\phi}\in\mathcal{H}_{\rm ext}}
    \bra{\phi}\hat{\pi}
    e^{-\hat{H}/T}
    \ket{\phi}\, , 
\end{align}
and hence, 
\begin{align}
    \mathrm{Tr}_{\mathcal{H}_{\rm inv}}e^{-\hat{H}/T}
    =
    \mathrm{Tr}_{\mathcal{H}_{\rm ext}}(\hat{\pi}e^{-\hat{H}/T})\, . 
\end{align}

The same argument applies to the SU($N$) matrix model. As $\ket{\phi}\in\mathcal{H}_{\rm ext}$ we can take a wave packet or Fock state, for example. The projection to $\mathcal{H}_{\rm inv}$ is defined by
\begin{align}
\ket{\phi}_{\rm inv}
\equiv
\mathcal{N}_{\ket{\phi}}^{-1/2}\hat{\pi}\ket{\phi}, 
\qquad
\mathcal{N}_{\ket{\phi}}\equiv\bra{\phi}\hat{\pi}\ket{\phi}\, , 
\end{align}
where $\mathcal{N}_{\ket{\phi}}$ is the inverse of the over-counting factor. An intuitive way to understand it is that $\int dg\bra{\phi}\hat{g}\ket{\phi}$ is, roughly speaking, the volume of the stabilizer of $\ket{\phi}$. This is somewhat imprecise because $G=\textrm{SU}(N)$ is continuous and an infinitesimal transformation away from $\ket{\phi}$ can have a nonzero contribution to $\int dg\bra{\phi}\hat{g}\ket{\phi}$. A more precise way is to notice that the over-counting factor $C_{\ket{\phi}}$ is determined so that all gauge orbits $\{\hat{g}\ket{\phi}\ |\ g\in G\}$ give equal contribution to the partition function at infinite temperature. Namely, 
\begin{align}
    C_{\ket{\phi}}\int dg\int dg' (\bra{\phi}\hat{g}^{-1})(\hat{g}'\ket{\phi})
    =
    C_{\ket{\phi}}\cdot(\textrm{Vol}G)^2\cdot\mathcal{N}_{\ket{\phi}}
\end{align}
should be independent of $\ket{\phi}$. Gauge-invariant states such as the confining vacuum satisfy $C_{\ket{\phi}}=\mathcal{N}_{\ket{\phi}}=1$, and hence,  $C_{\ket{\phi}}=\mathcal{N}_{\ket{\phi}}^{-1}$. 
%%%%%%%%%%%%%%%%%%%%%%%%%%%%%%%%
%%%%%%%%%%%%%%%%%%%%%%%%%%%%%%%%
\section{SU($N$) Haar random distribution}\label{sec:Haar-random-distribution}
%%%%%%%%%%%%%%%%%%%%%%%%%%%%%%%%
%%%%%%%%%%%%%%%%%%%%%%%%%%%%%%%%
In this appendix, we show the analytic formula for the Haar-random distribution of the phases in SU($N$),  
\begin{align}
\rho_{\rm Haar}(\theta)=\frac{1}{2\pi}\left(1-(-1)^N\cdot\frac{2}{N}\cos(N\theta)\right)\, . 
\label{eq:appendix:Haar-random}
\end{align}
For other details on random matrix theory, see also Ref.~\cite{Akemann:2011csh} and references therein.
(After this paper was submitted to the journal, Ref.~\cite{Nishigaki:2024phx} provided a rigorous proof of this formula for arbitrary $N$.)

\begin{figure}[hbtp]
\centering
\scalebox{0.8}{
\input{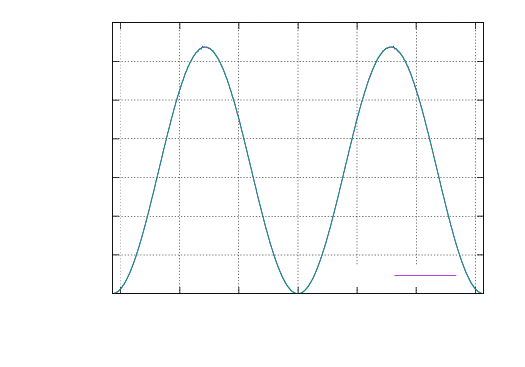}}
\scalebox{0.8}{
\input{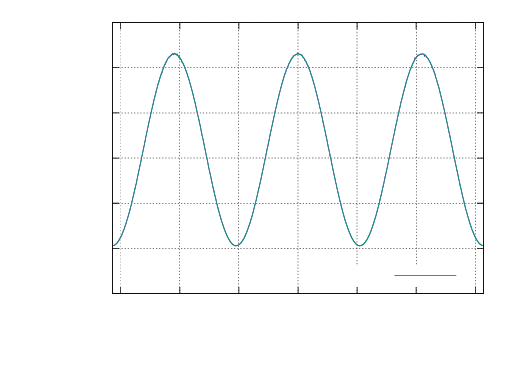}}
\scalebox{0.8}{
\input{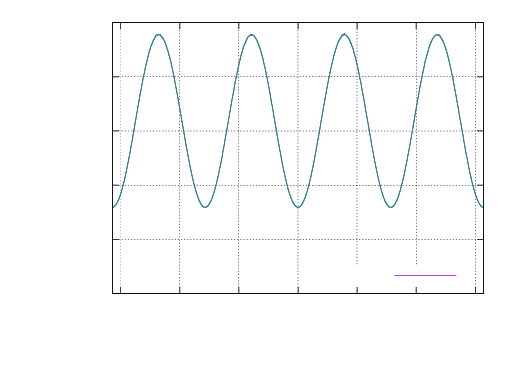}}
\scalebox{0.8}{
\input{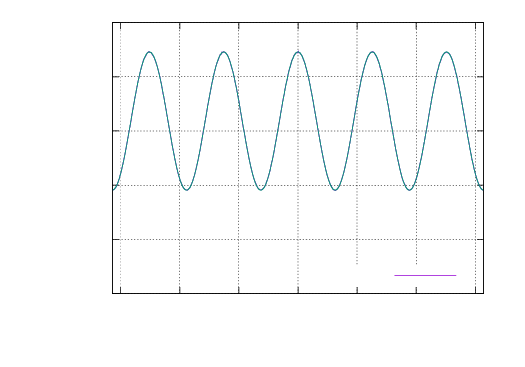}}
%\scalebox{0.8}{
%\input{Haar-SU6}}
%\scalebox{0.8}{
%\input{Haar-SU7}}
\scalebox{0.8}{
\input{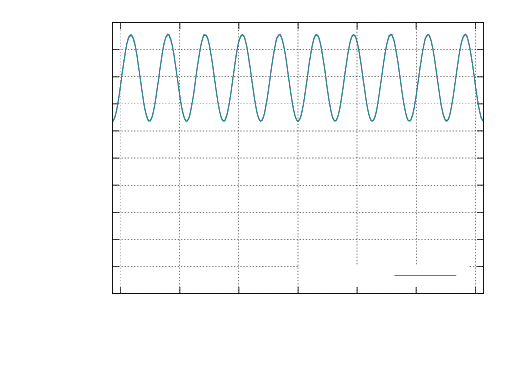}}
\scalebox{0.8}{
\input{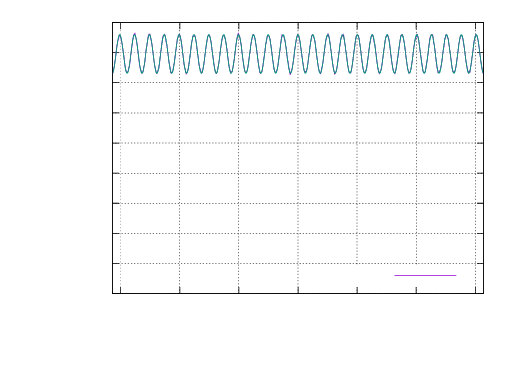}}
\caption{SU($N$) Haar-random distribution for $N=2,3,4,5,10,25$ (purple line) and $\frac{1}{2\pi}\left(1-(-1)^N\cdot\frac{2}{N}\cos(N\theta)\right)$ (green line) are plotted together. They overlap completely.}\label{fig:Haar-random}
\end{figure}

The starting point is 
\begin{equation}
    \qty[d U] 
    =
    \rho(\theta_1,\cdots,\theta_N)
    \prod_{j=1}^N d \theta_j\, ,
\end{equation}
where
\begin{align}
    \rho(\theta_1,\cdots,\theta_N)
    =
    C \prod_{j<k} \sin^2\left(\frac{\theta_j - \theta_k}{2}\right)
    \times
    \sum_{n=-\infty}^{\infty}
    \delta\left(
        \sum_{j=1}^N \theta_j -2\pi n
    \right)\, . 
    \label{eq:SU(N)_Haar_measure}
\end{align}
Here, $C$ is the normalization factor and $\prod_{j<k} \sin^2\left(\frac{\theta_j - \theta_k}{2}\right)$ is the Vandermonde determinant. 
Our task is to integrate out $\theta_2,\cdots,\theta_N$ and estimate
\begin{align}
    \rho(\theta_1)
    =
    \int d\theta_2\cdots\int d\theta_N\ 
    \rho(\theta_1,\theta_2,\cdots,\theta_N)\, . 
\end{align}
We can easily generate $\rho(\theta_1)$ by using the Monte Carlo method such as the Gibbs sampling algorithm; see Fig.~\ref{fig:Haar-random}.\footnote{
See Ref.~\cite{Fasi2020SamplingTE} for a numerical study in the past
and Ref.~\cite{Hanada-Matsuura} for an introduction to the Monte Carlo method.
} The agreement with \eqref{eq:appendix:Haar-random} can be confirmed numerically. When $N$ is not so large, we can perform the integral analytically and confirm \eqref{eq:appendix:Haar-random}, as we will do for $N=2$ and $N=3$ below. 

%%%%%%%%%%%%%%%%%%%%%%%%%%%
\subsubsection*{$N=2$ case}
%%%%%%%%%%%%%%%%%%%%%%%%%%%
If we fix $\theta_1$, $\theta_2$ is also fixed as $\theta_2 = -\theta_1$ up to the shift by $2\pi n$.
By substituting this into \eqref{eq:SU(N)_Haar_measure}, we obtain the distribution for fixed $\theta_1$ as
\begin{align}
    \rho(\theta_1)
    =
    C \sin^2\theta_1
    =
    C\cdot\frac{1-\cos (2\theta_1)}{2}\, .
\end{align}
The normalization constant $C$ can be determined as 
\begin{equation}
    \int_{-\pi}^\pi d \theta_1\,
    \rho(\theta_1)
    =
    1\, . 
\end{equation}
The final answer is 
\begin{align}
    \rho(\theta_1)
    =
    \frac{1 - \cos(2\theta_1)}{2\pi}\, . 
\end{align}

%%%%%%%%%%%%%%%%%%%%%%%%%%%
\subsubsection*{$N=3$ case}
%%%%%%%%%%%%%%%%%%%%%%%%%%%
Same as the $N=2$ case, $\theta_3 = -(\theta_1 + \theta_2)$ for fixed $\theta_1, \theta_2$ up to the shift by $2\pi n$.
Then, we obtain the distribution for fixed $\theta_1, \theta_2$ as
\begin{align}
    \rho(\theta_1,\theta_2)
    &=
    C
    \sin^2\left(\frac{\theta_1-\theta_2}{2}\right)
    \sin^2\left(\frac{\theta_1 + 2\theta_2}{2}\right)
    \sin^2\left(\frac{\theta_2 + 2\theta_1}{2}\right)
    \nonumber\\
    &=
    \frac{C}{8}
    \left(1-\cos(\theta_1-\theta_2)\right)
    \left(1-\cos(\theta_1+2\theta_2)\right)
    \left(1-\cos(2\theta_1+\theta_2)\right)
    \nonumber\\
    &=    
    \frac{C}{8}
    \left(
    \frac{3}{4}
    +
    \frac{1}{2}\cos(3\theta_1)
    +
    \cdots
    \right)\, ,          
\end{align}
where $\cdots$ are terms like $\cos(\theta_1-\theta_2)$ or $\cos(\theta_1+2\theta_2)$ which disappear after $\theta_2$ is integrated. To go from the second line to the third line, we used $\cos x \cdot\cos y = \frac{1}{2}\left(\cos(x+y)+\cos(x-y)\right)$ several times. Therefore, 
\begin{equation}
    \rho(\theta_1) 
    =
    \int_{-\pi}^\pi d \theta_2\, \rho(\theta_1,\theta_2)
    \propto
    1 + \frac{2}{3}\cos(3\theta_1)\, . 
\end{equation}
The overall factor is determined such that $\int_{-\pi}^\pi d \theta_1\, \rho(\theta_1)=1$. The final result is
\begin{align}
    \rho(\theta_1) 
    =
    \frac{1}{2\pi}\left(
    1 + \frac{2}{3}\cos(3\theta_1)
    \right).
\end{align}
%%%%%%%%%%%%%%%%%%%%%%%%%%%%%
\section{Characters of SU(3) group}\label{sec:character}
%%%%%%%%%%%%%%%%%%%%%%%%%%%%%
The character of $G$ is defined by the trace of the representation matrix $R_r$ in the representation $r$ as
\begin{equation}
    \chi_r(g) := \tr R_r(g),
\end{equation}
where $g \in G$ is the holonomy along the thermal circle.
For the irreducible representations, the orthonormal condition,
\begin{equation}
    \frac{1}{\mathrm{Vol}(G)} \int_G dg\, \chi_r(g)\left(\chi_s(g)\right)^\ast = \delta_{rs},
    \label{eq:orthogonality}
\end{equation}
is satisfied.

When we consider $G=$ SU(3), we can identify $g$ and its representation in the fundamental representation: $g$ is $3\times 3$ unitary matrix with a determinant equal to 1. 
The SU(3) characters are, therefore, functions only with respect to the eigenvalues of $g$, denoted by $\lambda_j$ $ (j=1,2,3)$. 
Namely, 
\begin{equation}
    \chi_r(g) = \chi_r(\{\lambda\}).
\end{equation}
Note that $\lambda_j = e^{i\theta_j}$ in terms of the Polyakov line phase $\theta_j$, and $\lambda_1\lambda_2\lambda_3=1$ due to the condition $\det g = 1$.
%%%%%%%%%%%%%%%%%%%%%%%%%%%%%%%%%%%%%
%%%%%%%%%%%%%%%%%%%%%%%%%%%%%%%%%%%%%
\subsection{List of characters for irreducible representations}
%%%%%%%%%%%%%%%%%%%%%%%%%%%%%%%%%%%%%
%%%%%%%%%%%%%%%%%%%%%%%%%%%%%%%%%%%%%
The character is a symmetric polynomial in $\lambda$’s. 
There should be a nice way to compute the character from, say, the Young tableaux.
Each Young tableau corresponds to an irreducible representation of SU(3).
In the following, the notation $(n,m)$ represents a Young tableau that has $n$ and $m$ boxes in the first and second rows, respectively.

\begin{description}
%%%%%%%%%%%%%%%%%%%%%%%
\item[(0,0)=1: trivial] \,\\ 
%%%%%%%%%%%%%%%%%%%%%%%
Trivial representation is $R_{(0,0)}(g)=1$. The character is
\begin{equation}
    \chi_{\rm trivial}
    \equiv
    \chi_{(0,0)} = 1\, . 
\end{equation}

%%%%%%%%%%%%%%%%%%%%%%%%%%%
\item[(1,0)=3: fundamental] \,\\ 
%%%%%%%%%%%%%%%%%%%%%%%%%%%
Fundamental representation is $3\times 3$ special unitary matrix $g$ itself: $R_{(1,0)}(g)=g$. Hence the character is $\mathrm{Tr}g$, which is the sum of three eigenvalues $\lambda_1$, $\lambda_2$, and $\lambda_3$:
\begin{equation}
    \chi_{\rm fund.}
    \equiv
    \chi_{(1,0)}
    =
    \lambda_1 + \lambda_2 + \lambda_3\, . 
\end{equation}
%%%%%%%%%%%%%%%%%%%%%%%%%%%%%%%%%%%%%%%%%
\item[(1,1)=$\bar{\bold 3}$: rank-two anti-symmetric, or anti-fundamental] \,\\ 
%%%%%%%%%%%%%%%%%%%%%%%%%%%%%%%%%%%%%%%%%
The (1,1)-representation acts on the three-dimensional vector space spanned by $\vec{e}_{[a}\otimes\vec{e}_{b]}=\vec{e}_a\otimes\vec{e}_b-\vec{e}_b\otimes\vec{e}_a$, where $[\ \ ]$ stands for the anti-symmetrization. Note that $(a,b)=(1,2)$, $(2,3)$, or $(3,1)$. $R_{(1,1)}(g)$ is a $3\times 3$ matrix that acts on $\vec{e}_{[a}\otimes\vec{e}_{b]}$ as $R_{(1,1)}(g):\vec{e}_{[a}\otimes\vec{e}_{b]}\mapsto \sum_{a'b'}g_{aa'}g_{bb'}\vec{e}_{[a'}\otimes \vec{e}_{b']}$. $R_{(1,1)}(g)$ can be diagonalized by diagonalizing $g$. Specifically, for eigenvectors $\vec{v}_a$ that satisfy $g\vec{v}_a=\lambda_a\vec{v}_a$, we have $R_{(1,1)}(g):\vec{v}_{[a}\otimes\vec{v}_{b]}\mapsto \lambda_a\lambda_b\vec{v}_{[a}\otimes \vec{v}_{b]}$. Therefore, the eigenvalues are $\lambda_1\lambda_2=\lambda_3^{-1}$, $\lambda_2\lambda_3=\lambda_1^{-1}$, and $\lambda_3\lambda_1=\lambda_2^{-1}$. The character is  
\begin{equation}
    \chi_{\rm antisym.}
    \equiv
    \chi_{(1,1)}
    =
    \lambda_1\lambda_2 + \lambda_2\lambda_3 + \lambda_3\lambda_1
    =
    \lambda_1^{-1}
    +
    \lambda_2^{-1}
    +
    \lambda_3^{-1}
    =
    (\chi_{\rm fund.})^\ast\, . 
\end{equation}
%%%%%%%%%%%%%%%%%%%%%%
\item[(2,0)=6: rank-two symmetric] \,\\ 
%%%%%%%%%%%%%%%%%%%%%%
The (2,0)-representation acts on the three-dimensional vector space spanned by $\vec{e}_{(a}\otimes\vec{e}_{b)}=\vec{e}_a\otimes\vec{e}_b+\vec{e}_b\otimes\vec{e}_a$, where $(\ \ )$ stands for the symmetrization. Note that $(a,b)=(1,2)$, $(2,3)$, $(3,1)$, $(1,1)$, $(2,2)$, or $(3,3)$. $R_{(2,0)}(g)$ is a $6\times 6$ matrix that acts on $\vec{e}_{(a}\otimes\vec{e}_{b)}$ as $R_{(2,0)}(g):\vec{e}_{(a}\otimes\vec{e}_{b)}\mapsto \sum_{a'b'}g_{aa'}g_{bb'}\vec{e}_{(a'}\otimes \vec{e}_{b')}$. $R_{(2,0)}(g)$ can be diagonalized by diagonalizing $g$. Specifically, for eigenvectors $\vec{v}_a$ that satisfy $g\vec{v}_a=\lambda_a\vec{v}_a$, we have $R_{(2,0)}(g):\vec{v}_{(a}\otimes\vec{v}_{b)}\mapsto \lambda_a\lambda_b\vec{v}_{(a}\otimes \vec{v}_{b)}$. Therefore, the eigenvalues are $\lambda_1\lambda_2$, $\lambda_2\lambda_3$, $\lambda_3\lambda_1$, $\lambda_1^2$, $\lambda_2^2$, and $\lambda_3^2$. The character is 
\begin{equation}
    \chi_{(2,0)}
    =
    \lambda_1^2 + \lambda_2^2 + \lambda_3^2 + 
    \lambda_1\lambda_2 + \lambda_2\lambda_3 + \lambda_3\lambda_1\, . 
\end{equation}

%%%%%%%%%%%%%%%%%%%%%%%
\item[(2,1)=8: adjoint] \,\\ 
%%%%%%%%%%%%%%%%%%%%%%%
The adjoint representation acts on $3\times 3$ traceless Hermitian matrices. A convenient basis to calculate the eigenvalues consist of $\vec{v}_{(a}\otimes\vec{v}_{b)}{}^\ast (a\neq b)$, $i\vec{v}_{[a}\otimes\vec{v}_{b]}{}^\ast (a\neq b)$, $\vec{v}_{1}\otimes\vec{v}_{1}{}^\ast-\vec{v}_{2}\otimes\vec{v}_{2}{}^\ast$, and $\vec{v}_{2}\otimes\vec{v}_{2}{}^\ast-\vec{v}_{3}\otimes\vec{v}_{3}{}^\ast$. The latter two are eigenvectors with eigenvalue $1$. Although the former two are not eigenvectors, we can get eigenvectors $\vec{v}_a\otimes\vec{v}_b{}^\ast (a\neq b)$ with eigenvalue $\lambda_a\lambda_b^{-1}$ by taking linear combinations.

\begin{equation}
    \chi_{\rm adj.}
    \equiv
    \chi_{(2,1)}
    =
    2 + 
    \sum_{a\neq b}
    \lambda_a\lambda_b^{-1}\, . 
\end{equation}

\end{description}
We list the characters for several other representations:
\begin{description}
%%%%%%%%%%%%%%%%%%%%%%%%%%%%%%%%%%%%%
\item[(3,0)=10: rank-three symmetric] 
%%%%%%%%%%%%%%%%%%%%%%%%%%%%%%%%%%%%%
\begin{align}
\chi_{\rm 3\mathchar`-sym.}
\equiv
    \chi_{(3,0)}
    =
    1+
    \lambda_1^3 + \lambda_2^3 + \lambda_3^3
    +
    \lambda_1^2\lambda_2 + \lambda_1^2\lambda_3 + \lambda_2^2\lambda_1 + \lambda_2^2\lambda_3 + \lambda_3^2\lambda_1 + 
    \lambda_3^2\lambda_2
\end{align}

%%%%%%%%%%%%%%%%%%%%%%%%%%%%%%%%%%%%
\item[(4,0)=15: rank-four symmetric]
%%%%%%%%%%%%%%%%%%%%%%%%%%%%%%%%%%%%
\begin{align}
\chi_{\rm 4\mathchar`-sym.}
\equiv
    \chi_{(4,0)}
    = &
    \lambda_1^4 + \lambda_2^4 + \lambda_3^4 
    +
    \lambda_1^3\lambda_2 + \lambda_1^3\lambda_3 + \lambda_2^3\lambda_1 + \lambda_2^3\lambda_3 + \lambda_3^3\lambda_1 + \lambda_3^3\lambda_2 
    \notag\\
    &+
    \lambda_1^2\lambda_2^2 + \lambda_2^2\lambda_3^2 + \lambda_3^2\lambda_1^2 +
    \lambda_1 + \lambda_2 + \lambda_3
\end{align}

\item[(3,1)=${\bf 15}'$]
\begin{align}
    \chi_{(3,1)}
    =&
    \lambda_1^3\lambda_2 + \lambda_1^3\lambda_3 + \lambda_2^3\lambda_1 + \lambda_2^3\lambda_3 + \lambda_3^3\lambda_1 + \lambda_3^3\lambda_2 
    \notag\\
    &+
    \lambda_1^2\lambda_2^2 + \lambda_2^2\lambda_3^2 + \lambda_3^2\lambda_1^2 
    +
    2(\lambda_1 + \lambda_2 + \lambda_3)
\end{align}

\item[(5,0)=21]
\begin{align}
\chi_{\rm 5\mathchar`-sym.}
\equiv
    \chi_{(5,0)}
    =&
        \lambda_1^5 + \lambda_2^5 + \lambda_3^5 
    +
    \lambda_1^4\lambda_2 + \lambda_1^4\lambda_3 + \lambda_2^4\lambda_1 + \lambda_2^4\lambda_3 + \lambda_3^4\lambda_1 + \lambda_3^4\lambda_2 
    \notag\\
    &+
    \lambda_1^3\lambda_2^2 + \lambda_1^3\lambda_3^2 + \lambda_2^3\lambda_1^2 + \lambda_2^3\lambda_3^2 + \lambda_3^3\lambda_1^2 + \lambda_3^3\lambda_2^2
    \notag\\
    &+
    \lambda_1^2 + \lambda_2^2 + \lambda_3^2 
    +
    \lambda_1\lambda_2 + \lambda_2\lambda_3 + \lambda_3\lambda_1 
\end{align}

\item[(4,1)=24]
\begin{align}
    \chi_{(4,1)}
    =&
    \lambda_1^4\lambda_2 + \lambda_1^4\lambda_3 + \lambda_2^4\lambda_1 + \lambda_2^4\lambda_3 + \lambda_3^4\lambda_1 + \lambda_3^4\lambda_2 
    \notag\\
    &+
    \lambda_1^3\lambda_2^2 + \lambda_1^3\lambda_3^2 + \lambda_2^3\lambda_1^2 + \lambda_2^3\lambda_3^2 + \lambda_3^3\lambda_1^2 + \lambda_3^3\lambda_2^2
    \notag\\
    &+
    2(\lambda_1^2 + \lambda_2^2 + \lambda_3^2 
    +
    \lambda_1\lambda_2 + \lambda_2\lambda_3 + \lambda_3\lambda_1 )
\end{align}

\item[(6,0)=28]
\begin{align}
\chi_{\rm 6\mathchar`-sym.}
\equiv
    \chi_{(6,0)}
    =&
    \lambda_1^6 + \lambda_2^6 + \lambda_3^6 
    +
    \lambda_1^5\lambda_2 + \lambda_1^5\lambda_3 + \lambda_2^5\lambda_1 + \lambda_2^5\lambda_3 + \lambda_3^5\lambda_1 + \lambda_3^5\lambda_2 
    \notag\\
    &+
    \lambda_1^4\lambda_2^2 + \lambda_1^4\lambda_3^2 + \lambda_2^4\lambda_1^2 + \lambda_2^4\lambda_3^2 + \lambda_3^4\lambda_1^2 + \lambda_3^4\lambda_2^2
    +
    \lambda_1^3\lambda_2^3 + \lambda_2^3\lambda_3^3 + \lambda_3^3\lambda_1^3
    \notag\\
    &+
    \lambda_1^3 + \lambda_2^3 + \lambda_3^3 
    +
    \lambda_1^2\lambda_2 + \lambda_1^2\lambda_3 + \lambda_2^2\lambda_1 + \lambda_2^2\lambda_3 + \lambda_3^2\lambda_1 + \lambda_3^2\lambda_2 
    + 
    1
\end{align}

\item[(5,1)=35]
\begin{align}
    \chi_{(5,1)}
    =&
    \lambda_1^5\lambda_2 + \lambda_1^5\lambda_3 + \lambda_2^5\lambda_1 + \lambda_2^5\lambda_3 + \lambda_3^5\lambda_1 + \lambda_3^5\lambda_2 
    \notag\\
    &+
    \lambda_1^4\lambda_2^2 + \lambda_1^4\lambda_3^2 + \lambda_2^4\lambda_1^2 + \lambda_2^4\lambda_3^2 + \lambda_3^4\lambda_1^2 + \lambda_3^4\lambda_2^2
    +
    \lambda_1^3\lambda_2^3 + \lambda_2^3\lambda_3^3 + \lambda_3^3\lambda_1^3
    \notag\\
    &+
    2(\lambda_1^3 + \lambda_2^3 + \lambda_3^3 
    +
    \lambda_1^2\lambda_2 + \lambda_1^2\lambda_3 + \lambda_2^2\lambda_1 + \lambda_2^2\lambda_3 + \lambda_3^2\lambda_1 + \lambda_3^2\lambda_2 
    + 
    1)
\end{align}

\end{description}
%%%%%%%%%%%%%%%%%%%%%%%%%%%%%%%%%%%%%
%%%%%%%%%%%%%%%%%%%%%%%%%%%%%%%%%%%%%
\subsection{Multiply-wound Polyakov loops and characters}\label{sec:winding-loop-vs-character}
%%%%%%%%%%%%%%%%%%%%%%%%%%%%%%%%%%%%%
%%%%%%%%%%%%%%%%%%%%%%%%%%%%%%%%%%%%%
Characters in the irreducible representation play significant roles as the basis of functions on the group manifold $G$. To see how the characters are related to the distribution of Polyakov line phases, we express the multiply-wound Polyakov loops in terms of characters. The multiply-wound loops are expressed as
\begin{align}
    u_n 
    \equiv
    \frac{1}{N}\mathrm{Tr}\mathcal{P}^n
    = 
    \frac{1}{N}\sum_{j=1}^N \lambda_j^n\, . 
    %=
    %\sum_r u_r^{(n)} \chi_r(\{\lambda\}).
\end{align}
For the $N=3$ case, we can check
\begin{align}
    3 u_1 &= \chi_{(1,0)} = \chi_{\rm fund.}\, ,
    \\
    3 u_2 
    &= 
    \chi_{(2,0)} - \chi_{(1,1)} = \chi_{\rm 2\mathchar`-sym.} - (\chi_{\rm fund.})^\ast\, ,
    \\
    3 u_3
    &=
    \chi_{(3,0)} - \chi_{(2,1)} + \chi_{(0,0)}
    =\chi_{\rm 3\mathchar`-sym.} - \chi_{\rm adj.}+ \chi_{\rm trivial}\, ,
    \\
    3 u_4
    &=
    \chi_{(4,0)} - \chi_{(3,1)} + \chi_{(1,0)}\, ,
    \\
    3 u_5
    &=
    \chi_{(5,0)} - \chi_{(4,1)} + \chi_{(2,0)}\, ,
    \\
    3 u_6
    &=
    \chi_{(6,0)} - \chi_{(5,1)} + \chi_{(3,0)}\, ,
    \\
  & \qquad\qquad\quad \vdots
   \nonumber
\end{align}
Note that only $u_3$ contains $\chi_\mathrm{trivial}$. We can see it from a relation 
\begin{align}
    \frac{1}{\mathrm{Vol}(G)} \int_G dg\, u_n(g) = 
    \frac{1}{\mathrm{Vol}(G)} \int_G d g\, \chi_\mathrm{trivial}(g)u_n(g) =
    \frac{\delta_{n,3}}{3}\, 
\end{align}
which immediately follows from the explicit form of the Haar-random distribution \eqref{eq:appendix:Haar-random} for $N=3$. Due to the orthogonality condition \eqref{eq:orthogonality}, this means only $u_3$ contains $\chi_\mathrm{trivial}$. 
In the same way, we can see that $u_N$ contains $\chi_\mathrm{trivial}$ for the SU($N$) theory, assuming \eqref{eq:appendix:Haar-random} is correct. 

Nonzero expectation values of $\chi_r$ for nontrivial representations $r$ characterize the deviation from the SU(3) Haar-random distribution.
We emphasize that the Haar randomness provides stronger restriction than that by the $\mathbb{Z}_3$ center symmetry: 
center symmetry allows nonzero values of $u_{3n}$.
The Haar randomness, however, leads to $\chi_\mathrm{trivial} = 1$ and $\chi_r = 0$ for any other representation $r$, which allows only $u_3$ to be nonzero.

%%%%%%%%%%%%%
%\bibliographystyle{unsrt}
\bibliographystyle{utphys}
\bibliography{PD-review}

\providecommand{\href}[2]{#2}\begingroup\raggedright\begin{thebibliography}{10}

\bibitem{Polyakov:1978vu}
A.~M. Polyakov, ``{Thermal Properties of Gauge Fields and Quark Liberation},''
  \href{http://dx.doi.org/10.1016/0370-2693(78)90737-2}{{\em Phys. Lett. B}
  {\bfseries 72} (1978) 477--480}.

\bibitem{Susskind:1979up}
L.~Susskind, ``{Lattice Models of Quark Confinement at High Temperature},''
  \href{http://dx.doi.org/10.1103/PhysRevD.20.2610}{{\em Phys. Rev. D}
  {\bfseries 20} (1979) 2610--2618}.

\bibitem{Aoki:2006we}
Y.~Aoki, G.~Endrodi, Z.~Fodor, S.~Katz, and K.~Szabo, ``{The Order of the
  quantum chromodynamics transition predicted by the standard model of particle
  physics},'' \href{http://dx.doi.org/10.1038/nature05120}{{\em Nature}
  {\bfseries 443} (2006) 675--678},
  \href{http://arxiv.org/abs/hep-lat/0611014}{{\ttfamily
  arXiv:hep-lat/0611014}}.

\bibitem{Aharony:2003sx}
O.~Aharony, J.~Marsano, S.~Minwalla, K.~Papadodimas, and M.~Van~Raamsdonk,
  ``{The Hagedorn - deconfinement phase transition in weakly coupled large N
  gauge theories},'' \href{http://dx.doi.org/10.4310/ATMP.2004.v8.n4.a1}{{\em
  Adv. Theor. Math. Phys.} {\bfseries 8} (2004) 603--696},
  \href{http://arxiv.org/abs/hep-th/0310285}{{\ttfamily arXiv:hep-th/0310285}}.

\bibitem{Sundborg:1999ue}
B.~Sundborg, ``{The Hagedorn transition, deconfinement and N=4 SYM theory},''
  \href{http://dx.doi.org/10.1016/S0550-3213(00)00044-4}{{\em Nucl. Phys. B}
  {\bfseries 573} (2000) 349--363},
  \href{http://arxiv.org/abs/hep-th/9908001}{{\ttfamily arXiv:hep-th/9908001}}.

\bibitem{Schnitzer:2004qt}
H.~J. Schnitzer, ``{Confinement/deconfinement transition of large N gauge
  theories with N(f) fundamentals: N(f)/N finite},''
  \href{http://dx.doi.org/10.1016/j.nuclphysb.2004.06.057}{{\em Nucl. Phys. B}
  {\bfseries 695} (2004) 267--282},
  \href{http://arxiv.org/abs/hep-th/0402219}{{\ttfamily arXiv:hep-th/0402219}}.

\bibitem{Hanada:2019rzv}
M.~Hanada, G.~Ishiki, and H.~Watanabe, ``{Partial deconfinement in gauge
  theories},'' \href{http://dx.doi.org/10.22323/1.363.0055}{{\em PoS}
  {\bfseries LATTICE2019} (2019) 055},
  \href{http://arxiv.org/abs/1911.11465}{{\ttfamily arXiv:1911.11465
  [hep-lat]}}.

\bibitem{Hanada:2016pwv}
M.~Hanada and J.~Maltz, ``{A proposal of the gauge theory description of the
  small Schwarzschild black hole in AdS$_5\times$S$^5$},''
  \href{http://dx.doi.org/10.1007/JHEP02(2017)012}{{\em JHEP} {\bfseries 02}
  (2017) 012}, \href{http://arxiv.org/abs/1608.03276}{{\ttfamily
  arXiv:1608.03276 [hep-th]}}.

\bibitem{Berenstein:2018lrm}
D.~Berenstein, ``{Submatrix deconfinement and small black holes in AdS},''
  \href{http://dx.doi.org/10.1007/JHEP09(2018)054}{{\em JHEP} {\bfseries 09}
  (2018) 054}, \href{http://arxiv.org/abs/1806.05729}{{\ttfamily
  arXiv:1806.05729 [hep-th]}}.

\bibitem{Hanada:2018zxn}
M.~Hanada, G.~Ishiki, and H.~Watanabe, ``{Partial Deconfinement},''
  \href{http://dx.doi.org/10.1007/JHEP03(2019)145}{{\em JHEP} {\bfseries 03}
  (2019) 145}, \href{http://arxiv.org/abs/1812.05494}{{\ttfamily
  arXiv:1812.05494 [hep-th]}}. [Erratum: JHEP 10, 029 (2019)].

\bibitem{Hanada:2019czd}
M.~Hanada, A.~Jevicki, C.~Peng, and N.~Wintergerst, ``{Anatomy of
  Deconfinement},'' \href{http://dx.doi.org/10.1007/JHEP12(2019)167}{{\em JHEP}
  {\bfseries 12} (2019) 167}, \href{http://arxiv.org/abs/1909.09118}{{\ttfamily
  arXiv:1909.09118 [hep-th]}}.

\bibitem{Hanada:2019kue}
M.~Hanada and B.~Robinson, ``{Partial-Symmetry-Breaking Phase Transitions},''
  \href{http://dx.doi.org/10.1103/PhysRevD.102.096013}{{\em Phys. Rev. D}
  {\bfseries 102} no.~9, (2020) 096013},
  \href{http://arxiv.org/abs/1911.06223}{{\ttfamily arXiv:1911.06223
  [hep-th]}}.

\bibitem{Hanada:2021ksu}
M.~Hanada, J.~Holden, M.~Knaggs, and A.~O'Bannon, ``{Global symmetries and
  partial confinement},'' \href{http://dx.doi.org/10.1007/JHEP03(2022)118}{{\em
  JHEP} {\bfseries 03} (2022) 118},
  \href{http://arxiv.org/abs/2112.11398}{{\ttfamily arXiv:2112.11398
  [hep-th]}}.

\bibitem{Bergner:2019rca}
G.~Bergner, N.~Bodendorfer, M.~Hanada, E.~Rinaldi, A.~Sch\"afer, and P.~Vranas,
  ``{Thermal phase transition in Yang-Mills matrix model},''
  \href{http://dx.doi.org/10.1007/JHEP01(2020)053}{{\em JHEP} {\bfseries 01}
  (2020) 053}, \href{http://arxiv.org/abs/1909.04592}{{\ttfamily
  arXiv:1909.04592 [hep-th]}}.

\bibitem{Watanabe:2020ufk}
H.~Watanabe, G.~Bergner, N.~Bodendorfer, S.~Shiba~Funai, M.~Hanada, E.~Rinaldi,
  A.~Sch\"afer, and P.~Vranas, ``{Partial deconfinement at strong coupling on
  the lattice},'' \href{http://dx.doi.org/10.1007/JHEP02(2021)004}{{\em JHEP}
  {\bfseries 02} (2021) 004}, \href{http://arxiv.org/abs/2005.04103}{{\ttfamily
  arXiv:2005.04103 [hep-th]}}.

\bibitem{Gautam:2022exf}
V.~Gautam, M.~Hanada, J.~Holden, and E.~Rinaldi, ``{Linear confinement in the
  partially-deconfined phase},''
  \href{http://dx.doi.org/10.1007/JHEP03(2023)195}{{\em JHEP} {\bfseries 03}
  (2023) 195}, \href{http://arxiv.org/abs/2208.14402}{{\ttfamily
  arXiv:2208.14402 [hep-th]}}.

\bibitem{Hanada:2020uvt}
M.~Hanada, H.~Shimada, and N.~Wintergerst, ``{Color confinement and
  Bose-Einstein condensation},''
  \href{http://dx.doi.org/10.1007/JHEP08(2021)039}{{\em JHEP} {\bfseries 08}
  (2021) 039}, \href{http://arxiv.org/abs/2001.10459}{{\ttfamily
  arXiv:2001.10459 [hep-th]}}.

\bibitem{Hanada:2021ipb}
M.~Hanada, ``{Bulk geometry in gauge/gravity duality and color degrees of
  freedom},'' \href{http://dx.doi.org/10.1103/PhysRevD.103.106007}{{\em Phys.
  Rev. D} {\bfseries 103} no.~10, (2021) 106007},
  \href{http://arxiv.org/abs/2102.08982}{{\ttfamily arXiv:2102.08982
  [hep-th]}}.

\bibitem{Hanada:2021swb}
M.~Hanada, ``{Large-N limit as a second quantization},''
  \href{http://dx.doi.org/10.22323/1.406.0260}{{\em PoS} {\bfseries CORFU2021}
  (2022) 260}, \href{http://arxiv.org/abs/2103.15873}{{\ttfamily
  arXiv:2103.15873 [hep-th]}}.

\bibitem{Gross:1980he}
D.~J. Gross and E.~Witten, ``{Possible Third Order Phase Transition in the
  Large N Lattice Gauge Theory},''
\href{http://dx.doi.org/10.1103/PhysRevD.21.446}{{\em Phys. Rev.} {\bfseries
  D21} (1980) 446--453}.
%%CITATION = PHRVA,D21,446;%%.

\bibitem{Wadia:2012fr}
S.~R. Wadia, ``{A Study of U(N) Lattice Gauge Theory in 2-dimensions},''
\href{http://arxiv.org/abs/1212.2906}{{\ttfamily arXiv:1212.2906 [hep-th]}}.
%%CITATION = ARXIV:1212.2906;%%.

\bibitem{Hanada:2023krw}
M.~Hanada, H.~Ohata, H.~Shimada, and H.~Watanabe, ``{A new perspective on
  thermal transition in QCD},''
  \href{http://arxiv.org/abs/2310.01940}{{\ttfamily arXiv:2310.01940
  [hep-th]}}.

\bibitem{Gautam:2022akq}
V.~Gautam, M.~Hanada, A.~Jevicki, and C.~Peng, ``{Matrix entanglement},''
  \href{http://dx.doi.org/10.1007/JHEP01(2023)003}{{\em JHEP} {\bfseries 01}
  (2023) 003}, \href{http://arxiv.org/abs/2204.06472}{{\ttfamily
  arXiv:2204.06472 [hep-th]}}.

\bibitem{Furuuchi:2003sy}
K.~Furuuchi, E.~Schreiber, and G.~W. Semenoff, ``{Five-brane thermodynamics
  from the matrix model},''
  \href{http://arxiv.org/abs/hep-th/0310286}{{\ttfamily arXiv:hep-th/0310286}}.

\bibitem{Kawahara:2006hs}
N.~Kawahara, J.~Nishimura, and K.~Yoshida, ``{Dynamical aspects of the
  plane-wave matrix model at finite temperature},''
  \href{http://dx.doi.org/10.1088/1126-6708/2006/06/052}{{\em JHEP} {\bfseries
  06} (2006) 052}, \href{http://arxiv.org/abs/hep-th/0601170}{{\ttfamily
  arXiv:hep-th/0601170}}.

\bibitem{Kogut:1974ag}
J.~B. Kogut and L.~Susskind, ``{Hamiltonian Formulation of Wilson's Lattice
  Gauge Theories},'' \href{http://dx.doi.org/10.1103/PhysRevD.11.395}{{\em
  Phys. Rev. D} {\bfseries 11} (1975) 395--408}.

\bibitem{Gupta:2007ax}
S.~Gupta, K.~Huebner, and O.~Kaczmarek, ``{Renormalized Polyakov loops in many
  representations},'' \href{http://dx.doi.org/10.1103/PhysRevD.77.034503}{{\em
  Phys. Rev. D} {\bfseries 77} (2008) 034503},
  \href{http://arxiv.org/abs/0711.2251}{{\ttfamily arXiv:0711.2251 [hep-lat]}}.

\bibitem{Mykkanen:2012ri}
A.~Mykkanen, M.~Panero, and K.~Rummukainen, ``{Casimir scaling and
  renormalization of Polyakov loops in large-N gauge theories},''
  \href{http://dx.doi.org/10.1007/JHEP05(2012)069}{{\em JHEP} {\bfseries 05}
  (2012) 069}, \href{http://arxiv.org/abs/1202.2762}{{\ttfamily arXiv:1202.2762
  [hep-lat]}}.

\bibitem{Datta:2015bzm}
S.~Datta, S.~Gupta, and A.~Lytle, ``{Using Wilson flow to study the SU(3)
  deconfinement transition},''
  \href{http://dx.doi.org/10.1103/PhysRevD.94.094502}{{\em Phys. Rev. D}
  {\bfseries 94} no.~9, (2016) 094502},
  \href{http://arxiv.org/abs/1512.04892}{{\ttfamily arXiv:1512.04892
  [hep-lat]}}.

\bibitem{Petreczky:2015yta}
P.~Petreczky and H.~P. Schadler, ``{Renormalization of the Polyakov loop with
  gradient flow},'' \href{http://dx.doi.org/10.1103/PhysRevD.92.094517}{{\em
  Phys. Rev. D} {\bfseries 92} no.~9, (2015) 094517},
  \href{http://arxiv.org/abs/1509.07874}{{\ttfamily arXiv:1509.07874
  [hep-lat]}}.

\bibitem{Vafa:1983tf}
C.~Vafa and E.~Witten, ``{Restrictions on Symmetry Breaking in Vector-Like
  Gauge Theories},'' \href{http://dx.doi.org/10.1016/0550-3213(84)90230-X}{{\em
  Nucl. Phys. B} {\bfseries 234} (1984) 173--188}.

\bibitem{tHooft:1979rat}
G.~'t~Hooft, ``{Naturalness, chiral symmetry, and spontaneous chiral symmetry
  breaking},'' \href{http://dx.doi.org/10.1007/978-1-4684-7571-5_9}{{\em NATO
  Sci. Ser. B} {\bfseries 59} (1980) 135--157}.

\bibitem{Gaiotto:2017yup}
D.~Gaiotto, A.~Kapustin, Z.~Komargodski, and N.~Seiberg, ``{Theta, Time
  Reversal, and Temperature},''
  \href{http://dx.doi.org/10.1007/JHEP05(2017)091}{{\em JHEP} {\bfseries 05}
  (2017) 091}, \href{http://arxiv.org/abs/1703.00501}{{\ttfamily
  arXiv:1703.00501 [hep-th]}}.

\bibitem{Chen:2020syd}
S.~Chen, K.~Fukushima, H.~Nishimura, and Y.~Tanizaki, ``{Deconfinement and
  $\mathcal {CP}$ breaking at $\theta=\pi$ in Yang-Mills theories and a novel
  phase for SU(2)},'' \href{http://dx.doi.org/10.1103/PhysRevD.102.034020}{{\em
  Phys. Rev. D} {\bfseries 102} no.~3, (2020) 034020},
  \href{http://arxiv.org/abs/2006.01487}{{\ttfamily arXiv:2006.01487
  [hep-th]}}.

\bibitem{Choi:2021lbk}
S.~Choi, S.~Jeong, and S.~Kim, ``{The Yang-Mills duals of small AdS black
  holes},'' \href{http://arxiv.org/abs/2103.01401}{{\ttfamily arXiv:2103.01401
  [hep-th]}}.

\bibitem{Buividovich:2015oju}
P.~V. Buividovich, G.~V. Dunne, and S.~N. Valgushev, ``{Complex Path Integrals
  and Saddles in Two-Dimensional Gauge Theory},''
  \href{http://dx.doi.org/10.1103/PhysRevLett.116.132001}{{\em Phys. Rev.
  Lett.} {\bfseries 116} no.~13, (2016) 132001},
  \href{http://arxiv.org/abs/1512.09021}{{\ttfamily arXiv:1512.09021
  [hep-th]}}.

\bibitem{Umeda:2012er}
{\bfseries WHOT-QCD} Collaboration, T.~Umeda, S.~Aoki, S.~Ejiri, T.~Hatsuda,
  K.~Kanaya, Y.~Maezawa, and H.~Ohno, ``{Equation of state in 2+1 flavor QCD
  with improved Wilson quarks by the fixed scale approach},''
  \href{http://dx.doi.org/10.1103/PhysRevD.85.094508}{{\em Phys. Rev. D}
  {\bfseries 85} (2012) 094508},
  \href{http://arxiv.org/abs/1202.4719}{{\ttfamily arXiv:1202.4719 [hep-lat]}}.

\bibitem{Sheikholeslami:1985ij}
B.~Sheikholeslami and R.~Wohlert, ``{Improved Continuum Limit Lattice Action
  for QCD with Wilson Fermions},''
  \href{http://dx.doi.org/10.1016/0550-3213(85)90002-1}{{\em Nucl. Phys. B}
  {\bfseries 259} (1985) 572}.

\bibitem{Iwasaki:1983iya}
Y.~Iwasaki, ``{Renormalization Group Analysis of Lattice Theories and Improved
  Lattice Action. II. Four-dimensional non-Abelian SU(N) gauge model},''
  \href{http://arxiv.org/abs/1111.7054}{{\ttfamily arXiv:1111.7054 [hep-lat]}}.

\bibitem{Taniguchi:2016tjc}
Y.~Taniguchi, K.~Kanaya, H.~Suzuki, and T.~Umeda, ``{Topological susceptibility
  in finite temperature ( 2+1 )-flavor QCD using gradient flow},''
  \href{http://dx.doi.org/10.1103/PhysRevD.95.054502}{{\em Phys. Rev. D}
  {\bfseries 95} no.~5, (2017) 054502},
  \href{http://arxiv.org/abs/1611.02411}{{\ttfamily arXiv:1611.02411
  [hep-lat]}}.

\bibitem{Luscher:2010iy}
M.~L\"uscher, ``{Properties and uses of the Wilson flow in lattice QCD},''
  \href{http://dx.doi.org/10.1007/JHEP08(2010)071}{{\em JHEP} {\bfseries 08}
  (2010) 071}, \href{http://arxiv.org/abs/1006.4518}{{\ttfamily arXiv:1006.4518
  [hep-lat]}}. [Erratum: JHEP 03, 092 (2014)].

\bibitem{Asakawa:1995zu}
M.~Asakawa and T.~Hatsuda, ``{What thermodynamics tells about QCD plasma near
  phase transition},'' \href{http://dx.doi.org/10.1103/PhysRevD.55.4488}{{\em
  Phys. Rev. D} {\bfseries 55} (1997) 4488--4491},
  \href{http://arxiv.org/abs/hep-ph/9508360}{{\ttfamily arXiv:hep-ph/9508360}}.

\bibitem{Glozman:2022zpy}
L.~Y. Glozman, ``{Chiral spin symmetry and hot/dense QCD},''
  \href{http://dx.doi.org/10.1016/j.ppnp.2023.104049}{{\em Prog. Part. Nucl.
  Phys.} {\bfseries 131} (2023) 104049},
  \href{http://arxiv.org/abs/2209.10235}{{\ttfamily arXiv:2209.10235
  [hep-lat]}}.

\bibitem{Cohen:2023hbq}
T.~D. Cohen and L.~Y. Glozman, ``{Large $N_c$ QCD phase diagram at $\mu_B =
  0$},'' \href{http://arxiv.org/abs/2311.07333}{{\ttfamily arXiv:2311.07333
  [hep-ph]}}.

\bibitem{Bergner:2023rpw}
G.~Bergner, V.~Gautam, and M.~Hanada, ``{Color Confinement and Random Matrices
  -- A random walk down group manifold toward Casimir scaling --},''
  \href{http://arxiv.org/abs/2311.14093}{{\ttfamily arXiv:2311.14093
  [hep-th]}}.

\bibitem{Gross:1993hu}
D.~J. Gross and W.~Taylor, ``{Two-dimensional QCD is a string theory},''
  \href{http://dx.doi.org/10.1016/0550-3213(93)90403-C}{{\em Nucl. Phys. B}
  {\bfseries 400} (1993) 181--208},
  \href{http://arxiv.org/abs/hep-th/9301068}{{\ttfamily arXiv:hep-th/9301068}}.

\bibitem{Gross:1993yt}
D.~J. Gross and W.~Taylor, ``{Twists and Wilson loops in the string theory of
  two-dimensional QCD},''
  \href{http://dx.doi.org/10.1016/0550-3213(93)90042-N}{{\em Nucl. Phys. B}
  {\bfseries 403} (1993) 395--452},
  \href{http://arxiv.org/abs/hep-th/9303046}{{\ttfamily arXiv:hep-th/9303046}}.

\bibitem{Berenstein:2023srv}
D.~Berenstein and K.~Yan, ``{The endpoint of partial deconfinement},''
  \href{http://dx.doi.org/10.1007/JHEP12(2023)030}{{\em JHEP} {\bfseries 12}
  (2023) 030}, \href{http://arxiv.org/abs/2307.06122}{{\ttfamily
  arXiv:2307.06122 [hep-th]}}.

\bibitem{Yamada:2022imq}
M.~Yamada and K.~Yonekura, ``{Cosmic strings from pure Yang\textendash{}Mills
  theory},'' \href{http://dx.doi.org/10.1103/PhysRevD.106.123515}{{\em Phys.
  Rev. D} {\bfseries 106} no.~12, (2022) 123515},
  \href{http://arxiv.org/abs/2204.13123}{{\ttfamily arXiv:2204.13123
  [hep-th]}}.

\bibitem{Fujikura:2023fbi}
K.~Fujikura, Y.~Nakai, R.~Sato, and Y.~Wang, ``{Cosmological phase transitions
  in composite Higgs models},''
  \href{http://dx.doi.org/10.1007/JHEP09(2023)053}{{\em JHEP} {\bfseries 09}
  (2023) 053}, \href{http://arxiv.org/abs/2306.01305}{{\ttfamily
  arXiv:2306.01305 [hep-ph]}}.

\bibitem{Maldacena:1997re}
J.~M. Maldacena, ``{The Large N limit of superconformal field theories and
  supergravity},'' \href{http://dx.doi.org/10.1023/A:1026654312961,
  10.4310/ATMP.1998.v2.n2.a1}{{\em Int. J. Theor. Phys.} {\bfseries 38} (1999)
  1113--1133}, \href{http://arxiv.org/abs/hep-th/9711200}{{\ttfamily
  arXiv:hep-th/9711200 [hep-th]}}.
[Adv. Theor. Math. Phys.2,231(1998)].
%%CITATION = HEP-TH/9711200;%%.

\bibitem{Gubser:1998bc}
S.~S. Gubser, I.~R. Klebanov, and A.~M. Polyakov, ``{Gauge theory correlators
  from noncritical string theory},''
  \href{http://dx.doi.org/10.1016/S0370-2693(98)00377-3}{{\em Phys. Lett. B}
  {\bfseries 428} (1998) 105--114},
  \href{http://arxiv.org/abs/hep-th/9802109}{{\ttfamily arXiv:hep-th/9802109}}.

\bibitem{Witten:1998qj}
E.~Witten, ``{Anti-de Sitter space and holography},''
  \href{http://dx.doi.org/10.4310/ATMP.1998.v2.n2.a2}{{\em Adv. Theor. Math.
  Phys.} {\bfseries 2} (1998) 253--291},
  \href{http://arxiv.org/abs/hep-th/9802150}{{\ttfamily arXiv:hep-th/9802150}}.

\bibitem{Aharony:1999ti}
O.~Aharony, S.~S. Gubser, J.~M. Maldacena, H.~Ooguri, and Y.~Oz, ``{Large N
  field theories, string theory and gravity},''
  \href{http://dx.doi.org/10.1016/S0370-1573(99)00083-6}{{\em Phys. Rept.}
  {\bfseries 323} (2000) 183--386},
\href{http://arxiv.org/abs/hep-th/9905111}{{\ttfamily arXiv:hep-th/9905111
  [hep-th]}}.
%%CITATION = HEP-TH/9905111;%%.

\bibitem{Rinaldi:2021jbg}
E.~Rinaldi, X.~Han, M.~Hassan, Y.~Feng, F.~Nori, M.~McGuigan, and M.~Hanada,
  ``{Matrix-Model Simulations Using Quantum Computing, Deep Learning, and
  Lattice Monte Carlo},''
  \href{http://dx.doi.org/10.1103/PRXQuantum.3.010324}{{\em PRX Quantum}
  {\bfseries 3} no.~1, (2022) 010324},
  \href{http://arxiv.org/abs/2108.02942}{{\ttfamily arXiv:2108.02942
  [quant-ph]}}.

\bibitem{Akemann:2011csh}
G.~Akemann, J.~Baik, and P.~Di~Francesco, {\em {The Oxford Handbook of Random
  Matrix Theory}}.
\newblock Oxford Handbooks in Mathematics. Oxford University Press, 9, 2011.

\bibitem{Nishigaki:2024phx}
S.~Nishigaki, ``{Eigenphase distributions of unimodular circular ensembles},''
  \href{http://arxiv.org/abs/2401.09045}{{\ttfamily arXiv:2401.09045
  [math-ph]}}.

\bibitem{Fasi2020SamplingTE}
M.~Fasi and L.~Robol, ``Sampling the eigenvalues of random orthogonal and
  unitary matrices,'' {\em ArXiv} {\bfseries abs/2009.11515} (2020) .
  \url{https://api.semanticscholar.org/CorpusID:221879010}.

\bibitem{Hanada-Matsuura}
M.~Hanada and S.~Matsuura, {\em {MCMC from scratch --- a practical introduction
  to Markov Chain Monte Carlo method}}.
\newblock Springer, 2022.

\end{thebibliography}\endgroup

\end{document}